\documentclass[aip,jcp,reprint,superscriptaddress,footinbib,floatfix]{revtex4-2}

\usepackage{graphicx}
\usepackage{bm}
\usepackage{amssymb}
\usepackage{amsbsy}
\usepackage{amsmath}
\usepackage{mathtools}
\usepackage{xcolor}
\usepackage{stmaryrd}
\usepackage{mathrsfs}
\usepackage{tikz}
\usetikzlibrary{shapes}
\usepackage{cancel}
\usepackage[normalem]{ulem}
\usepackage{hyperref}

\makeatletter
\let\cat@comma@active\@empty
\makeatother

\makeatletter 
\gdef\@ptsize{2}
\let\@currsize\normalsize 
\makeatother 

\begin{document}
\title{Universal scaling of the diffusivity of dendrimers in a semidilute solution of linear polymers}
\author{Silpa Mariya}
\email{silpa.mariya@monash.edu}
\affiliation{IITB-Monash Research Academy, Indian Institute of Technology Bombay, Mumbai, 400076, India.}
\affiliation{Department of Chemical Engineering,
  Indian Institute of Technology Bombay, Mumbai, 400076, India.}
\affiliation{Department of Chemical Engineering, Monash University,
Melbourne, VIC 3800, Australia}
\author{Jeremy J. Barr}
\email{jeremy.barr@monash.edu}
\affiliation{School of Biological Sciences, Monash University, Clayton, VIC 3800, Australia.}
\author{P. Sunthar}
\email{p.sunthar@iitb.ac.in}
\affiliation{Department of Chemical Engineering,
  Indian Institute of Technology Bombay, Mumbai, 400076, India.}
\author{J. Ravi Prakash}
\email{ravi.jagadeeshan@monash.edu}
\affiliation{Department of Chemical Engineering, Monash University,
Melbourne, VIC 3800, Australia}

\begin{abstract}
The static and dynamic properties of dendrimers in semidilute solutions of linear chains of comparable size are investigated using Brownian dynamics simulations. The radius of gyration and diffusivity of a wide variety of low generation dendrimers and linear chains in solution follow universal scaling laws independent of their topology. Analysis of the shape functions and internal density of dendrimers shows that they are more spherical than linear chains and have a dense core. At intermediate times, dendrimers become subdiffusive, with an exponent higher than that previously reported for nanoparticles in semidilute polymer solutions. The long-time diffusivity of dendrimers does not follow theoretical predictions for nanoparticles. We propose a new scaling law for the long-time diffusion coefficients of dendrimers which accounts for the fact that, unlike nanoparticles, dendrimers shrink with an increase in background solution concentration. Analysis of the properties of a special case of a higher functionality dendrimer shows a transition from polymer-like to nanoparticle-like behaviour.
\end{abstract}

\maketitle

\section{\label{sec:intro} Introduction}

\begin{figure*}[pth!]
	\begin{center}
		\begin{tabular}{ccc}
			\resizebox{7cm}{!} {\includegraphics[width=9cm, height=7cm]{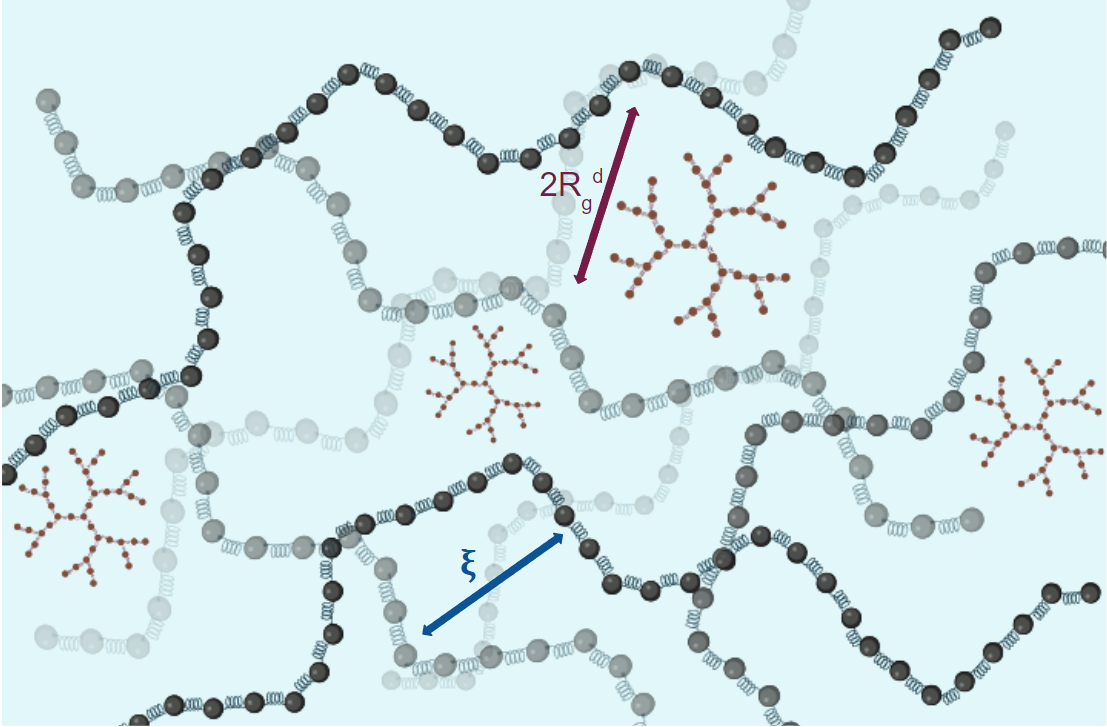}} &	
			&	\hspace{40pt}	
			\resizebox{5.9cm}{!} {\includegraphics[width=3.0cm]{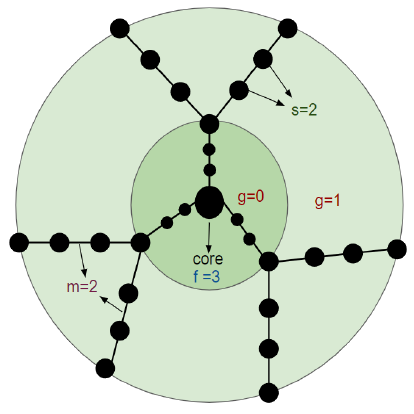}}\\
			(a) & \hspace{30pt}	& (b)  
		\end{tabular}
	\end{center}
        \caption{ (Color online) (a) Schematic representation of dendrimers in a solution of linear chains. The correlation length of the solution ($\xi$) and the size of dendrimer($2R_{\textrm{g}}^{\textrm{d}}$) are indicated. (b) A generation one ($g=1$) dendrimer with functionality three ($f=3$) and two spacer beads ($s=2$). The order of dendra ($m=f-1$) is two ($m=2$). Beads belonging to each generation are included in a concentric circle.} 
    \label{fig:dendrimer}
\end{figure*}

The movement of tracer particles through crowded environments is an area of active research due to its applications in medicine and nanotechnology~\cite{amblard1996subdiffusion,alai2015application,salata2004applications,de2008applications,gomez2016dynamics,patteson2015running}. Most research till date has focused on employing rigid nanoparticles as the tracer particle~\cite{stark2015industrial,kesharwani2018nanotechnology,aghebati2020nanoparticles,mitchell2021engineering}. However, recent biotechnological advances have enabled the use of dendrimers as drug delivery agents~\cite{liu1999designing,gillies2005dendrimers}. Polymers of varying topology and soft colloids have also been used in biology where they diffuse in complex environments such as semidilute polymer solutions and networks of polymer chains~\cite{baghbanbashi2022polymersomes,lee2005designing}. The movement of such tracers through crowded media has not been studied extensively. In this work, we use dendrimers as prototypical soft colloids, and examine their structural and dynamic properties when dissolved in a semidilute unentangled polymer solution shown schematically in Fig.~\ref{fig:dendrimer}(a). By systematically varying the dendrimer architecture and the concentration of the background solution, the similarities and differences with previously reported behaviour of rigid nanoparticles in a similar environment are studied.

Dendrimers are branched polymeric macromolecules with a tree-like structure and a central core monomer with branches emanating from it. The simplest case of a generation zero dendrimer is a star polymer with a core and $f$ linear chains attached to it. Subsequent generations are built by successively adding layers of short linear chains with multifunctional units to previous generations. The size of dendrimers can be tuned using four parameters: functionality ($f$), generation number ($g$), spacer length ($s$), and the order of dendra ($m$). Fig. ~\ref{fig:dendrimer}(b) shows a first generation ($g=1$) symmetric dendrimer with functionality three ($f=3$), two spacer beads ($s=2$) and order of dendra two ($m=2$). Dendrimers are often referred to as soft colloids as they act like a bridge between the floppy linear chains and hard spheres~\cite{harreis2003can}. The transition from linear chain-like behaviour to hard spheres is controlled by $f$, $s$, $g$ and $m$~\cite{vlassopoulos2004colloidal,likos1998star}. The molecular mass dependency of the self-diffusivity and intrinsic viscosity of low-generation dendrimers of fixed architecture follows the same power law scaling as for linear chains in dilute solution\cite{bosko2011universal}. 

The diffusivity of probe particles in simple viscous fluids is given by the Stoke-Einstein (SE) equation, which relates the diffusion coefficient of the particle to the viscous drag experienced by it. In a complex fluid, like a polymer solution, the viscoelastic effects of the fluid are accounted for by the generalized Stokes-Einstein (GSE) equation~\cite{mason1995optical,squires2010fluid}. However, the underlying assumption that the fluid is a continuum breaks down when the size of the particle is comparable to the characteristic length scales in the solution like the radius of gyration of the background chains, or the correlation length of the solution. As a result, the SE and GSE equations fail to describe the dynamics of tracer particles~\cite{mackay2003nanoscale,ye1998transport,maldonado2016breakdown}. In semidilute polymer solutions, the radius of gyration of a chain and the correlation length of the solution are concentration-dependent quantities. 

Several theories have been developed over the past few years to describe the dynamics of nanoparticles in semidilute polymer solutions~\cite{ge2023scaling}. Early theoretical studies of the diffusion of hard spheres in linear polymer solutions proposed a stretched exponential dependence of the reduced diffusion coefficient of the probe on the concentration of solution\cite{langevin1978sedimentation,cukier1984diffusion,altenberger1984theory}. This was modified by \citet{phillies1989macroparticle} to include the effect of size of the probe and the molecular weight of the linear polymers to obtain the generalized scaling relation $D/D_0=\textrm{exp}\left(-bR^uM^xc^y\right)$, where $D$ and $D_0$ are the diffusion coefficients of the probe in the polymer solution and in the pure solvent, $R$ is the size of the probe, $M$ is the molecular weight of the polymer, $c$ is the solution concentration, with exponents $u=0 \pm 0.2, x=0.8$ and $y$ ranging from $0.5$ to $1.0$. However, these models do not account for the fluctuations in the mesh size and the size of the probe relative to the correlation length of the solution.Based on the works by \citet{de1979scaling} and \citet{tong1997sedimentation} and \citet{tong1997sedimentation} which highlighted these factors,  a scaling relation was proposed for probe diffusion through semidilute polymer solutions 
$D/D_0=\textrm{exp}\left[-\beta \left(R/\xi\right)^{\delta}\right]$ by \citet{cheng2002diffusion}. Here $\beta$ is $2.2$, $\delta$ is $0.95$, and $\xi$ is the correlation length of the polymer solution. Even though this theory could explain several experimental results, the regime with the size of the nanoparticle being comparable to that of the radius of gyration of the background polymers could not be captured. 

Holyst and coworkers~\cite{holyst2009scaling,kalwarczyk2015motion} introduced an effective size for the nanoparticle, which is a combination of its radius and the hydrodynamic radius of the polymers that determines its dynamics in this crossover regime. The scaling law for the long-time diffusivity proposed by them is given by:
\begin{align}
    \dfrac{D}{D_0}=\textrm{exp}\left[\dfrac{-\gamma}{RT} \left( \dfrac{R_{\textrm{eff}}}{\xi}\right)^{a}\right]
    \label{eq:Holyst_model}
\end{align}
where $R_{\textrm{eff}}^{-2} = R_{\textrm{H}}^{-2} + R^{-2}$, with $R_{\textrm{H}}$ being the hydrodynamic radius of the polymer and $R$ is the radius of the nanoparticle. $\gamma \, (\gamma>0)$ and $a \, (a>0)$ are system-dependent parameters and are reported in terms of an effective excess diffusion energy, $\Delta E_{\rm{a}}=\gamma \left( {R_{\textrm{eff}}}/{\xi} \right)^a$, compared to that in pure solvent. According to this model, the length-scale of hydrodynamic flow in the solution determines the viscosity experienced by the probe particle and is equal to $R_{\textrm{eff}}$. The Holyst model has been validated by experiments\cite{kalwarczyk2011comparative,sozanski2013activation} and simulations\cite{chen2017effect}. 

Coupling theory, on the other hand, takes into account the coupling between the probe particle dynamics and the relaxation of the surrounding polymers~\cite{cai2011mobility}. According to this theory, probes larger than the correlation length of the solution get trapped in a polymer cage, thus leading to its subdiffusion at intermediate time scales. At longer times, the particle is set free as the polymer relaxes, with the diffusion coefficient scaling as $d_{\textrm{NP}}/\xi ^{-2}$, where $d_{\textrm{NP}}$ is the diameter of the nanoparticle. Recent experiments on nanoparticle diffusion in a partially hydrolyzed polyacrylamide solution indicate that the long-time diffusivity and short-time dynamics of particles larger and smaller than the correlation length show agreement with the predictions of the coupling theory~\cite{poling2015size}. However, the diffusion exponents of particles with a size comparable to the characteristic length scale of the solution were much higher than the predictions of the coupling theory. This discrepancy was resolved in a recent Multi-Particle Collision Dynamics (MPCD) study in which the dynamics of nanoparticle was found to be coupled to the dynamics of the centre of mass of the polymers, thus giving it an additional mechanism to move through the polymer solution~\cite{chen2018coupling}. The diffusion exponents of the nanoparticle and the polymer centre of mass were found to be correlated even in the absence of many-body hydrodynamic interactions. Coupling theory has also been supported by simulations\cite{chen2018coupling} and experiments\cite{poling2015size} on various systems.

The behaviour of dendrimers in a semidilute solution of linear polymers has not been studied extensively. Experiments have shown that the size of higher generation dendrimers does not change significantly with concentration~\cite{cheng2002diffusion} and, therefore, can be considered as a hard sphere. The dynamics of such dendrimers were shown to follow the Holyst model in semidilute solutions of linear chain polymers~\cite{cheng2002diffusion,de2015scaling}. However, the question of whether these theories can be applied to a low generation dendrimer, whose size is a function of the solution concentration, has not been addressed so far. In the current study, we have simulated low generation dendrimers in unentangled semidilute solutions of linear chain polymers in athermal solvent conditions using Brownian dynamics simulations. Dendrimer parameters like the generation number, functionality, and number of spacer beads were varied to understand their effects on the shape and internal bead arrangements of dendrimers in dilute solutions. We also studied the effect of the concentration of linear chains on the size and dynamics of dendrimers in semidilute solutions. It should be noted that the focus of this work is semidilute unentangled polymer solutions in which hydrodynamic interactions (HI) are significant and the HOOMD-Blue simulation package with the Positive Split Ewald algorithm has been used to implement hydrodynamic interactions. To understand its effect on the dynamics of dendrimers, we have performed simulations with and without HI. The algorithm used in this work is not appropriate for entangled concentrated systems since entanglements have not been taken into account. In the concentrated-entangled regime, hydrodynamic interactions are screened and there is no need to incorporate them in the simulation algorithm. Several studies on nanoparticles in entangled and concentrated polymer solutions\cite{cai2011mobility} and networks\cite{cai2015hopping} reported characteristic features which cannot be derived directly from the results in this work. However, there is very little literature on the dynamics of these systems, especially for dendrimers, where HI plays an important role.

The paper is organized as follows: The governing equations and the various intra- and intermolecular interactions included in the model, the procedure to choose various dendrimer and linear chain parameters to simulate specific architectures and formulas for the properties studied are discussed in section~\ref{sec:Model}. The static and dynamic properties of dendrimers and linear chains obtained from simulations are discussed in section~\ref{sec:results}, and concluding remarks are given in section~\ref{sec:conclusions}.

\section{\label{sec:Model} Model and Method}
\subsection{\label{sec:SM} Governing equations}

Polymers in this study were modelled using the coarse-grained bead-spring chain model, with $N_b$ beads connected by  $N_b-1$ springs~\cite{Bird}. The simulated semidilute solution contains $N_{\textrm{c}}^{\textrm{lc}}$ linear chain molecules with $N_{\textrm{b}}^{\textrm{lc}}$ beads and $N_{\textrm{c}}^{\textrm{d}}$ dendrimers with $N_{\textrm{b}}^{\textrm{d}}$ beads, immersed in an incompressible Newtonian fluid. This system is contained in a cubic, periodic simulation box of length $L$ and volume $V$, where $V=L^3$. The total monomer concentration in a simulation box is given by $c=N/V$, where $N$ is the total number of monomers in the box given by $N= (N_{\textrm{c}}^{\textrm{lc}} \times N_{\textrm{b}}^{\textrm{lc}})+(N_{\textrm{c}}^{\textrm{d}} \times N_{\textrm{b}}^{\textrm{d}})$.
 The It$\hat{\textrm{o}}$ stochastic differential equation, which is the governing equation in Brownian dynamics simulations, provides the bead position vectors $\mathbf{r}_\mu (\mu = 1, 2, ..., N)$ as a function of time. The Euler integration algorithm form of this equation is given below:
\begin{align}\label{gov-eqn}
\begin{aligned}
\mathbf{r}_\mu(t + \Delta t) =\, & \mathbf{r}_\mu(t) + \frac{\Delta t}{4} \sum\limits_{\nu=1}^N\mathbf D_{\mu\nu}\cdot(\mathbf F_\nu^{\textrm{s}}+ \mathbf F_\nu^{\textrm{SDK}}) \\ & +\frac{1}{\sqrt{2}}\sum\limits_{\nu=1}^N \mathbf B_{\mu\nu}\cdot\Delta \mathbf W_\nu
\end{aligned}
\end{align}
Here the length and time non-dimensionalization factors are $l_H=\sqrt{k_BT/H}$ and $\lambda_H=\zeta/4H$, respectively, where  $k_B$ is the Boltzmann constant, $T$ is temperature, $H$ is the spring constant, and $\zeta=6\pi\eta_s a$ is the Stokes friction coefficient of a spherical bead with radius $a$ and $\eta_s$ is the solvent viscosity. In eqn~\eqref{gov-eqn}, $\pmb D_{\nu\mu}$ is the diffusion tensor, defined as $\pmb D_{\nu\mu} = \delta_{\mu\nu} \pmb \delta + \pmb \Omega_{\mu\nu}$, where $\delta_{\mu\nu}$ is the Kronecker delta, $\pmb \delta$ is the unit tensor, and $\pmb{\Omega}_{\mu\nu}$ is the hydrodynamic interaction tensor. $\pmb{B}_{\mu\nu }$ is a non-dimensional tensor whose evaluation requires the decomposition of the diffusion tensor and $\Delta\pmb W_\nu$ is a non-dimensional Wiener process with mean zero and variance $\Delta t$. If $\mathcal{D}$ and $\mathcal{B}$ are block matrices consisting of $N \times N$ blocks each having dimensions of $3 \times 3$, with the $(\mu,\nu)$-th block of $\mathcal{D}$ containing the components of the diffusion tensor $\pmb{D}_{\mu\nu }$, and the corresponding block of $\mathcal{B}$ being $\pmb{B}_{ \mu\nu}$, the decomposition rule for obtaining $\mathcal{B}$ is then given by $\mathcal{B} \cdot {\mathcal{B}}^\textsc{t} = \mathcal{D} \label{decomp}$.

The spring force exerted on individual beads is represented by $\mathbf F_\nu^{\textrm{s}}$. We have used finitely extensible nonlinear elastic (FENE) springs and the spring potential is given by,
\begin{align}\label{eq-fraenkel}
U_{\textrm{FENE}}= -\frac{1}{2}Q_0^2\ln \left( 1-\frac{r^{2}}{Q_0^2}\right)
\end{align}
where $Q_0$ is the dimensionless maximum stretchable length of a single spring, and $k_B T$ is used to non-dimensionalize energy. 

The force due to the excluded volume interactions between bead pairs is denoted by $\mathbf F_\nu^{\textrm{SDK}}$ and is obtained from the Soddemann-D\"{u}nweg-Kremer (SDK) potential, $U_{\text{SDK}}$, ~\cite{SDK}
\begin{align}\label{eq:SDK}
U_{\textrm{SDK}}=\left\{
\begin{array}{l l l}
&4\left[ \left( \dfrac{\sigma}{r} \right)^{12} - \left( \dfrac{\sigma}{r} \right)^6 + \dfrac{1}{4} \right] - \epsilon;  & r\leq 2^{1/6}\sigma \vspace{0.5cm} \\
& \dfrac{1}{2} \epsilon \left[ \cos \,(\alpha \left(\dfrac{r}{\sigma}\right)^2+ \beta) - 1 \right] ;& 2^{1/6}\sigma \leq r \leq r_c \vspace{0.5cm} \\
& 0; &  r \geq r_c
\end{array}\right.
\end{align}
Here, $\epsilon$ is the well depth of the potential which controls the interaction strength between bead pairs and the non-dimensional distance $\sigma$ is fixed as $1$ in this study. The repulsive part of this potential is modelled by a truncated Lennard-Jones (LJ) potential and the attractive part is modelled using a cosine function. The constants $\alpha$ and $\beta$ are determined from boundary conditions, $U_{\textrm{SDK}} = 0$ at $r=r_c$ and $U_{\textrm{SDK}} = -\epsilon$ at $r=2^{1/6}\sigma$ which is the minima of the potential. The cut-off radius $r_c$ was taken to be $1.82 \sigma$, following the discussion in recent work by \citet{santra2019universality}. The advantage of using the SDK potential is that by varying a single parameter $\epsilon$, a range of solvent qualities can be studied and it affects the attractive part of the potential without altering the repulsive force. Also, the short-ranged attractive tail of this potential smoothly approaches zero, unlike the LJ potential. At $\epsilon=0$, the SDK potential reduces to the purely attractive Weeks-Chandler-Anderson (WCA) potential, and solvent quality can be varied by changing the value of $\epsilon$. Note that in this study, we have fixed $\epsilon=0$ since we restrict our simulations to athermal solvent conditions.

We use the regularized Rotne-Prager-Yamakawa (RPY) tensor to compute hydrodynamic interactions (HI),
\begin{equation}
{\pmb{\Omega}_{\mu \nu}} = {\pmb{\Omega}} ( {\mathbf{r}_{\mu \nu}})
\end{equation}
where ${\mathbf{r}_{\mu \nu}} = {\mathbf{r}_{\mu}} - {\mathbf{r}_{\nu}}$ and the function $\pmb{\Omega}$ is
\begin{equation}
\pmb{\Omega}(\mathbf{r}) =  {\Omega_1{ \pmb \delta} +\Omega_2\frac{\mathbf{r r}}{{r}^2}}
\end{equation}
with
\begin{equation*}
\Omega_1 = \begin{cases} \dfrac{3\sqrt{\pi}}{4} \dfrac{h^*}{r}\left({1+\dfrac{2\pi}{3}\dfrac{{h^*}^2}{{r}^2}}\right) & \text{for} \quad r\ge2\sqrt{\pi}h^* \\
 1- \dfrac{9}{32} \dfrac{r}{h^*\sqrt{\pi}} & \text{for} \quad r\leq 2\sqrt{\pi}h^* 
\end{cases}
\end{equation*}
and 
\begin{equation*}
\Omega_2 = \begin{cases} \dfrac{3\sqrt{\pi}}{4} \dfrac{h^*}{r} \left({1-\dfrac{2\pi}{3}\dfrac{{h^*}^2}{{r}^2}}\right) & \text{for} \quad r\ge2\sqrt{\pi}h^* \\
 \dfrac{3}{32} \dfrac{r}{h^*\sqrt{\pi}} & \text{for} \quad r\leq 2\sqrt{\pi}h^* 
\end{cases}
\end{equation*}
The hydrodynamic interaction parameter $h^*$ gives the dimensionless bead radius in the bead-spring model and is defined as $h^* = a/(\sqrt{\pi k_BT/H})$. The Brownian dynamics (BD) simulation code used in this work is based on the GPU-accelerated Python package named HOOMD-Blue, developed in Michigan University \cite{anderson2020hoomd} for the study of colloidal suspensions. It has been modified recently to study the dynamics of associative polymer solutions \cite{robe2023evanescent} along the lines of an earlier in-house BD code based on the Molecular Modelling ToolKit \cite{jain2012optimization,jain2015brownian,santra2021universal}. The decomposition of the diffusion tensor, which is computationally demanding, has been efficiently implemented recently using the Positive Split Ewald algorithm (PSE)\cite{fiore2017rapid}. This is available as a plugin to HOOMD-Blue.

\subsection{\label{sec:SD} Details of the simulation algorithm}

\begin{figure}[!ptbh]
    \centering
    \begin{tabular}{ccc}		    
			\resizebox{2.7cm}{!} {\includegraphics[width=0.16\textwidth]{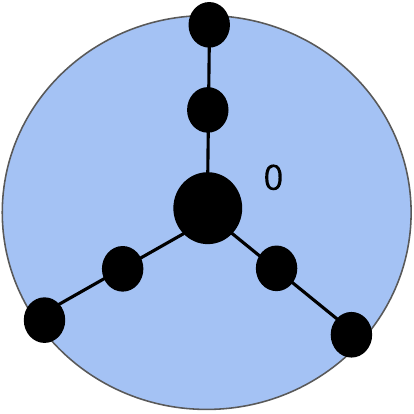}} &        
			\resizebox{2.7cm}{!} {\includegraphics[width=0.16\textwidth]{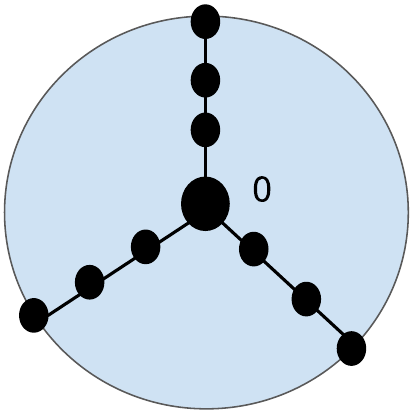}} &			
			\resizebox{2.7cm}{!} {\includegraphics[width=0.16\textwidth]{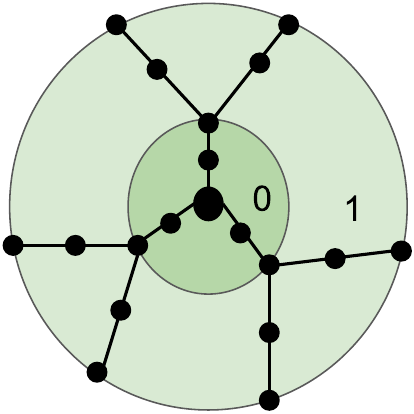}}\\
			(a) & (b) & (c) \\			
			\resizebox{2.7cm}{!} {\includegraphics[width=0.16\textwidth]{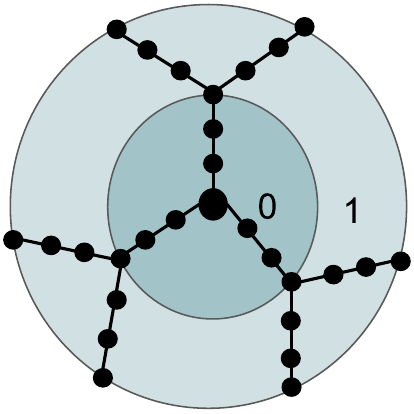}} &                
			\resizebox{2.7cm}{!} {\includegraphics[width=0.16\textwidth]{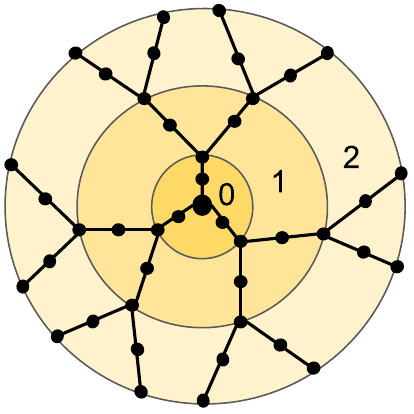}} &			
			\resizebox{2.7cm}{!} {\includegraphics[width=0.16\textwidth]{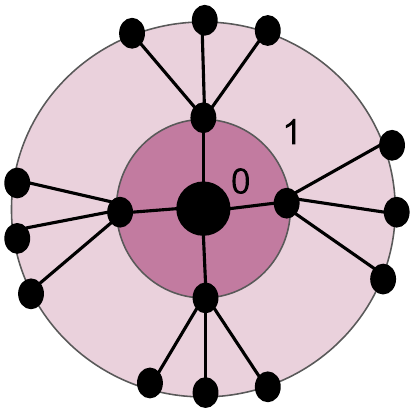}}\\
			(d) & (e) & (f) \\
		\end{tabular}
    \caption{ (Color online) The various dendrimer architectures simulated in this study. The specific simulation parameters corresponding to the topologies (a) to (f) are given in Table ~\ref{tab:parameter_space} }
    \label{fig:Dendrimer_schematic}
\end{figure}

\begin{table}[th]
    \caption{List of dendrimer parameters and linear chain lengths used in each system. $g$ is the generation number, $f$ is the functionality, $s$ is the number of spacer beads, $N_{\textrm{b}}^{\textrm{d}}$ is the number of beads in a dendrimer molecule, $R_{\textrm{g0}}^{\textrm{d}}$ is the radius of gyration of the dendrimer in the solvent, $\chi$ is the ratio between the radius of gyration of dendrimers and linear chains in dilute limit, $N_{\textrm{b}}^{\textrm{lc}}$ is the number of beads in the background linear chain, $R_{\textrm{g0}}^{\textrm{lc}}$ is the radius of gyration of the linear chain. Note that dendrimer topologies in (c) and (h) are identical, but have different $\chi$.}
    \label{tab:parameter_space}
    \centering
    \setlength{\tabcolsep}{8.5pt}
    \renewcommand{\arraystretch}{1.5}
    \begin{tabular}{|c|c|c|c|c|c|c|c|c|}
    \hline
    $ $     & $f$
            & $s$ 
            & $g$ 
            & $N_{\textrm{b}}^{\textrm{d}}$ 
            & $R_{\textrm{g0}}^{\textrm{d}}$
            & $\chi$ 
            & $N_{\textrm{b}}^{\textrm{lc}}$ 
            & $R_{\textrm{g0}}^{\textrm{lc}}$
\\                
\hline
\hline
            a
            & $3$
            & $1$
            & $0$
            & $7$  
            & $1.90$
            & $0.50$
            & $19$
            & $3.90$
\\
            b
            & $3$
            & $2$
            & $0$
            & $10$ 
            & $2.40$
            & $0.50$
            & $26$
            & $4.80$
\\
            c
            & $3$
            & $1$
            & $1$
            & $19$ 
            & $3.12$
            & $0.46$
            & $43$
            & $6.23$
\\
            d
            & $3$
            & $2$
            & $1$
            & $28$ 
            & $3.88$
            & $0.46$
            & $61$
            & $7.77$
\\
            e
            & $3$
            & $1$
            & $2$
            & $43$ 
            & $4.34$
            & $0.50$
            & $74$
            & $8.68$
\\
            f
            & $4$
            & $0$
            & $1$
            & $17$ 
            & $2.49$
            & $1.0$
            & $9$
            & $2.49$
\\
            h
            & $3$
            & $1$
            & $1$
            & $19$ 
            & $3.12$
            & $1.0$
            & $13$
            & $3.12$          
\\
\hline
    \end{tabular}
    \end{table}
        
To choose the simulation system, we fixed the dendrimer architecture first and calculated the number of beads ($N_{\textrm{b}}^{\textrm{d}}$) on each dendrimer molecule. The various architectures chosen are given in Fig.~\ref{fig:Dendrimer_schematic}. The radius of gyration of dendrimers was determined by carrying out simulations in the dilute limit and the values are reported in Table ~\ref{tab:parameter_space}. Depending on the choice of the ratio between the radius of gyration of the dendrimer and linear chain, $\chi$, we calculated the radius of gyration of the linear chain. The number of beads on the linear polymer was then estimated by running simulations in the dilute limit and reading off the value of $N_{\textrm{b}}^{\textrm{lc}}$ from the $R_{\textrm{g}}^{\textrm{lc}}$ vs $N_{\textrm{b}}^{\textrm{lc}}$ plot (see Fig.~S1 of the Supplementary Material). The simulation box length, $L$ was chosen to be $L \geq 2R_e$, where $R_e$ is the end-to-end distance of linear chains in the solution. This ensures that the molecules do not wrap around themselves. Simulations were carried out keeping the dendrimer concentration fixed and adding linear chains to the simulation box to increase the concentration of the solution. Hence the resulting solution is always dilute in dendrimer concentration and semidilute in linear chains. The list of chain lengths, dendrimer parameters, and number of chains in each simulation are given in Table ~\ref{tab:parameter_space}. All the dendrimers simulated in this study were symmetric, $m=f-1$. 
\begin{figure}[t]
	\begin{center}
	 {\includegraphics[width=8cm,height=!]{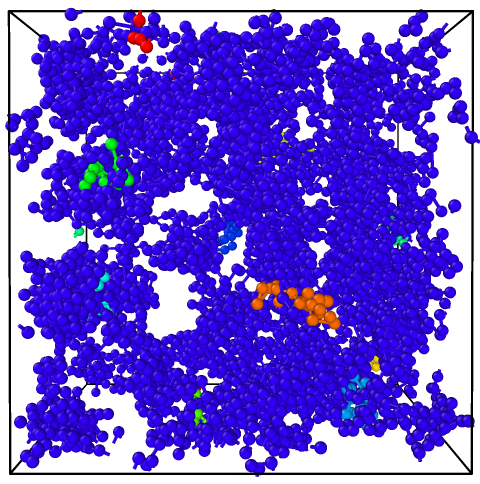}}
	\end{center}
	\vspace{-10pt}
	\caption{(Color online) Snapshots from simulation of $f = 3,s = 1,g = 1$ dendrimer with $\chi=0.46$ at $c/c_{\textrm{lc}}^{\ast}=1$. The blue beads belong to linear chains and other coloured beads belong to dendrimers.}
\label{fig:Snaps}
\vspace{-10pt}
\end{figure}
One special case we studied is a $f=4,g=1,s=0$ dendrimer. By increasing the number of arms and reducing the spacer beads, we expect it to be more closely packed compared to the rest of the dendrimers, and thus behave qualitatively more like a nanoparticle. The concentration of linear chains was varied from $0.5$ to $6 c^{\ast}$. All simulations were carried out in athermal solvent conditions ($\epsilon = 0$) and the hydrodynamic interaction parameter, $h^{\ast}=0.2$. When hydrodynamic interactions are not present (the free draining case), $h^{\ast}$ is set equal to zero, so the RPY tensor is not invoked and the diffusion tensor simplifies to $\pmb D_{\mu\nu}= \delta_{\mu\nu} \pmb \delta$. The maximum stretchable length of each spring, $Q_0^2 = 50.0$, for all simulations. Every independent trajectory had an equilibration phase of 8 relaxation times followed by a production phase of $10$ relaxation times with non-dimensional time step, $\Delta t= 10^{-3} \ \text{to} \ 10^{-4}$. The dynamic properties were calculated as a function of time in the production phase of individual trajectories. Ensemble averages and error estimates were calculated over $700$ to $2000$ independent trajectories. Fig.~\ref{fig:Snaps} shows a snapshot of a simulation box containing generation 1 dendrimers and linear chains at $c/c_{\textrm{lc}}^{\ast} = 1$. 

\subsection{\label{sec:PC} Definition of static and dynamic properties}

\subsubsection{\label{sec:Rg} Radius of gyration and correlation length}
The radius of gyration of dendrimers and linear chains in the solution is calculated using the expression 
\begin{align}
\label{Rg}
R_{\textrm{g}}^2 &=  \dfrac{1}{N_{\textrm{b}}} \langle \sum\limits_{i=1}^{N_{\textrm{b}}} \left( \mathbf R_i - \mathbf R_{\textrm{CM}} \right)^2 \rangle \ ; \ N_{\textrm{b}} \in \{ N_{\textrm{b}}^{\textrm{d}},N_{\textrm{b}}^{\textrm{lc}} \}
\end{align} 
where $R_{\textrm{CM}}$ is the centre of mass of the molecule given by
\begin{align}
    \mathbf R_{\textrm{CM}}=\dfrac{1}{N_{\textrm{b}}} \sum\limits_{i=1}^{N_{\textrm{b}}} \mathbf R_i 
\end{align}
We introduce a quantity $\chi$, which is the ratio of the radius of gyration of dendrimer ($R_{\textrm{g0}}^{\textrm{d}}$) to that of the linear chain ($R_{\textrm{g0}}^{\textrm{lc}}$) in the background, both in the dilute limit, given by $\chi=\left({R_{\textrm{g0}}^{\textrm{d}}}/{R_{\textrm{g0}}^{\textrm{lc}}}\right)$. The overlap concentration, $c^*$, of a polymer solution is defined in terms of the radius of gyration of the polymer in the dilute limit, given by $c^{\ast} = \left({N_{\textrm{b}}}/{({4}/{3})\pi R_{\textrm{g0}}^3}\right)$. In this study, the solution contains polymers of two architectures. Therefore, $c^*$ can be estimated separately for each of them  ($c^*_\textrm{d}$ for dendrimers and $c^*_\textrm{lc}$ for linear chains) given by:
\begin{align}
    c^{\ast}_{\textrm{d}}  &= \dfrac{N_{\textrm{b}}^{\textrm{d}}}{\dfrac{4}{3}\pi (R_{\textrm{g0}}^{\textrm{d}})^3} \label{eq:cstar_d} \\
    c^{\ast}_{\textrm{lc}}  &= \dfrac{N_{\textrm{b}}^{\textrm{lc}}}{\dfrac{4}{3}\pi (R_{\textrm{g0}}^{\textrm{lc}})^3} \label{eq:cstar_lc}
\end{align}

Another important length scale in the solution is the correlation length ($\xi$). It is defined in terms of the radius of gyration in the dilute limit, the overlap concentration of linear chains, and the total monomer concentration in the solution\cite{colby2003polymer}, 
\begin{align}\label{eq:xi}
     \xi=R_{\textrm{g0}}^{\textrm{lc}} \left(\frac{c}{c^{\ast}_{\textrm{lc}}}\right)^{\frac{-\nu}{\left(3\nu -1 \right)}}
\end{align}
\noindent where $\nu$ is the Flory exponent, assumed here to be equal to $0.588$. Note that the total monomer concentration $c$, includes monomers of both species in the system. 

\subsubsection{Relative shape anisotropy and the universal ratio $U_{\textrm{RD}}$}

This work aims to compare the dynamics of dendrimers to nanoparticles, and therefore, it is essential to understand the influence of dendrimer parameters on their shape and compactness. Two important measures that can be used to probe these are the relative shape anisotropy ($\kappa^2$) and the dimensionless ratio $U_{\textrm{RD}}$.

The relative shape anisotropy, one of the shape functions, has been used to understand the asymmetry in the shape of polymer molecules. It is defined in terms of the eigenvalues, $\lambda_1^2,\lambda_2^2,\lambda_3^2$, of the gyration tensor $\mathbf G$ given by 
\begin{align}
    \mathbf G = \dfrac{1}{2N_b^2} \sum\limits_{\mu=1}^{N_b} \sum\limits_{\nu=1}^{N_b} \mathbf r_{\mu\nu} \mathbf r_{\mu\nu}
\end{align}
\noindent Here $\lambda_1^2,\lambda_2^2,\lambda_3^2$ are arranged in ascending order. The relative shape anisotropy $\kappa^2$ is defined as~\cite{theodorou1985shape,bishop1986polymer,kumari2020computing}:
\begin{align} \label{eq:shape_anisotropy}
    \kappa^2 = 1 - 3 \left[ \frac{\langle I_2 \rangle}{\langle I_1^2\rangle} \right ]
\end{align}
\noindent where $I_1=\lambda_1^2 + \lambda_2^2 + \lambda_3^2$ and $I_2= \lambda_1^2 \lambda_2^2 + \lambda_2^2 \lambda_3^2 + \lambda_1^2 \lambda_3^2$. $\kappa^2 = 0$ for a spherically symmetric molecule while it is equal to one for a rod-shaped molecule. 

$U_{\textrm{RD}}$, which is a measure of the hard sphere-like behaviour of molecules, is given by 
\begin{align} \label{eq:U_rd}
    U_{\textrm{RD}} = \dfrac{R_{\textrm{g}}}{R_{\textrm{H}}} = \dfrac{4 \, D \, R_{\textrm{g}}}{h^{\ast} \sqrt{\pi}}
\end{align}
where $R_{\textrm{H}}$ is the hydrodynamic radius of the molecule related to its long time diffusivity, $D$, by $R_{\textrm{H}}=\dfrac{h^{\ast} \sqrt{\pi}}{4 D}$. For a polymer chain, $U_{\textrm{RD}}$ is approximately 1.4~\cite{kroger2000variance,sunthar2006dynamic,pan2014viscosity} while for a hard sphere, it is 0.77~\cite{guinier1955small}. 

\subsubsection{\label{sec:density} Intra-molecular bead density}
The internal bead density, obtained from the arrangement of beads about the centre of mass of the molecule, leads to an understanding of its internal structure. It is calculated by counting the number of beads along the major axis of the gyration tensor in intervals of fixed length and binning them. This requires all polymer configurations in the simulation ensemble to be aligned along their respective major axes. The linear bead number density obtained is given by:
\begin{align}
    \rho_{\rm{l}}(r)=\dfrac{n_{\textrm{b}}(x+ \Delta x)-n_{\textrm{b}}(x)}{\Delta x}
\end{align}
where $n_{\textrm{b}}(x)$ is the number of beads of a molecule within a distance $x$ along the major axis, and $\Delta x$ is the length of the fixed interval. 

\subsubsection{\label{sec:MSD} Mean squared displacement and diffusivity}

The mean squared displacement of the centre of mass of the polymer molecules in solution is calculated using the following expression
\begin{align}\label{eq:MSD}
    \textrm{MSD}(\Delta t) = \langle | \mathbf R_{\textrm{CM}}(t+\Delta t)-\mathbf R_{\textrm{CM}}(t) |^2 \rangle = 6 \,D \, t^{\alpha}
\end{align}
where $\mathbf R_{\textrm{CM}}(t)$ and $\mathbf R_{\textrm{CM}}(t+\Delta t)$ are the position vectors of the centre of mass of the molecule at times $t$ and $t+\Delta t$ respectively, $D$ is the diffusion coefficient and $\alpha$ is the diffusion exponent. $\alpha=1$ for normal diffusion and the molecule exhibits subdiffusion if $\alpha<1$. The long time diffusivity, $D$ is estimated from the slope of the mean squared displacement versus time plots at longer times.

\begin{figure}[t]
	\begin{center}
	 {\includegraphics[width=7.1cm,height=6.5cm]{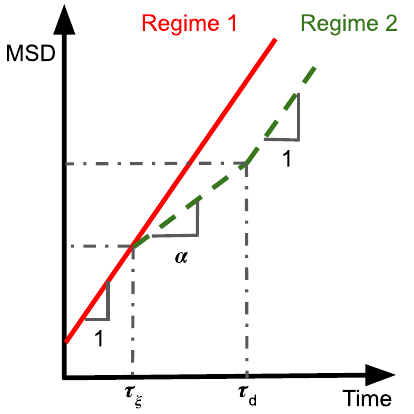}}
	\end{center}
	\caption{(Color online) Schematic representation of the diffusive nature of dendrimers belonging to different size regimes. $\tau_{\xi}$ and $\tau_{\textrm{d}}$, defined by eqn~\eqref{eq:tau_xi} and eqn~\eqref{eq:tau_d}, are the transition time scales to subdiffusive and diffusive behaviour respectively, for dendrimers with size larger than the solution correlation length (Regime-2). The exponents of time are indicated in the figure.}
\label{fig:schematic}
\end{figure}

\citet{cai2011mobility} have identified several regimes for nanoparticles in entangled systems based on their size relative to the solution correlation length and the tube diameter. However, in the current study of dendrimers in semidilute polymer solution, there are only two regimes based on the relative sizes of the dendrimer and correlation length: Regime-1 in which the dendrimer is smaller than the solution correlation length ($2R_{\textrm{g}}^{\textrm{d}} < \xi$) and Regime-2 in which the dendrimer size is larger than the correlation length ($2R_{\textrm{g}}^{\textrm{d}} > \xi$). It is important to note that, unlike a nanoparticle that has a constant size at all polymer concentrations, the size of a dendrimer is a concentration-dependent quantity. 

The dynamics of dendrimers in the two regimes are different due to the influence of the surrounding linear chains as shown schematically in Fig ~\ref{fig:schematic}. Following the arguments proposed by \citet{cai2011mobility}, the following dynamics are expected in the two regimes:
\begin{enumerate}
    \item Regime-1: When the size of the dendrimer is less than the correlation length ($2R_{\textrm{g}}^{\textrm{d}} < \xi$), it can diffuse freely through the solution, without being affected by the background linear chains. The diffusivity of a dendrimer is expected to be that experienced by it in the dilute limit ($D_0$) and exhibit normal diffusion at all time scales with the diffusion exponent, $\alpha = 1$. As a result, the mean squared displacement, $\textrm{MSD} = 6 \, D_0 \, t$.
    
    \item Regime-2: When the dendrimer is larger than correlation length, ($2R_{\textrm{g}}^{\textrm{d}} > \xi$), its dynamics is influenced by the presence of linear chains. The mean squared displacement shows three scaling regimes based on the time scale at which it is probed. At short times, the dendrimer does not experience any hindrance to its motion and therefore, exhibits normal diffusion. This continues up to a time of the order of the relaxation time of the correlation blob, $\tau_{\xi}$ given by:
    \begin{align}\label{eq:tau_xi}
    \tau_{\xi} \equiv \eta_s \xi^3/k_BT
    \end{align}
    At times higher than $\tau_{\xi}$, the dendrimer is trapped in cages formed by the linear polymers, and its motion is expected to become subdiffusive. Its dynamics are coupled to the motion of the centre of mass of the polymer and the polymer segmental relaxation times. The mean squared displacement of dendrimers at these intermediate times is given by $\textrm{MSD} = 6 \, D_{\alpha} \, t^{\alpha}$, where $\alpha < 1$. The subdiffusive regime extends until $t \sim \tau_{\textrm{d}}$ which is equal to the relaxation time of a chain segment with a size equal to the size of the dendrimer given by:
    \begin{align}\label{eq:tau_d}
    \tau_{\textrm{d}} \equiv \tau_{\xi} \left( 2R_{\textrm{g}}^{\textrm{d}} / \xi \right)^4
    \end{align}
    Beyond $\tau_{\textrm{d}}$, the dendrimer is set free due to the relaxation of linear chains and therefore the mean squared displacement is given by $\textrm{MSD} = 6 \, D \, t $. The diffusivity, $D$, in this regime, is referred to as the long-time diffusivity of the dendrimer.
\end{enumerate} 

\subsubsection{\label{sec:PDD} Probability distribution function of displacement}

Apart from the long-time diffusivity, the probability distribution function of displacement gives additional insight into the dynamics of the polymer molecules. The probability of a molecule displacing by a distance $\Delta x$ in a time $\Delta t$ is given by
\begin{align}\label{eq:PDD}
    \textrm{P}(\Delta x,\Delta t) = \langle \delta \left(\Delta x -  |x_{\textrm{CM}}(t+\Delta t)- x_{\textrm{CM}}(t)|\right) \rangle
\end{align}
where $ x_{\textrm{CM}}(t)$ and $ x_{\textrm{CM}}(t+\Delta t)$ are the x-components of the centre of mass of the molecule at times $t$ and $t+\Delta t$ respectively.

\begin{figure}[!tbph]
	\begin{center}
	 {\includegraphics[width=8.0cm,height=!]{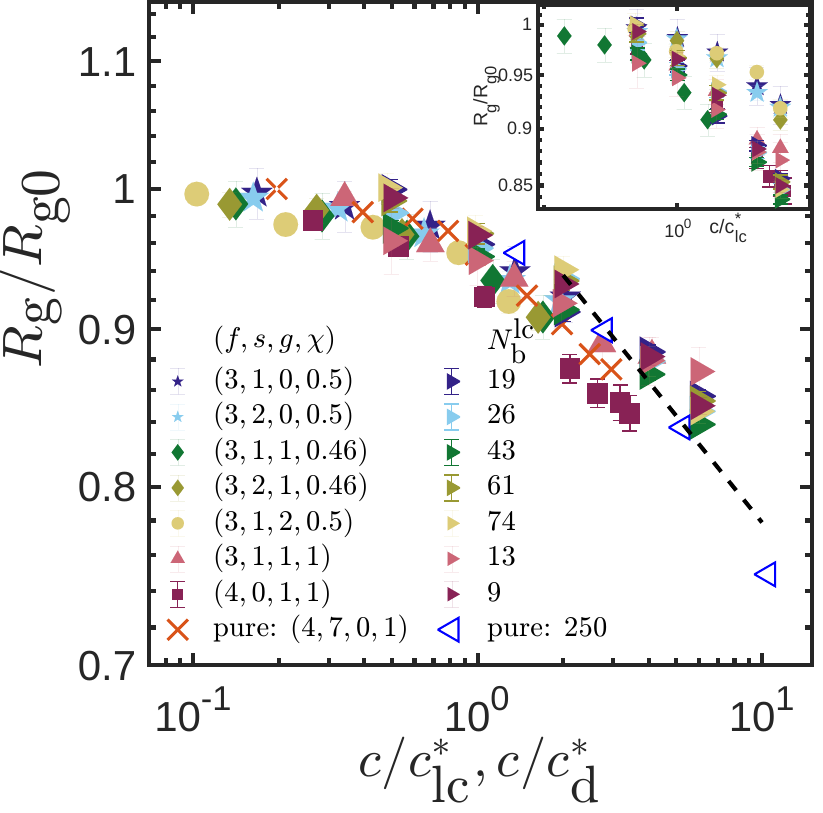}}
	\end{center}
	    \vspace{-10pt}
	\caption{(Color online) Universal behaviour of radius of gyration of dendrimers and linear chains. The normalized radius of gyration of polymers is plotted as a function of the ratio of monomer concentration normalized by the respective $c^{\ast}$ of each species. The dendrimer architecture is included in the order ($f,s,g,\chi$) and $N_{\textrm{b}}^{\textrm{lc}}$ is the number of beads on a linear chain. The same coloured symbols belong to one simulated system. Data from work on pure linear chain solutions by \citet{huang2010semidilute} (empty triangles) and our simulation data for pure star polymer solutions (orange x-mark) are included.  The dashed line is the scaling law given by eqn~\eqref{eq:rg_scalinglaw}. Inset: Normalised radius of gyration of dendrimers and linear chains in the solution as a function of the ratio of monomer concentration normalized by the overlap concentration of linear chains, $c^{\ast}_{\textrm{lc}}$.}
\label{fig:rg_scaling}
    \vspace{-10pt}
\end{figure}

\section{\label{sec:results}Results and Discussion}

\subsection{\label{sec:Rg_results}{Radius of gyration}}
According to the blob theory for polymers, the radius of gyration of linear chain polymers reduces with an increase in concentration and follows the scaling law~\cite{daoud1975solutions,doi1988theory,huang2010semidilute}:
\begin{align}\label{eq:rg_scalinglaw}
     R_{\textrm{g}}^{\textrm{lc}} = R_{\textrm{g0}}^{\textrm{lc}}\left(\frac{c}{c_{\textrm{lc}}^{\ast}}\right)^{\frac{1-2\nu}{2\left(3\nu -1 \right)}}
\end{align}
\begin{figure*}[tbph]
	\begin{center}
		\begin{tabular}{ccc}
			\resizebox{8cm}{!} {\includegraphics[width=8cm]{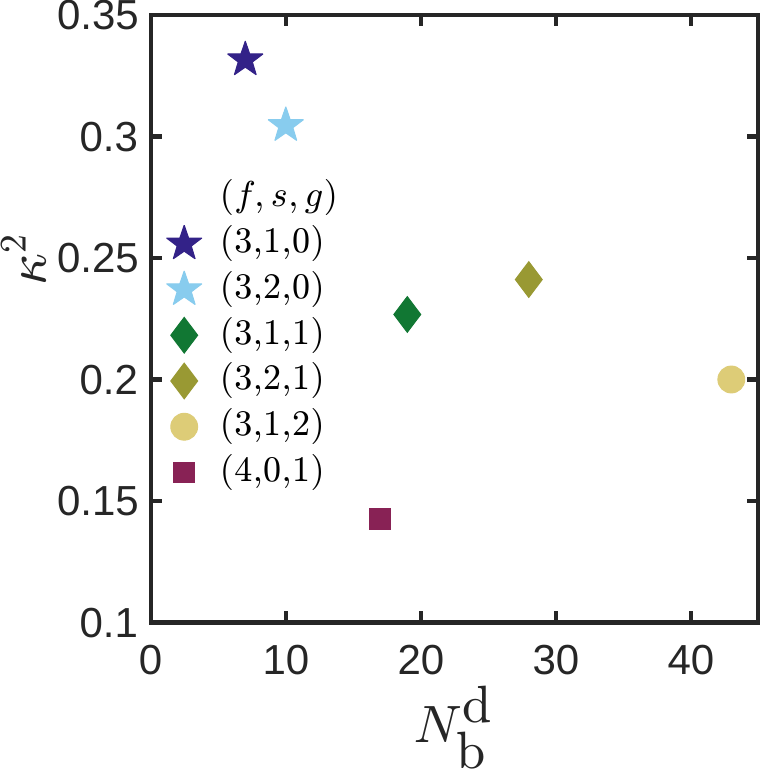}} &
			\hspace{1cm} &
			\resizebox{8cm}{!} {\includegraphics[width=8cm]{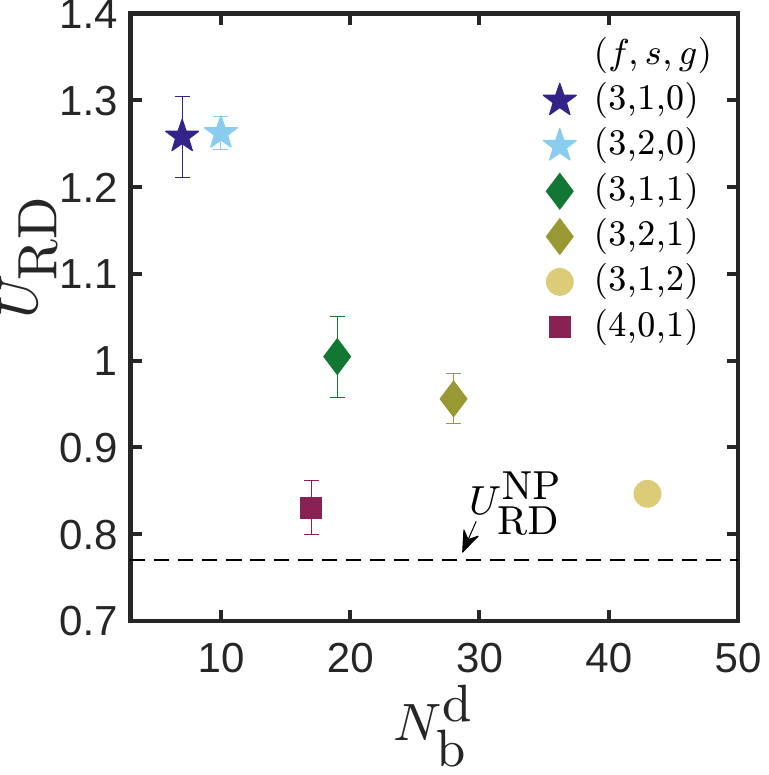}}\\
			(a) & \hspace{1cm} & (b) \\
		\end{tabular}
	\end{center}
    \vspace{-10pt}
	\caption{ (Color online) Relative shape anisotropy and $U_{\textrm{RD}}$ of dendrimers.(a) Effect of different architectures on the relative shape anisotropy of dendrimers in dilute solution as a function of the number of beads in a dendrimer molecule. The symbols represent dendrimer topology given by the combination $f, s, g$. (b) $U_{\textrm{RD}}$ of dendrimers in dilute solution as a function of the number of beads in a dendrimer molecule. The dashed line is the value of $U_{\textrm{RD}}$ for a nanoparticle. The symbols represent dendrimer topology given by the combination $f, s, g$.}
    \label{fig:shape_fun}
    \vspace{-10pt}
\end{figure*}

\noindent On plotting the ratio $R_{\textrm{g}}/R_{\textrm{g0}}$ of dendrimers and linear chains as a function of the total monomer concentration, $c$, in solution normalized by the overlap concentration of linear chains, $c^{\ast}_{{\textrm{lc}}}$, dendrimers seem to shrink at a slower rate compared to the linear chains (inset to Fig.~\ref{fig:rg_scaling}). However, if the normalizing factor for monomer concentration is the overlap concentration for each species, the normalised radius of gyration of dendrimers of all architectures and linear chains in the background collapse onto a universal curve as shown in Fig. ~\ref{fig:rg_scaling}. Thus, the factor that determines the size of a polymer molecule is the number of monomers it interacts with in its neighbourhood. In other words, the concentration of monomers relative to its overlap concentration if all the monomers in the solution had belonged to its own architecture is what determines its shape. The fact that the solution has molecules of different architectures is unimportant.

At lower concentrations, the correlation length of the solution is large compared to the size of polymers (Regime-1). With an increase in concentration, the size of dendrimers, linear chains, and the correlation length decreases given by eqn~\eqref{eq:xi} and \eqref{eq:rg_scalinglaw}. However, the correlation length $\xi$ decreases faster than the diameter of the dendrimer $2 R_{\textrm{g}}^{\textrm{d}}$ and after a certain concentration, the dendrimers in solution become bigger than correlation length, $2 R_{\textrm{g}}^{\textrm{d}} > \xi$ (Regime-2). Thus there is a crossover from Regime-1 to Regime-2 with increasing concentration (see Fig.~S2 of the Supplementary Material).

\subsection{Relative shape anisotropy and $U_{\textrm{RD}}$}

In Fig.~\ref{fig:shape_fun}(a), the relative shape anisotropy of dendrimers of different architectures in dilute solution, calculated using eqn~\eqref{eq:shape_anisotropy}, is plotted as a function of the number of beads per molecule. As the generation number ($g$) increases at constant $f$ and $s$, $\kappa^2$ decreases. The number of beads inside and on the outer shell of a molecule is more for a dendrimer with higher $g$. Due to the excluded volume interactions, the internal beads spread out, and the molecule with higher $g$ becomes more spherical compared to its lower-generation counterpart. The relative shape anisotropy of a functionality $4$ dendrimer with $g=1$ and $s=0$ (Fig.~\ref{fig:Dendrimer_schematic}(f)) is much lower than that of the other architectures. With no spacers, every internal bead in the molecule is a branching point connected to 4 other beads. This gives it less chance of conformational fluctuations and hence behaves more like a compact sphere\cite{cannon1991equilibrium}. This is clear from the distribution of radius of gyration of dendrimers in dilute solution (see Fig.~S3 of the Supplementary Material). The number of spacer beads appears not to have a significant effect on the shape of dendrimers. Along with the relative shape anisotropy, asphericity ($B$) is also a measure of the shape and compactness of molecules. It shows a similar trend as that of the $\kappa^2$ for dendrimers in our simulations supporting the argument of a spherical and compact structure (see Fig.~S4(a) of the Supplementary Material). Interestingly, $\kappa^2$ and $B$ are unaffected by the concentration of the solution as shown in Fig~.S4(b) and (c) of the Supplementary Material. 
 
In Fig.~\ref{fig:shape_fun} (b), the ratio $U_{\textrm{RD}}$ for dendrimers of different architectures in the dilute limit is plotted as a function of the number of beads in the molecule. Similar to the behaviour of relative shape anisotropy, $U_{\textrm{RD}}$ decreases with increasing $g$ when $f$ and $s$ are fixed. Our simulation results are in line with the previous experiments~\cite{tande2001viscosimetric} and simulations~\cite{bosko2011universal} for dendrimers in dilute solution. The $f=4, g=1, s=0$ (Fig.~\ref{fig:Dendrimer_schematic}(f)) dendrimer has a value of $U_{\textrm{RD}}$ very close to that of a hard sphere even though the number of beads in it is almost the same as that of a $f=3,g=1,s=1$ (Fig.~\ref{fig:Dendrimer_schematic}(c)) dendrimer. Thus, by increasing $g$ and $f$, dendrimers can be seen to transition from `fractal-like' to a `nanoparticle-like' structure. 

\begin{figure*}[t]
	\begin{center}
		\begin{tabular}{ccc}
			\resizebox{8cm}{!} {\includegraphics[width=8cm,height=!]{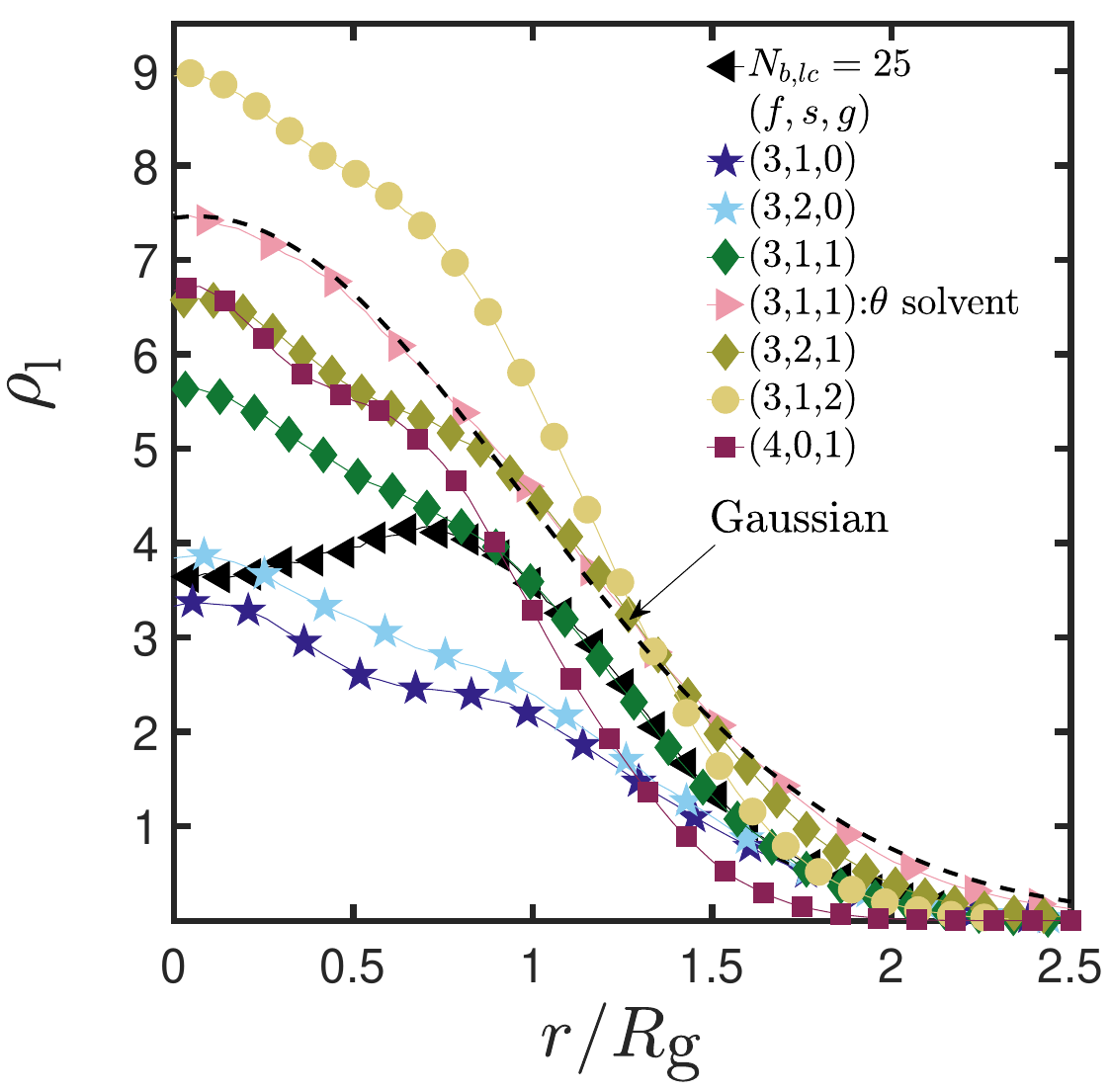}} &
			   \hspace{1cm} &
			\resizebox{8cm}{!} {\includegraphics[width=8cm,height=!]{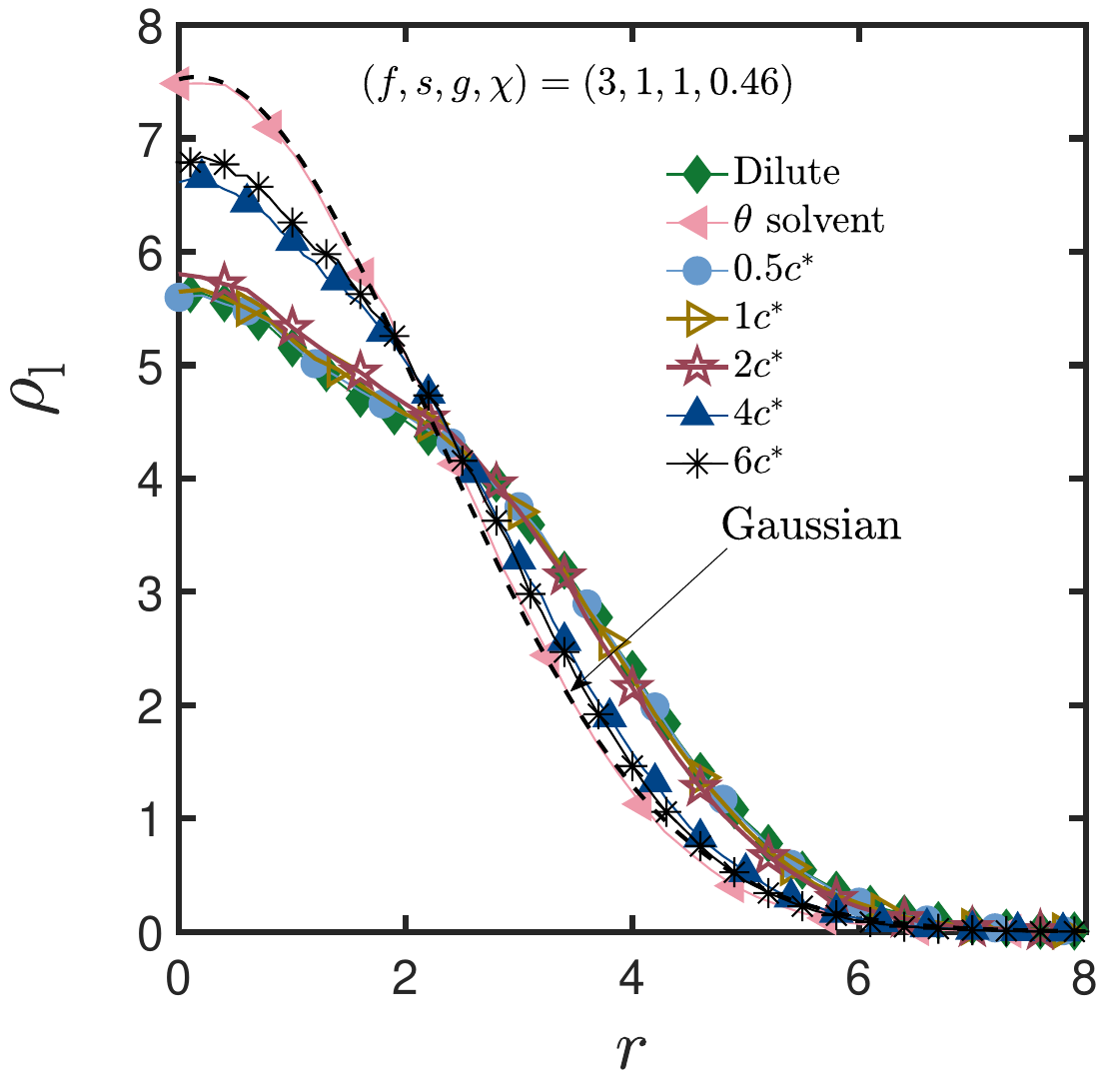}} \\
			(a) &    \hspace{1cm} & (b) 
		\end{tabular}
	\end{center}
	\vspace{-5pt}
        \caption{ (Color online) Linear bead density distribution of dendrimers (a) Effect of architecture on the internal bead density distribution of dendrimers along the major axis in dilute solution. Black triangles represent linear chain with 25 beads and dendrimers are represented using the combination $f, s, g$. The pink triangle represents a $f=3,s=1,g=1$ dendrimer in theta solvent conditions. The dashed line is a Gaussian fit to this data. (b) Effect of concentration on the bead density distribution of $f=3,s=1,g=1$ dendrimer in a semidilute solution of linear chains with $\chi=0.46$. The pink triangles represent dendrimer in theta solvent conditions and the dashed line is a Gaussian fit to this data.}
    \label{fig:bead density}
    	\vspace{-10pt}
\end{figure*}

\subsection{Bead density distribution}

Fig.~\ref{fig:bead density}(a) shows the linear bead density distribution along the major axis of the radius of the gyration tensor of linear chains and dendrimers in a dilute solution. Linear chains have a dip in the density distribution at the centre of the molecule \cite{kumari2020computing} implying that in an athermal solution, beads near the centre have a higher effective excluded volume repulsion. The density distributions of dendrimer molecules, on the other hand, are peaked at the centre of mass. The linear number density is the density of the projected coordinates of the beads along the major axis. A higher functionality $f$ of a dendrimer leads to the localisation of beads closer to the core, increasing the linear core density. The other factor responsible for higher core linear density is the presence of a larger number of beads in the molecule (the projected coordinates of which are likely to be higher near the core). The other characteristic feature of the linear number density of dendrimers is the presence of a non-Gaussian shoulder region midway between the core and periphery. We do not have a simple geometrical explanation for this behaviour. The linear bead density distribution of $f=3,s=1,g=1$ in theta solvent conditions shows a Gaussian distribution, with a significantly high core density compared to that in athermal conditions. This could be due to the folding back of the dendrimer arms to the core due to the absence of the excluded volume interactions.

The density distribution is different along the 3 axes of the gyration tensor, with the internal bead distribution being more clearly non-Gaussian along the major axis (see Fig.~S5(a) of the Supplementary Material) and Gaussian-like along the minor axes. The effect of increasing solution concentration on the bead density distribution along the major axis of generation-1 dendrimer is plotted in Fig.~\ref{fig:bead density}(b). Due to the excluded volume interactions in an athermal solution, dendrimer molecules tend to swell, resulting in a lower core density. The shoulder that is present in the density distribution in the dilute case decreases as the concentration of the background linear chains increases. As is well known, at higher concentrations, the excluded volume interactions are screened in the molecule, and the onset of this is revealed in the figure. Thus, the molecule is shrinking, bringing beads closer to the centre of mass. After $2c^*$, the distribution approaches Gaussian, which corresponds to the theta solvent conditions. It should be noted that all simulation results reported in this manuscript for theta solvents have been obtained with the well depth set to $\epsilon=0.45$ in the SDK potential.

The bead density distribution in concentric shells of equal volume, referred to as volumetric bead density, was also calculated (see Fig.S5(b) and (c) of the Supplementary Material). All dendrimer architectures show a high core density in dilute solutions, which monotonically decreases towards the periphery. As the solution concentration increases, the volumetric bead density approaches that in theta solvent conditions, similar to linear bead density.

\subsection{Mean squared displacement}

\begin{figure*}[tbph]
	\begin{center}
		\begin{tabular}{ccc}
			\resizebox{7.8cm}{!} {\includegraphics[width=7cm]{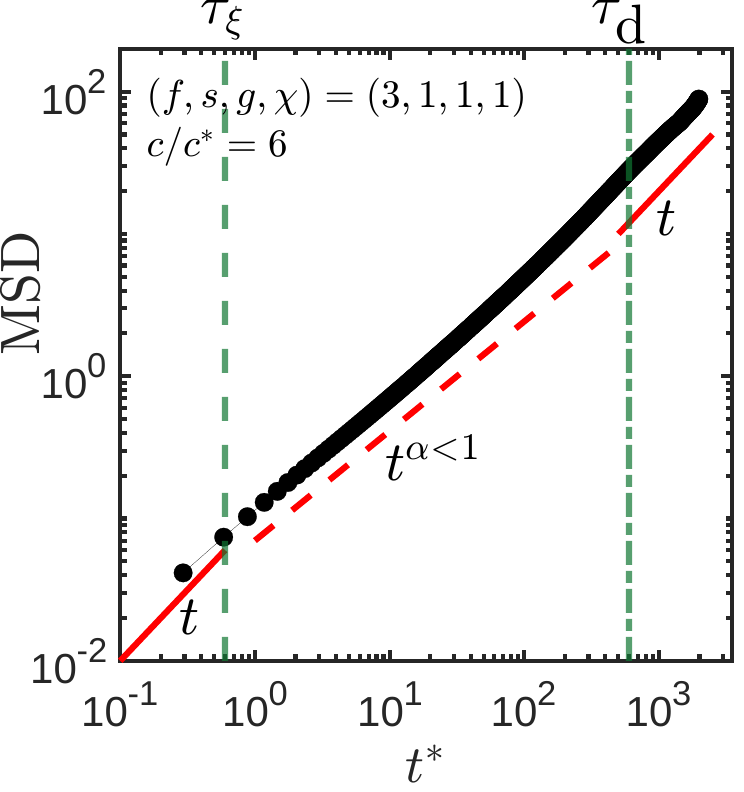}} &
			\hspace{1cm} &
			\resizebox{8cm}{!} {\includegraphics[width=8cm]{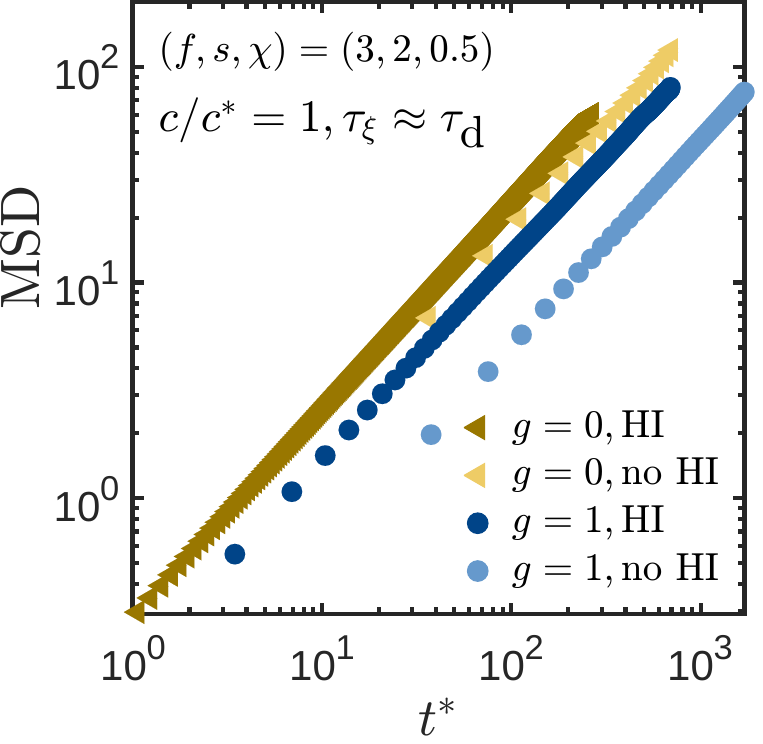}}\\
			(a) & \hspace{1cm}& (b)  \\
		\end{tabular}
	\end{center}
		\vspace{-10pt}
	\caption{ (Color online) Mean squared displacement of dendrimers as a function of time (a) The mean squared displacement of a dendrimer with $f=3,s=1,g=0, \chi=0.5$ in a solution of linear chains of concentration $6c^{\ast}$. The vertical dashed and dashed-dotted lines are $\tau_{\xi}$ and $\tau_{d}$, respectively. (b) Effect of hydrodynamic interaction on the mean squared displacement of dendrimers. The same symbol shapes represent identical systems. The concentration is fixed at $1c^{\ast}$ and $f=3$ for all dendrimers. For the star polymer, $s=1, \chi=0.5$ and for generation one dendrimer,  $s=2, \chi=0.46$ .}
    \label{fig:MSD}
    	\vspace{-10pt}
\end{figure*}

According to the coupling theory~\cite{cai2011mobility} for the diffusion of nanoparticles in a  polymer solution, the particle dynamics is dependent on its size relative to the solution correlation length. Nanoparticles in Regime-1 pass smoothly through the polymer mesh, whereas particles larger than $\xi$ get locally trapped and become subdiffusive at intermediate time scales, with the diffusion exponent $\alpha^{\textrm{NP}}=0.5$. However, a recent MPCD study~\cite{chen2018coupling} has shown that $\alpha^{\textrm{NP}}$ gradually decreases from unity to 0.5, with an increase in solution concentration due to its coupling to the polymer centre of mass motion. To understand the behaviour of soft dendrimers in polymer solutions, the mean squared displacement of dendrimers was plotted as a function of time as shown in Fig.~\ref{fig:MSD}(a). The system in the figure corresponds to Regime-2 ($2R_{\textrm{g}}^{\textrm{d}}>\xi$). The dendrimer is observed to be diffusive ($\alpha^{\textrm{d}}=1$) at times less than $\tau_{\xi}$. It then becomes subdiffusive at intermediate times ($\tau_{\xi} < t < \tau_{\textrm{d}}$) with $\alpha^{\textrm{d}}<1$, followed by normal diffusion at long times ( $t > \tau_{\textrm{d}}$), similar to the behaviour of a nanoparticle in the polymer solution. With increasing concentration, the mean squared displacement of dendrimers decreases and the subdiffusive regime increases(Fig.~S6(a) of the Supplementary Material). This is because $\tau_{\xi}$ (eqn~\eqref{eq:tau_xi}) decreases and $\tau_{\textrm{d}}$ (eqn~\eqref{eq:tau_d}) increases with increasing concentration (see Tables~S1, S2 and S3 in of the Supplementary Material). Therefore, dendrimers remain subdiffusive for a longer span at higher concentrations.   
	
\begin{figure*}[tbph]
	\begin{center}
		\begin{tabular}{cc}		    
			\resizebox{8.5cm}{!} {\includegraphics[width=4cm]{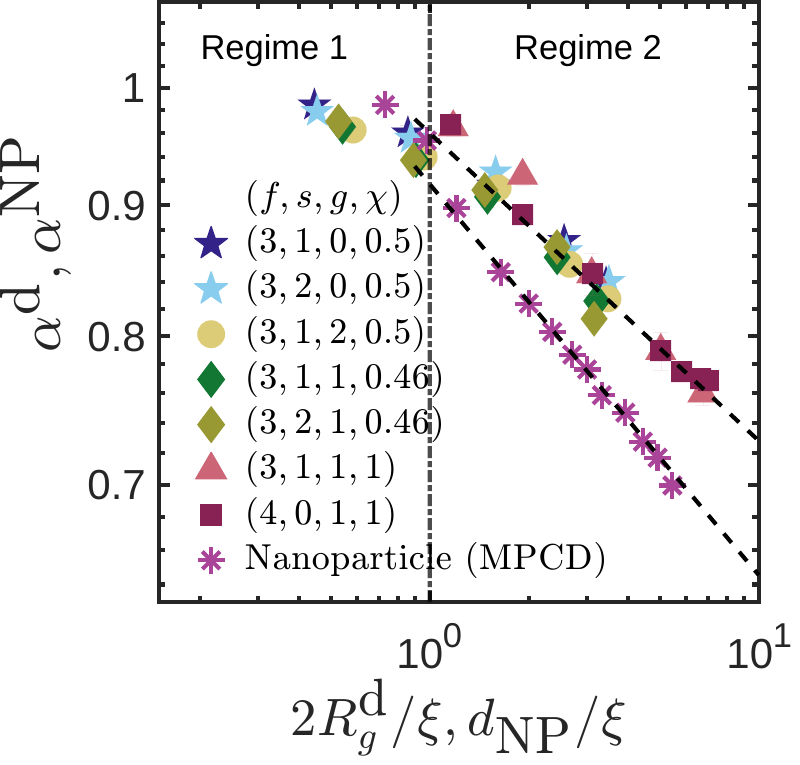}} &
			\resizebox{8.3cm}{!} {\includegraphics[width=3.9cm]{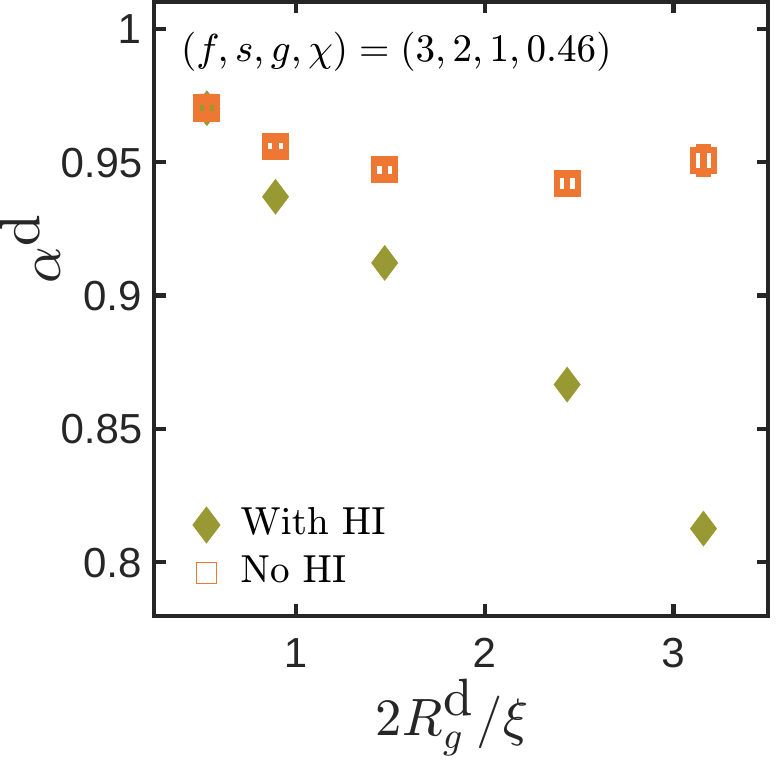}}\\
			(a) & (b)  \\
			\resizebox{8.5cm}{!} {\includegraphics[width=4cm]{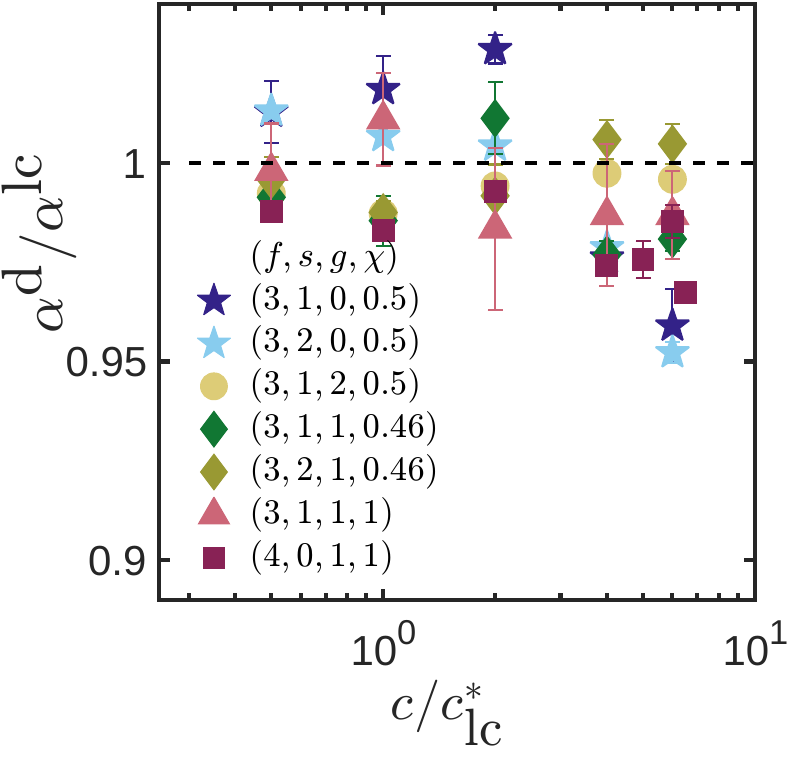}} &
			\resizebox{8.5cm}{!} {\includegraphics[width=4cm]{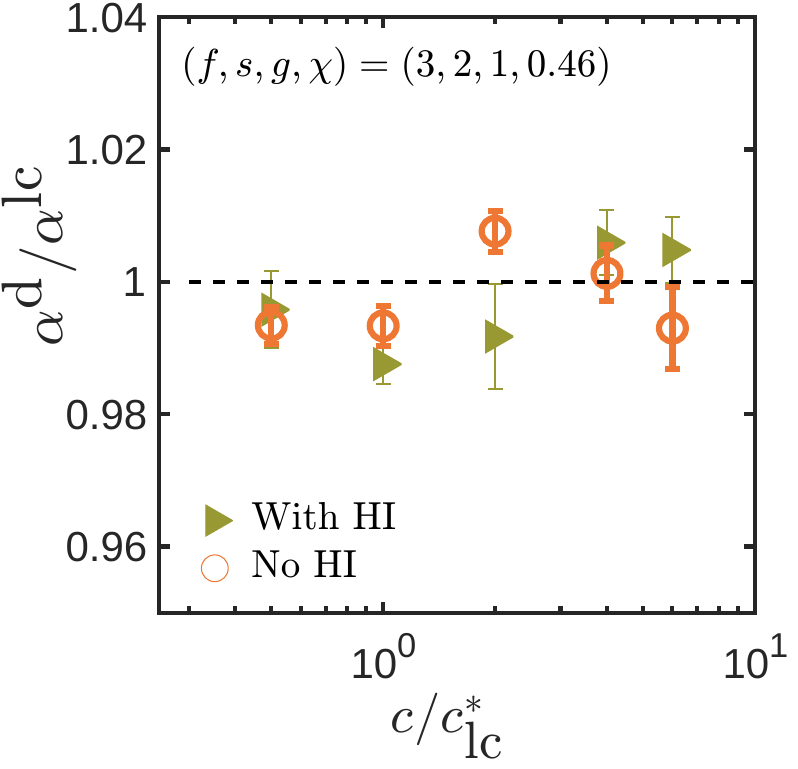}}\\
			(c) & (d)  \\
		\end{tabular}
	\end{center} 
	\vspace{-10pt}
	\caption{ (Color online) The diffusion exponents (a) The diffusion exponent of dendrimers of different architectures is plotted as a function of the ratio of its size to the correlation length in the two regimes. Data from work by \citet{chen2018coupling} for a nanoparticle in semidilute polymer solution of linear chains is included. The dashed lines are a guide to the eye and the vertical line demarcates the two regimes. (b) The diffusion exponent of $f=3,s=2,g=1$ dendrimers in solutions with and without hydrodynamic interactions with $\chi=0.46$. (c) The ratio of the diffusion exponents of dendrimers to that of linear chains in its background as a function of concentration for dendrimers of different architectures. The dashed line is equal to unity. The symbols represent dendrimers in the solution with topological parameters $f,s,g,\chi,$. (d) The ratio of the diffusion exponents of dendrimers to that of linear chains in solutions with and without hydrodynamic interactions. The system considered is the same as that in (b).}
    \label{fig:alpha}
    	\vspace{-10pt}
\end{figure*}

To understand the role of hydrodynamic interactions (HI) in determining the dynamics of dendrimers, simulations were performed with and without HI. From the mean squared displacement versus time plots, it is observed that HI promotes the movement of the dendrimers in the solution, thus leading to a higher MSD (Fig.~\ref{fig:MSD}(b)). This is in line with earlier observations for nanoparticles in polymer solutions\cite{chen2017effect}. 
The architecture of dendrimers also affects its mean squared displacement because its size increases with generation number, keeping $f$ and $s$ fixed. At a constant concentration, a generation 2 dendrimer will experience more hindrance to its motion compared to a simple star polymer ($g=0$), causing the former to diffuse slower than the latter (see Fig.~S6(b) of the Supplementary Material). However, we can collapse $\textrm{MSD}$ for different architectures by following a simple scaling relation as shown in Fig.~S6(c) of the Supplementary Material. 

\subsection{Diffusion exponent}

As mentioned earlier, dendrimers that are larger than the solution correlation length become subdiffusive at intermediate time scales. We have extracted the diffusion exponents, $\alpha_{\textrm{d}}$, for dendrimers of different architectures at intermediate times ($\tau_{\xi} < t < \tau_{\textrm{d}}$) and plotted them as a function of the ratio of their diameter to the correlation length (Fig.~\ref{fig:alpha}(a)). Dendrimers smaller than correlation length (Regime-1) exhibit diffusion with $\alpha^{\textrm{d}}=1$, whereas dendrimers larger than correlation length (Regime-2) experience hindrance to its movement and becomes subdiffusive $\alpha^{\textrm{d}}<1$. It is remarkable that the diffusion exponents of all dendrimers collapse onto a universal curve independent of the architecture of dendrimers. The figure also contains the diffusion exponent of nanoparticles from MPCD simulations carried out by ~\citet{chen2018coupling}. Dendrimers have a higher diffusion exponent compared to a hard sphere with the same $d_{\textrm{NP}} / \xi$ value in regime-2, where  $d_{\textrm{NP}}$ is the diameter of the nanoparticle. This may be due to the concentration-dependent size of the dendrimer, unlike that of a nanoparticle, which is a hard sphere. A dendrimer trapped in a polymer mesh can move out of it due to its own fluctuation along with fluctuations of the mesh-forming strands. It is consequently more diffusive compared to a nanoparticle of comparable size. Instantaneous diffusion exponents also can be obtained by taking the time derivatives of mean squared displacement. It was observed that $\alpha$ thus obtained at intermediate times is similar to that calculated from the time exponent of $\textrm{MSD}$ between $\tau_{\xi}$ and $\tau_{\textrm{d}}$. It starts increasing beyond $\tau_{\textrm{d}}$ and tends to unity (see Fig.~S8 of the Supplementary Material).

Fig.~\ref{fig:alpha}(b) shows the influence of hydrodynamic interactions on the diffusion exponents of dendrimers in a solution. The exponent $\alpha^{\textrm{d}}=1$ for both cases as dendrimers exhibit pure diffusion at low concentrations. As concentration increases, the diffusion exponent of dendrimers with HI decreases monotonically, while that without HI levels off. The diffusion exponent of nanoparticles in semidilute solutions has also been studied using MPCD simulations. While an earlier study~\cite{chen2017effect} indicated behaviour different from that observed here, the observations of a more recent study by \citet{chen2018coupling} are in agreement with our results. Paradoxically, HI seems to slow down the motion of dendrimers in the intermediate times as shown by the decreasing values of the diffusion exponent, while, as seen earlier, it enhances the long-time diffusivity. This is perhaps due to the backflow caused by HI that prevents the escape of dendrimers from polymer `cages' at intermediate times. However, at long times, HI enhances its motion through polymer-mediated motion as shown by the MSD plots (Fig.~\ref{fig:MSD}(b)). A better insight into the effect of HI at different time scales is obtained from the velocity field due to HI about the dendrimer molecule (see Fig.~S9 of the Supplementary Material). At a length scale comparable to dendrimer size, which corresponds to intermediate times, each monomer on a dendrimer molecule experiences a force of different magnitude and direction at every instant in time due to its configuration. This causes the molecule to change shape and move in random directions, thus slowing its diffusion process. Whereas, when viewed from larger length scales, which corresponds to $t>\tau_d$, the dendrimer is like a particle placed at its centre of mass. Its diffusion is enhanced by the unidirectional flow due to HI.

We also examined the dynamics of linear polymer chains in the background solution at intermediate time scales. Similar to dendrimers, linear chains also enter a subdiffusive regime at intermediate times, and the diffusion exponents of both species are highly correlated as shown in Fig.~\ref{fig:alpha}(c). This is similar to the observation made by \citet{chen2018coupling} for nanoparticles. The correlation has been attributed to the coupling of nanoparticle motion to the centre of mass of the linear chains in the solution along with the segmental relaxation time of polymer, which results in the gradual decrease of diffusion exponent from $1$ to $0.5$ (Fig.~\ref{fig:alpha}(a)), unlike the predictions of coupling theory~\cite{cai2011mobility} which predicts a constant value of $0.5$ as the diffusion exponent of nanoparticle. It appears that the correlation is lost for some of the architectures at higher concentrations, however, the scatter is due to insufficient statistics. Since the difference across the entire range of concentrations considered is less than $5\%$, the correlation exists at all concentrations and across all architectures. 

The coupling of the motion of dendrimers and linear chains would at first sight be attributable to the presence of long-range hydrodynamic interactions. However, the strong correlation between the diffusion exponents of both species exists even in the absence of HI as shown in Fig.~\ref{fig:alpha}(d). This confirms that the correlation is due to comparable time scales of relaxation between the dendrimers and linear chains, and not due to the many-body HI\cite{chen2018coupling}.  

\begin{figure}[b]   
    \begin{center}
    {\includegraphics[width=7.0cm,height=!]{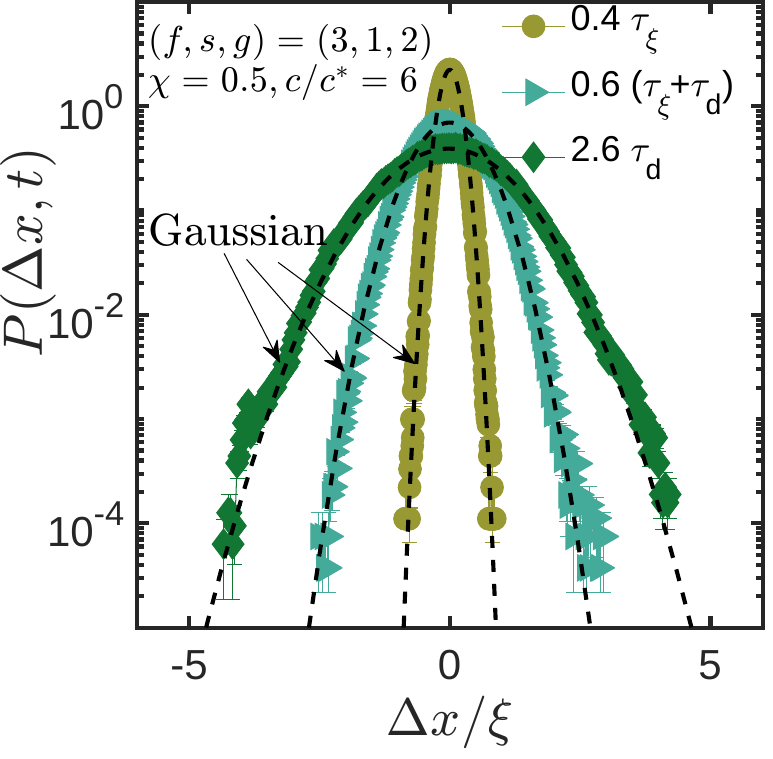}}
	\end{center}
        \vspace{-10pt}
	\caption{(Color online) Probability distribution function of displacement for a generation two dendrimer ($f=3, s=1, \chi=0.5$) at concentration $c=6c^{\ast}$ at different times ( at $t = 0.4 \tau_{\xi} \, , \,  0.6(\tau_{\xi}+\tau_{\textrm{d}}) \, , \,  2.6 \tau_{\textrm{d}}$).} 
        \label{fig:PDD}
        \vspace{-10pt}
\end{figure}

\subsection{Probability distribution function of displacement}

To get a better insight into the characteristic features of dendrimer dynamics, we analyzed its probability distribution function of displacement at various time intervals (Fig.~\ref{fig:PDD}). The displacements are normalized by the correlation length as it is the length scale at which the dendrimer experiences the presence of linear chains. It is clear from the distribution function that even though dendrimers become subdiffusive at intermediate time scales, they demonstrate Gaussian probability distributions. On the other hand, many systems of nanoparticles in complex environments exhibit non-Gaussian distributions~\cite{phillies2015complex,kumar2019transport}. There has been an increasing interest in studying Fickian systems with non-Gaussian probability distributions over the past few years~\cite{wang2012brownian,wang2009anomalous,slim2020local,smith2021dynamics}. In these systems, the observed deviation from Gaussian behaviour has been attributed to the temporal~\cite{lampo2017cytoplasmic,he2016dynamic} or spatial~\cite{jee2014nanoparticle,guan2014even} heterogeneities in the complex environment leading to a distribution of diffusivities, due to the shape anisotropy of the tracer~\cite{smith2021dynamics} or due to activated hopping mechanisms~\cite{xue2016probing}. 

Subdiffusive motion has been explained using two mathematical models: the continuous-time random walk model (CTRW)\cite{montroll1965random} and fractional Brownian motion (FBM)\cite{mandelbrot1968fractional}. While the former has a series of trapping and discrete jumps leading to subdiffusion and non-Gaussian probability distribution functions, the latter is a Gaussian process in which successive steps are correlated (non-Markovian). An analysis of the motion of the centre of mass of the dendrimer molecule confirms the absence of long waiting times and hopping (shown in Fig.~S7 of the Supplementary Material). Therefore, it appears from the $\textrm{MSD}$ plots and the probability distribution of displacements that the subdiffusive yet Gaussian behaviour of dendrimers mimics fractional Brownian motion rather than continuous time random walk motion. Similar results were obtained for polymer-grafted nanoparticles in a recent molecular dynamics study \cite{chen2022diffusion}. 

Fig.~\ref{fig:PDD} shows the probability distribution functions (PDD) for a $f=3, s=1, g=2 $ (Fig.~\ref{fig:Dendrimer_schematic}(e)) dendrimer at different time scales at a fixed concentration. When $t<\tau_{\xi}$ and $t>\tau_{d}$, dendrimers exhibit normal diffusion and hence the PDD is expected to be Gaussian. However, in the intermediate time scales when the dendrimer is subdiffusive, they have a Gaussian PDD. Note that the correlation length, $\tau_{\xi}$ and $\tau_{\textrm{d}}$ are concentration and architecture-dependent quantities. The probability distributions remain Gaussian at all concentrations irrespective of the architecture and the time (see Fig.~S10 of the Supplementary Material).

\subsection{Diffusivity as a function of concentration}

\begin{figure}[t]   
    \begin{center}
    {\includegraphics[width=8cm,height=!]{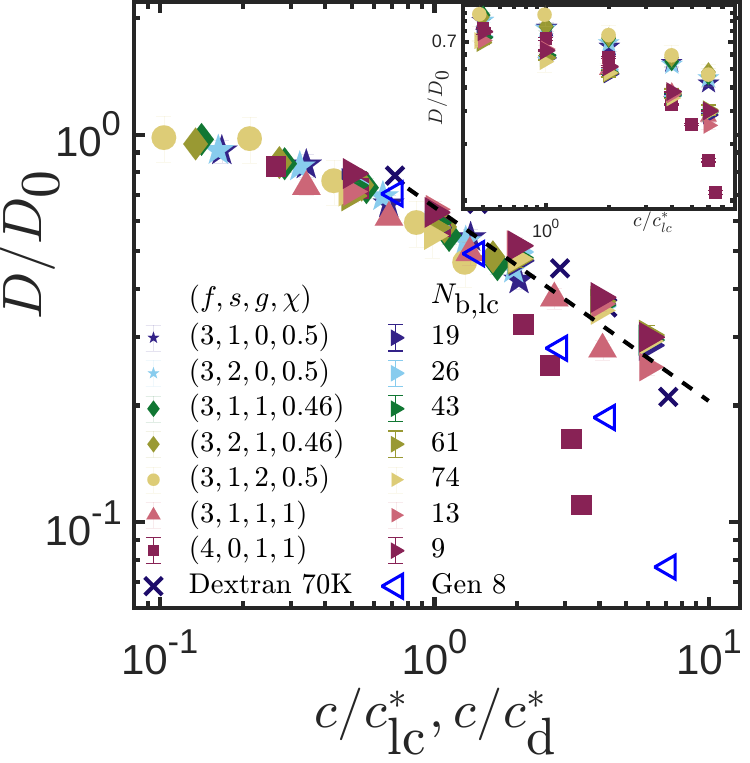}}
	\end{center}
        \vspace{-10pt}
	\caption{(Color online) Universal behaviour of normalised diffusivity 
        of dendrimers and linear chains as a function of the ratio of monomer concentration normalized by the respective $c^{\ast}$ of each species. The dendrimer architecture is included in the order ($f,s,g,\chi$) and $N_{\textrm{b}}^{\textrm{lc}}$ is the number of beads on a linear chain.  Symbols in the same colour belong to the same system, with right triangles representing linear chains. The dashed line is the scaling law given by eqn~\eqref{eq:diffusivity_scaling}. Experimental data for generation 8 dendrimers and dextrans from \citet{cheng2002diffusion} is included. Inset: Normalised diffusivity of dendrimers and linear chains in the solution as a function of the ratio of monomer concentration normalized by the overlap concentration of linear chains, $c^{\ast}_{\textrm{lc}}$.} 
        \label{fig:Diffusivityvsconcentration}
        \vspace{-10pt}
\end{figure} 

The diffusivity of linear chains in semidilute polymer solutions is known to follow a universal scaling law given by~\cite{doi1988theory}
\begin{align}\label{eq:diffusivity_scaling}
    D^{\textrm{lc}} = D_{\textrm{0}}^{\textrm{lc}}\left(\frac{c}{c^{\ast}_{\textrm{lc}}}\right)^{\frac{\nu-1}{3\nu -1}}
\end{align}

\begin{figure*}[t]
	\begin{center}
		\begin{tabular}{ccc}
			\resizebox{8cm}{!} {\includegraphics[width=8cm]{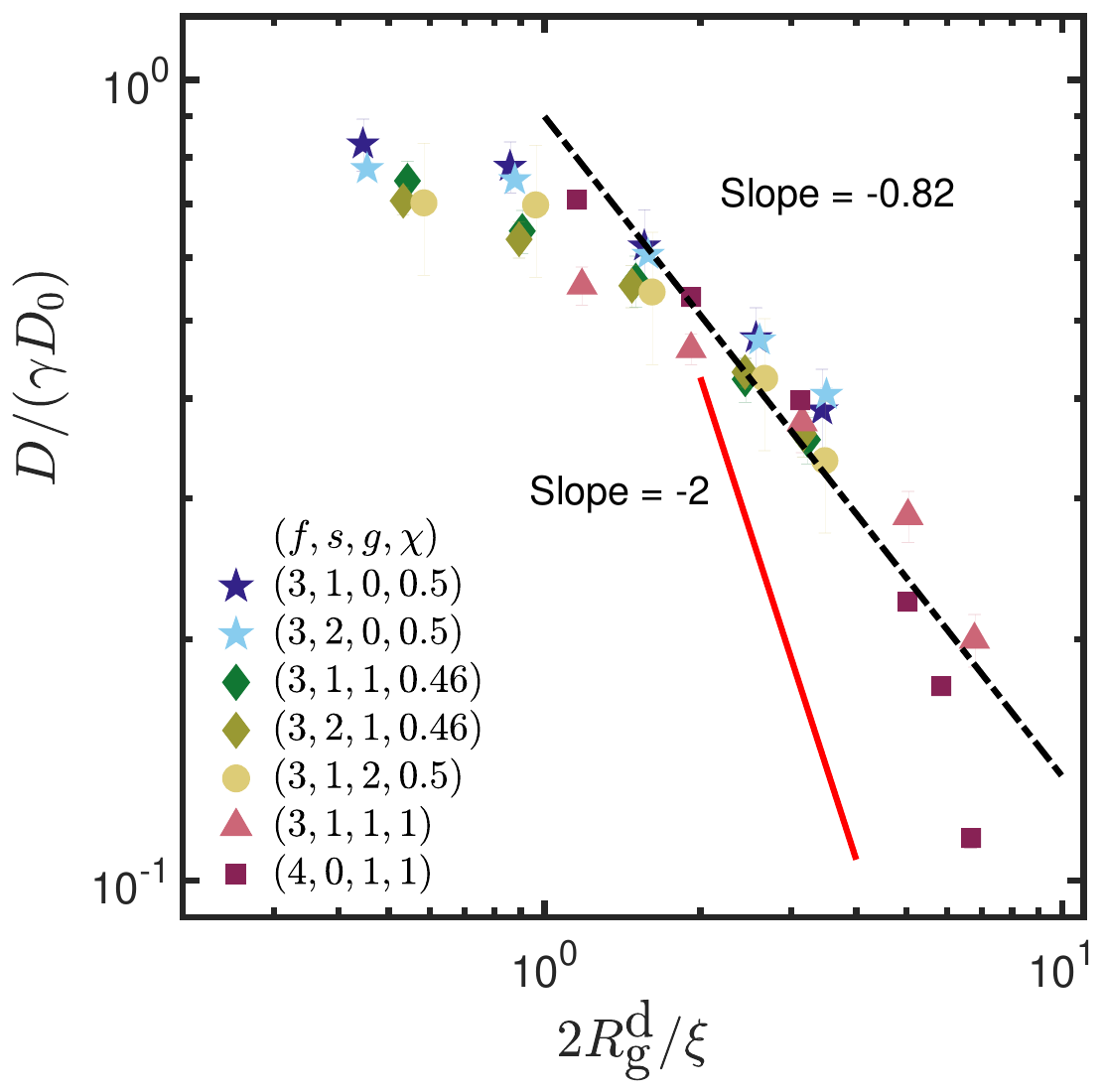}} &	
        \hspace{1cm}		&
			\resizebox{8cm}{!} {\includegraphics[width=8cm]{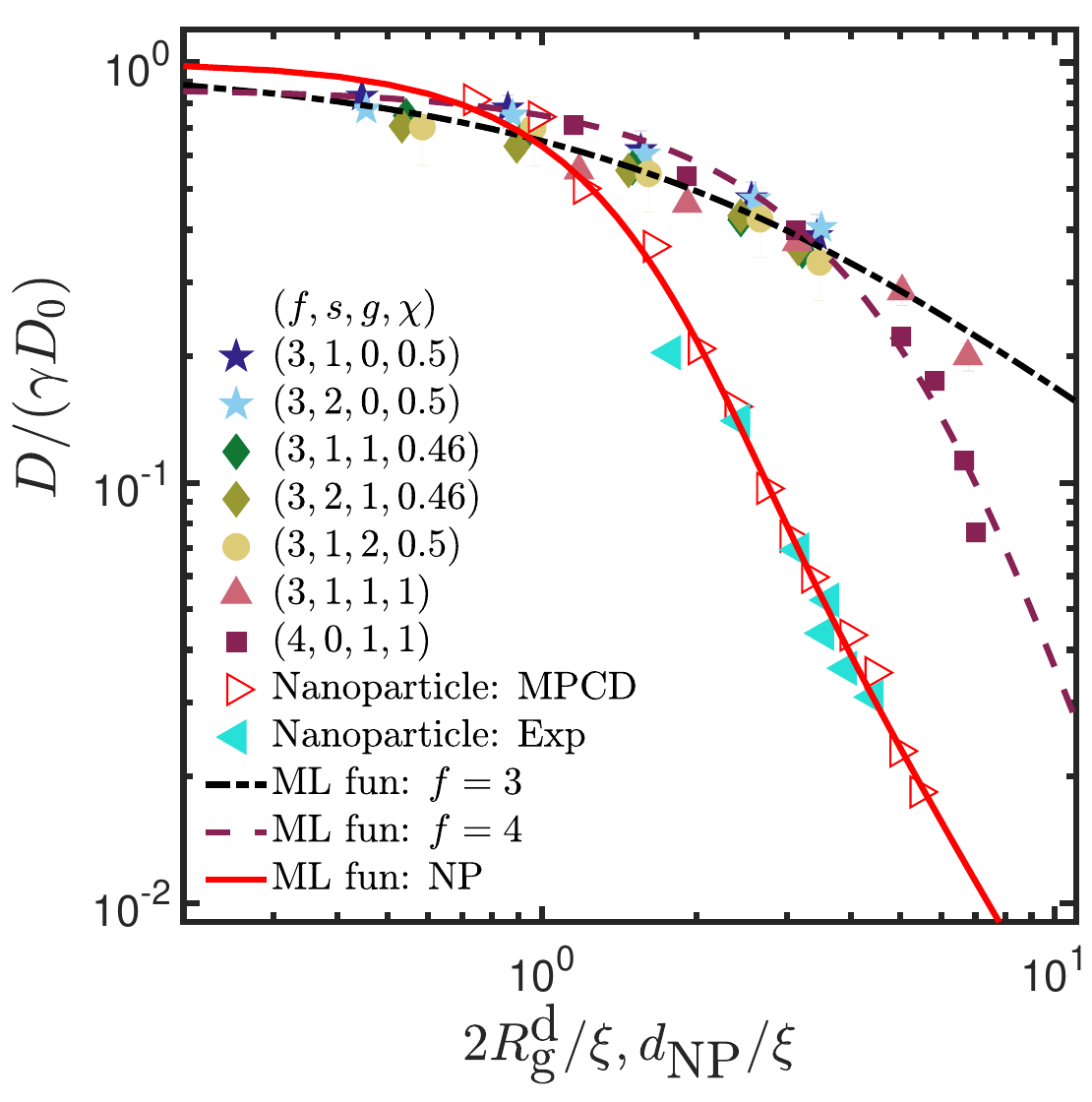}}\\
			(a)  &   \hspace{1cm} & (b)  
		\end{tabular}
	\end{center}
	            \vspace{-10pt}
        \caption{ (Color online) (a) Normalised diffusivity for dendrimers of all architectures in the solution of linear chains as a function of the ratio of its size to the correlation length of the solution. Symbols represent dendrimers with topological parameters in the order $f, s, g, \chi$. The dashed line represents the scaling law derived for dendrimers (eqn~\eqref{eq:new_scaling}) and the solid line represents the predictions of coupling theory~\cite{cai2011mobility}. (b) Normalized diffusivity for dendrimers of all architectures along with MPCD~\cite{chen2018coupling} (red empty triangles) and experimental data~\cite{kohli2012diffusion} (cyan filled triangles) from literature for nanoparticles. The lines are Mittag-Leffler fits (ML fun) to various systems with parameters $m$,$a$ and $p$ given in Table~\ref{tab:ML_params}.} 
    \label{fig:ML_fits}
	            \vspace{-10pt}
\end{figure*}

\noindent To understand the long-time dynamics of the simulated hybrid system, we calculated the long-time diffusivity of dendrimers and linear chains in the solution. The diffusivity, normalized by the respective dilute diffusivity, of dendrimers and linear chains decreases with an increase in concentration as shown in Fig.~\ref{fig:Diffusivityvsconcentration}. Dendrimers of functionality $f=3$ and linear chains in the solution follow the universal scaling law for linear chains when the monomer concentration ($c$) is normalized with respect to the overlap concentration of each species. However, the diffusivity of a $f=4$ dendrimer does not follow eqn~\eqref{eq:diffusivity_scaling}. Rather, it drops rapidly after $c/c^{\ast}_{\textrm{d}}=1$. 

There is a remarkable collapse of diffusivity of dendrimers of several architectures in spite of having different anisotropies. Therefore it would appear that anisotropy does not affect their dynamics. We have also calculated the distribution of the radius of gyration of dendrimers in a dilute solution (shown in Fig.~S3(a) of the Supplementary Material) and examined the effect of concentration (Fig.~S3(b) and (c)). This was obtained using the $R_g^d$ from many trajectories at different instances in time and binning them. The variance of the distribution of $R_g^d$ gives a measure of the size fluctuations. The $f=4$ dendrimer has a significant decrease in variance compared to the $f=3$ dendrimers in dilute solution. Note that the $(f,s,g)=(3,1,1)$ dendrimer has a similar number of beads per molecule to that of the $f=4$ dendrimer, and indeed the $(f,s,g)=(3,1,2)$ and $(f,s,g)=(3,2,1)$ dendrimers have a higher number of beads. Yet the $f=4$ dendrimer has a smaller variance of the distribution of $R_g$.  This suggests that a decrease in the fluctuations in size might be a more important factor in determining the difference observed in the behaviour of the functionality four dendrimer, rather than its anisotropy.  The variance in $R_g^d$ seems to decrease rapidly with concentration for both the architectures that are displayed in Fig.~S3(b) and (c) of the Supplementary Material.
Our simulation results are compared with experimental data for a generation 8 dendrimer and Dextran molecules reported by \citet{cheng2002diffusion}. A generation 8 dendrimer is a dense and compact molecule with fractal dimension $\approx 3.1$ whereas dextrans are slightly branched molecules with fractal dimension $2.3$~\cite{cheng2002diffusion}. The diffusivities of $f=3$ dendrimers and linear chains in our simulations behave similarly to Dextran molecules while the $f=4$ dendrimer behaves more like the generation 8 dendrimer, confirming that the former has a more compact structure. 

While dendrimers are just one example of soft colloids, there are other systems like microgels, multiarm stars, and hairy colloids that may also be considered as soft colloids. Our search of the literature seems to suggest that a systematic study of the dependence of diffusivity on the concentration of the semidilute solution has not been reported for this variety of soft colloids, rather the effect of temperature and pH has been well studied. Therefore, we are unable to compare our results directly with data for a range of these systems. On the other hand, there is data by \citet{poling2019soft} on polymer-grafted nanoparticles (PGNP) or hairy colloids on the dependency of its diffusivity on concentration. They have observed that the normalised diffusivity of a variety of PGNP-linear chain systems can be collapsed when represented in terms of $\left( c/c^{\ast}_{\textrm{lc}} \right) \left( M_{w,f}/M_{w,g}\right)^{-1/8}$. In our simulations, since, $c^{\ast}_{{\textrm{d}}}$ can also be represented in terms of $c^{\ast}_{{\textrm{lc}}}$ (for a detailed derivation see the Supplementary Material), the overlap concentration of dendrimers, $c^{\ast}_{\textrm{d}}$, can be written as $c^{\ast}_{{\textrm{d}}}=c^{\ast}_{{\textrm{lc}}}\left(a^{\textrm{lc}}/a^{\textrm{d}}\right)^{3}\left(N_{\textrm{b}}^{\textrm{d}}/N_{\textrm{b}}^{\textrm{lc}}\right)^{1-3\nu}$, where the constants $a^{\textrm{d}}$ and $a^{\textrm{lc}}$ are defined by $a^{\textrm{d}}=R_{\textrm{g0}}^{\textrm{d}}/(N_{\textrm{b}}^{\textrm{d}})^{\nu}$ and $a^{\textrm{lc}}=R_{\textrm{g0}}^{\textrm{lc}}/(N_{\textrm{b}}^{\textrm{lc}})^{\nu}$. Using this relation, the diffusivity of all the dendrimer architectures can be collapsed by plotting the data in Fig.~11 in terms of $c^{\ast}_{{\textrm{lc}}} \left ( N_{\textrm{b}}^{\textrm{d}}/N_{\textrm{b}}^{\textrm{lc}} \right )^{-0.76}$. Note that the exponent of the normalisation factor is different from that reported by \citet{poling2019soft}. A comparison of the results displayed in Fig.~11 for dendrimers with the data of \citet{poling2019soft} for one of their systems, is shown in Fig.~S11 of the Supplementary Material.

\subsection{Scaling law for dendrimer diffusivity}

Dendrimers are soft polymers, the size of which changes with the concentration of the background solution (Fig.~\ref{fig:rg_scaling}), unlike nanoparticles that have a fixed size. The theories that govern nanoparticle dynamics in homopolymer solutions are consequently not directly applicable to dendrimers. Combining eqn\eqref{eq:rg_scalinglaw} and \eqref{eq:diffusivity_scaling} for dendrimer radius of gyration and diffusivity, and eqn\eqref{eq:xi} for correlation length, a new scaling law for the dynamics of dendrimers in semidilute polymer solutions can be derived. The normalized diffusivity of dendrimers as a function of the ratio of its size to the correlation length of the solution is given by (for detailed derivation see the Supplementary Material): 
\begin{align}\label{eq:new_scaling}
    \frac{D^{\textrm{d}}}{D_{\textrm{0}}^{\textrm{d}}} = \gamma \left( \frac{2 R_{\textrm{g}}^{\textrm{d}}}{\xi}\right) ^{2(\nu-1)}
\end{align}
where $\gamma=\left( \dfrac{\beta^{\mu}}{2\chi^{(1+3\mu)}} \right)^{2(\nu-1)}$ with $\beta={N^{\textrm{d}}_{\textrm{b}}}/{N^{\textrm{lc}}_{\textrm{b}}}$. 
\vspace{5pt}

Fig.~\ref{fig:ML_fits}(a) shows our simulation data fitted with the scaling law for dendrimer diffusivity given by eqn~\eqref{eq:new_scaling}. As expected, the normalized long-time diffusivity is equal to 1 in the dilute limit and a crossover to the semidilute regime happens when the size of the dendrimer is comparable to the correlation length ($2 R_{\textrm{g}}^{\textrm{d}} \approx \xi$). Data for functionality 3 dendrimers collapse to a universal curve when divided by the factor $\gamma$, which depends on $\chi$ and $\beta$. However, the $f=4$ (Fig.~\ref{fig:Dendrimer_schematic}(f)) dendrimer does not follow this power law at sufficiently high concentrations. In eqn~\eqref{eq:new_scaling}, the exponent $2 \left( \nu -1\right)=-0.82$ when $\nu=0.588$. The $-2$ power law from coupling theory for nanoparticles is also displayed for comparison.

\begin{table}[b]
    \caption{Values of parameters used in the Mittag-Leffler fits to various data sets in Fig. ~\ref{fig:ML_fits}(b)}
    \label{tab:ML_params}
        \centering
    \setlength{\tabcolsep}{10pt}
    \renewcommand{\arraystretch}{1.5}
        \begin{tabular}{|c|c|c|c|}
         \hline
 $\textrm{System}$     & $m$ 
            & $a$  & $p$
\\                
\hline
\hline
            $f=3$
            & $-0.43$
            & $0.59$
            & $0.82$
\\
            $f=4$
            & $-0.11$
            & $0.79$
            & $1.40$
\\
            $\textrm{Nanoparticle}$
            & $0.45$
            & $0.78$
            & $2.00$
\\
\hline
    \end{tabular}
    \end{table}

In the case of nanoparticles, two competing theories predict different scaling behaviour for diffusivities: the hydrodynamic model gives a stretched exponential dependence of the normalized long-time diffusivity of nanoparticles on its size relative to the system correlation length, whereas the coupling theory predicts a power law dependence with an exponent $-2$. These theories have been well supported by experiments and simulations ~\cite{chen2017effect,chen2018coupling,kohli2012diffusion,de2015scaling,poling2015size}. Since the diffusivity is normalized by its dilute value, in both cases, the ratio tends to one at low concentrations. It is interesting to consider the disparity in scaling predictions with the help of the Mittag-Leffler (ML) function, which is unique as it becomes a stretched exponential at small values of its argument $x$, while it is a power law at large values of $x$~\cite{mainardi2020mittag,jaishankar2013power}. In this study, we are using a modified version of the ML function ($ E_{a,b}\left(mx^p\right)$), given by:
\begin{align}\label{eq:ML_fun}
    E_{a,b}\left(-mx^p\right)=\sum\limits_{k=0}^{\infty}\frac{(-m)^k x^{pk}}{\Gamma \left( ak+b\right)}
\end{align}
\noindent where $m$, $p$, $a$ and $b$ are arbitrary constants. At small $x$, eqn~\eqref{eq:ML_fun} is given by 
\begin{align}\label{eq:smallx_ML}
     E_{a,b}\left(-mx^p\right)=\textrm{exp} \left( \frac{-m x^p}{\Gamma \left( a+b\right)} \right)
\end{align}
 and at large values of $x$ it becomes a power law, with $p$ being the magnitude of the power law exponent:
 \begin{align}\label{eq:largex_ML}
    E_{a,b}\left(-mx^p\right)=\frac{x^{-p}}{m \Gamma \left( b-a\right)}
\end{align}
The diffusivity as a function of the ratio of the size to the correlation length of dendrimers of all architectures and nanoparticles can be fitted using different values of $m$, $a$ and $p$ of the Mittag-Leffler function (details of the code can be found in the Supplementary Material). Since $D/D_0=1$ in the dilute regime, $b=1$, and the expression is given by:
\begin{align}\label{eq:Diffusivity_ML_fun}
    D/D_0=E_{a,1}\left(-m \left( \dfrac{2R_\textrm{g}^{\textrm{d}}}{\xi} \right)^p\right)
\end{align}

\noindent All dendrimers with $f=3$ can be fitted using the modified Mittag-Leffler function (eqn~\eqref{eq:Diffusivity_ML_fun}) with a power law exponent $p=0.82$ as shown in Fig.~\ref{fig:ML_fits}(b). The normalized diffusivity of nanoparticles from MPCD simulations by \citet{chen2018coupling}, which follows the scaling law from coupling theory, is fitted with a power law exponent of $p=2$. Experimental results for nanoparticles that were thought to have a stretched exponential dependence~\cite{kohli2012diffusion} are also included (cyan triangles) for comparison. Interestingly, both these data sets can be fitted well with the same Mittag-Leffler function. The $f=4$ dendrimers that do not follow the scaling law for diffusivity of linear polymers (eqn ~\eqref{eq:diffusivity_scaling}) cannot be fitted with the same power law exponent of $f=3$ dendrimers. Fitting this data using eqn~\eqref{eq:ML_fun} gives $p=1.4$, which is between the power law exponents for $f=3$ dendrimers and hard spheres.  Hence, there is a transition from the polymer-like behaviour to a hard-sphere due to the increased functionality, leading to a compact structure. The values of the various parameters used for fitting in Fig.~\ref{fig:ML_fits}(b) are given in Table ~\ref{tab:ML_params}.

\section{\label{sec:conclusions}Conclusions}

Through Brownian Dynamics simulations, we have examined a hybrid system of soft dendrimers in a semidilute solution of linear polymers. Our investigation focused on the effects of solution concentration, dendrimer architecture, and hydrodynamic interactions on the static and dynamic properties of the dendrimers. We compared our findings to those of a rigid sphere in a semidilute polymer solution. Dendrimers with a mean $R_{\textrm{g}}$ comparable to that of linear chains in the background were chosen for all simulations.    

Our results showed that, unlike hard spheres, the size of dendrimers decreases as the solution concentration increases and follows the universal scaling law for linear chains represented by eqn~\eqref{eq:rg_scalinglaw}. In the dilute state, the shape functions of dendrimers decreased with increasing functionality and generation number but remained constant with concentration. Additionally, the internal bead density of all dendrimers was highest at the core and rapidly decreased towards the periphery of the molecule in a dilute solution. The higher functionality dendrimer showed a higher internal bead density compared to its counterpart with similar bead numbers. At higher concentrations, the bead density distribution approached a Gaussian distribution, similar to that of a polymer in theta solvent conditions. This is due to the screening of excluded volume interactions, which allows the overlap of dendrimer branches.

In the study, it was observed that dendrimers exhibit subdiffusion during intermediate times as they become trapped in cages formed by linear chains. However, the diffusion exponent for dendrimers was found to be higher than that of their hard sphere counterparts at higher concentrations. This can be attributed to the additional fluctuation of dendrimers, which facilitates their motion out of the polymer cages. It was also observed that dendrimers and linear chains have similar diffusion exponents at all concentrations, regardless of the presence or absence of hydrodynamic interactions (HI). As time progresses, dendrimers become diffusive as a result of polymer relaxation. The study also found that HI enhances polymer dynamics at longer times and increases its mean squared displacement. However, the diffusion exponents decrease with concentration when HI is present, unlike the free-draining case where there is no variation.

The long-time diffusivity of dendrimers decreased with increasing solution concentration, similar to that of the linear chains in the background, and collapsed onto a universal curve.  However, the diffusivity of dendrimers does not follow theoretical predictions for nanoparticle diffusivity as a function of its size relative to the solution correlation length. From the analysis of the effect of concentration on dendrimer size and diffusivity, as well as the correlation length of the solution, we developed a scaling law for dendrimer diffusivity based on the ratio of its size to solution correlation length. We found that dendrimers with functionality three follow this scaling law, but more dense dendrimers with higher functionality ($f=4$) do not obey it at higher concentrations. These dendrimers have a power law exponent that falls between that of $f=3$ dendrimers and hard spheres. This shows with increasing functionality, dendrimers shift from polymer-like to hard sphere-like structures. Moreover, using the Mittag-Leffler function, the crossover from the stretched exponential character at low concentrations to the power law behaviour at higher concentrations has been captured for both our simulation results and data from the literature for nanoparticles.

\section*{\label{sec:supp_mat} Supplementary Material}
Additional details of the simulation parameters, some properties calculated, comparison with other soft colloidal systems and derivation of the scaling law for dendrimer diffusivity are given in the Supplementary Material. Section~{SI} of the supplementary material includes the scaling law for the radius of gyration of polymers of different topologies. Section~{SII} contains the regimes based on the relative sizes of dendrimers, linear chains and the correlation length. Other static properties like the distribution of the radius of gyration, asphericity and bead density distributions of dendrimers of different architectures in dilute and semidilute solutions are given in Sections~{SIII}, {SIV} and {SV} respectively. Section~{SVI} contains the mean squared displacement plots and the values of time scales $\tau_{\xi}$ and $\tau_{\rm{d}}$ for various architectures used in this study. The calculation of diffusion exponent as a function of time is given in Section~{SVII} and the velocity field due to hydrodynamic interactions is given in Section~{SVIII}. The probability distribution functions of displacement of various architectures and concentrations are given in Section~{SIX}. Section~{SX} shows the comparison of our simulation results with other soft colloidal systems from the literature. Section~{SXI} gives a detailed derivation for the new scaling law proposed for dendrimer diffusivity as a function of its size relative to the correlation length of the solution. Section.~{SXII} contains the details of the code used for Mittag-Leffler fits and Section.~{SXIII} gives the scaling of diffusivity as a function of the radius of gyration of dendrimers.

\begin{acknowledgments}
We appreciate the funding and support from the IITB-Monash Research Academy and the Department of Biotechnology (DBT), India. We gratefully acknowledge the computational resources provided at the NCI National Facility systems at the Australian National University through the National Computational Merit Allocation Scheme and Adapter Scheme supported by the Australian Government, the MonARCH, and MASSIVE facilities maintained by Monash University.
\end{acknowledgments}


\bibliography{main}

\begin{thebibliography}{81}%
\makeatletter
\providecommand \@ifxundefined [1]{%
 \@ifx{#1\undefined}
}%
\providecommand \@ifnum [1]{%
 \ifnum #1\expandafter \@firstoftwo
 \else \expandafter \@secondoftwo
 \fi
}%
\providecommand \@ifx [1]{%
 \ifx #1\expandafter \@firstoftwo
 \else \expandafter \@secondoftwo
 \fi
}%
\providecommand \natexlab [1]{#1}%
\providecommand \enquote  [1]{``#1''}%
\providecommand \bibnamefont  [1]{#1}%
\providecommand \bibfnamefont [1]{#1}%
\providecommand \citenamefont [1]{#1}%
\providecommand \href@noop [0]{\@secondoftwo}%
\providecommand \href [0]{\begingroup \@sanitize@url \@href}%
\providecommand \@href[1]{\@@startlink{#1}\@@href}%
\providecommand \@@href[1]{\endgroup#1\@@endlink}%
\providecommand \@sanitize@url [0]{\catcode `\\12\catcode `\$12\catcode
  `\&12\catcode `\#12\catcode `\^12\catcode `\_12\catcode `\%12\relax}%
\providecommand \@@startlink[1]{}%
\providecommand \@@endlink[0]{}%
\providecommand \url  [0]{\begingroup\@sanitize@url \@url }%
\providecommand \@url [1]{\endgroup\@href {#1}{\urlprefix }}%
\providecommand \urlprefix  [0]{URL }%
\providecommand \Eprint [0]{\href }%
\providecommand \doibase [0]{https://doi.org/}%
\providecommand \selectlanguage [0]{\@gobble}%
\providecommand \bibinfo  [0]{\@secondoftwo}%
\providecommand \bibfield  [0]{\@secondoftwo}%
\providecommand \translation [1]{[#1]}%
\providecommand \BibitemOpen [0]{}%
\providecommand \bibitemStop [0]{}%
\providecommand \bibitemNoStop [0]{.\EOS\space}%
\providecommand \EOS [0]{\spacefactor3000\relax}%
\providecommand \BibitemShut  [1]{\csname bibitem#1\endcsname}%
\let\auto@bib@innerbib\@empty
\bibitem [{\citenamefont {Amblard}\ \emph {et~al.}(1996)\citenamefont
  {Amblard}, \citenamefont {Maggs}, \citenamefont {Yurke}, \citenamefont
  {Pargellis},\ and\ \citenamefont {Leibler}}]{amblard1996subdiffusion}%
  \BibitemOpen
  \bibfield  {author} {\bibinfo {author} {\bibfnamefont {F.}~\bibnamefont
  {Amblard}}, \bibinfo {author} {\bibfnamefont {A.~C.}\ \bibnamefont {Maggs}},
  \bibinfo {author} {\bibfnamefont {B.}~\bibnamefont {Yurke}}, \bibinfo
  {author} {\bibfnamefont {A.~N.}\ \bibnamefont {Pargellis}},\ and\ \bibinfo
  {author} {\bibfnamefont {S.}~\bibnamefont {Leibler}},\ }\href@noop {}
  {\bibfield  {journal} {\bibinfo  {journal} {Phys. Rev. Lett.}\ }\textbf
  {\bibinfo {volume} {77}},\ \bibinfo {pages} {4470} (\bibinfo {year}
  {1996})}\BibitemShut {NoStop}%
\bibitem [{\citenamefont {Alai}, \citenamefont {Lin},\ and\ \citenamefont
  {Pingale}(2015)}]{alai2015application}%
  \BibitemOpen
  \bibfield  {author} {\bibinfo {author} {\bibfnamefont {M.~S.}\ \bibnamefont
  {Alai}}, \bibinfo {author} {\bibfnamefont {W.~J.}\ \bibnamefont {Lin}},\ and\
  \bibinfo {author} {\bibfnamefont {S.~S.}\ \bibnamefont {Pingale}},\
  }\href@noop {} {\bibfield  {journal} {\bibinfo  {journal} {J. Food Drug
  Anal.}\ }\textbf {\bibinfo {volume} {23}},\ \bibinfo {pages} {351} (\bibinfo
  {year} {2015})}\BibitemShut {NoStop}%
\bibitem [{\citenamefont {Salata}(2004)}]{salata2004applications}%
  \BibitemOpen
  \bibfield  {author} {\bibinfo {author} {\bibfnamefont {O.~V.}\ \bibnamefont
  {Salata}},\ }\href@noop {} {\bibfield  {journal} {\bibinfo  {journal} {J.
  Nanobiotechnol.}\ }\textbf {\bibinfo {volume} {2}},\ \bibinfo {pages} {1}
  (\bibinfo {year} {2004})}\BibitemShut {NoStop}%
\bibitem [{\citenamefont {De}, \citenamefont {Ghosh},\ and\ \citenamefont
  {Rotello}(2008)}]{de2008applications}%
  \BibitemOpen
  \bibfield  {author} {\bibinfo {author} {\bibfnamefont {M.}~\bibnamefont
  {De}}, \bibinfo {author} {\bibfnamefont {P.~S.}\ \bibnamefont {Ghosh}},\ and\
  \bibinfo {author} {\bibfnamefont {V.~M.}\ \bibnamefont {Rotello}},\
  }\href@noop {} {\bibfield  {journal} {\bibinfo  {journal} {Adv. Mater.}\
  }\textbf {\bibinfo {volume} {20}},\ \bibinfo {pages} {4225} (\bibinfo {year}
  {2008})}\BibitemShut {NoStop}%
\bibitem [{\citenamefont {Gomez-Solano}, \citenamefont {Blokhuis},\ and\
  \citenamefont {Bechinger}(2016)}]{gomez2016dynamics}%
  \BibitemOpen
  \bibfield  {author} {\bibinfo {author} {\bibfnamefont {J.~R.}\ \bibnamefont
  {Gomez-Solano}}, \bibinfo {author} {\bibfnamefont {A.}~\bibnamefont
  {Blokhuis}},\ and\ \bibinfo {author} {\bibfnamefont {C.}~\bibnamefont
  {Bechinger}},\ }\href@noop {} {\bibfield  {journal} {\bibinfo  {journal}
  {Phys. Rev. Lett.}\ }\textbf {\bibinfo {volume} {116}},\ \bibinfo {pages}
  {138301} (\bibinfo {year} {2016})}\BibitemShut {NoStop}%
\bibitem [{\citenamefont {Patteson}\ \emph {et~al.}(2015)\citenamefont
  {Patteson}, \citenamefont {Gopinath}, \citenamefont {Goulian},\ and\
  \citenamefont {Arratia}}]{patteson2015running}%
  \BibitemOpen
  \bibfield  {author} {\bibinfo {author} {\bibfnamefont {A.}~\bibnamefont
  {Patteson}}, \bibinfo {author} {\bibfnamefont {A.}~\bibnamefont {Gopinath}},
  \bibinfo {author} {\bibfnamefont {M.}~\bibnamefont {Goulian}},\ and\ \bibinfo
  {author} {\bibfnamefont {P.}~\bibnamefont {Arratia}},\ }\href@noop {}
  {\bibfield  {journal} {\bibinfo  {journal} {Sci. Rep.}\ }\textbf {\bibinfo
  {volume} {5}},\ \bibinfo {pages} {15761} (\bibinfo {year}
  {2015})}\BibitemShut {NoStop}%
\bibitem [{\citenamefont {Stark}\ \emph {et~al.}(2015)\citenamefont {Stark},
  \citenamefont {Stoessel}, \citenamefont {Wohlleben},\ and\ \citenamefont
  {Hafner}}]{stark2015industrial}%
  \BibitemOpen
  \bibfield  {author} {\bibinfo {author} {\bibfnamefont {W.~J.}\ \bibnamefont
  {Stark}}, \bibinfo {author} {\bibfnamefont {P.~R.}\ \bibnamefont {Stoessel}},
  \bibinfo {author} {\bibfnamefont {W.}~\bibnamefont {Wohlleben}},\ and\
  \bibinfo {author} {\bibfnamefont {A.}~\bibnamefont {Hafner}},\ }\href@noop {}
  {\bibfield  {journal} {\bibinfo  {journal} {Chem. Soc. Rev}\ }\textbf
  {\bibinfo {volume} {44}},\ \bibinfo {pages} {5793} (\bibinfo {year}
  {2015})}\BibitemShut {NoStop}%
\bibitem [{\citenamefont {Kesharwani}\ \emph {et~al.}(2018)\citenamefont
  {Kesharwani}, \citenamefont {Gorain}, \citenamefont {Low}, \citenamefont
  {Tan}, \citenamefont {Ling}, \citenamefont {Lim}, \citenamefont {Chin},
  \citenamefont {Lee}, \citenamefont {Lee}, \citenamefont {Ooi}, \citenamefont
  {Choudhury},\ and\ \citenamefont {Pandey}}]{kesharwani2018nanotechnology}%
  \BibitemOpen
  \bibfield  {author} {\bibinfo {author} {\bibfnamefont {P.}~\bibnamefont
  {Kesharwani}}, \bibinfo {author} {\bibfnamefont {B.}~\bibnamefont {Gorain}},
  \bibinfo {author} {\bibfnamefont {S.~Y.}\ \bibnamefont {Low}}, \bibinfo
  {author} {\bibfnamefont {S.~A.}\ \bibnamefont {Tan}}, \bibinfo {author}
  {\bibfnamefont {E.~C.~S.}\ \bibnamefont {Ling}}, \bibinfo {author}
  {\bibfnamefont {Y.~K.}\ \bibnamefont {Lim}}, \bibinfo {author} {\bibfnamefont
  {C.~M.}\ \bibnamefont {Chin}}, \bibinfo {author} {\bibfnamefont {P.~Y.}\
  \bibnamefont {Lee}}, \bibinfo {author} {\bibfnamefont {C.~M.}\ \bibnamefont
  {Lee}}, \bibinfo {author} {\bibfnamefont {C.~H.}\ \bibnamefont {Ooi}},
  \bibinfo {author} {\bibfnamefont {H.}~\bibnamefont {Choudhury}},\ and\
  \bibinfo {author} {\bibfnamefont {M.}~\bibnamefont {Pandey}},\ }\href@noop {}
  {\bibfield  {journal} {\bibinfo  {journal} {Diabetes Res. Clin. Pr.}\
  }\textbf {\bibinfo {volume} {136}},\ \bibinfo {pages} {52} (\bibinfo {year}
  {2018})}\BibitemShut {NoStop}%
\bibitem [{\citenamefont {Aghebati-Maleki}\ \emph {et~al.}(2020)\citenamefont
  {Aghebati-Maleki}, \citenamefont {Dolati}, \citenamefont {Ahmadi},
  \citenamefont {Baghbanzhadeh}, \citenamefont {Asadi}, \citenamefont
  {Fotouhi}, \citenamefont {Yousefi},\ and\ \citenamefont
  {Aghebati-Maleki}}]{aghebati2020nanoparticles}%
  \BibitemOpen
  \bibfield  {author} {\bibinfo {author} {\bibfnamefont {A.}~\bibnamefont
  {Aghebati-Maleki}}, \bibinfo {author} {\bibfnamefont {S.}~\bibnamefont
  {Dolati}}, \bibinfo {author} {\bibfnamefont {M.}~\bibnamefont {Ahmadi}},
  \bibinfo {author} {\bibfnamefont {A.}~\bibnamefont {Baghbanzhadeh}}, \bibinfo
  {author} {\bibfnamefont {M.}~\bibnamefont {Asadi}}, \bibinfo {author}
  {\bibfnamefont {A.}~\bibnamefont {Fotouhi}}, \bibinfo {author} {\bibfnamefont
  {M.}~\bibnamefont {Yousefi}},\ and\ \bibinfo {author} {\bibfnamefont
  {L.}~\bibnamefont {Aghebati-Maleki}},\ }\href@noop {} {\bibfield  {journal}
  {\bibinfo  {journal} {J. Cell. Physiol.}\ }\textbf {\bibinfo {volume}
  {235}},\ \bibinfo {pages} {1962} (\bibinfo {year} {2020})}\BibitemShut
  {NoStop}%
\bibitem [{\citenamefont {Mitchell}\ \emph {et~al.}(2021)\citenamefont
  {Mitchell}, \citenamefont {Billingsley}, \citenamefont {Haley}, \citenamefont
  {Wechsler}, \citenamefont {Peppas},\ and\ \citenamefont
  {Langer}}]{mitchell2021engineering}%
  \BibitemOpen
  \bibfield  {author} {\bibinfo {author} {\bibfnamefont {M.~J.}\ \bibnamefont
  {Mitchell}}, \bibinfo {author} {\bibfnamefont {M.~M.}\ \bibnamefont
  {Billingsley}}, \bibinfo {author} {\bibfnamefont {R.~M.}\ \bibnamefont
  {Haley}}, \bibinfo {author} {\bibfnamefont {M.~E.}\ \bibnamefont {Wechsler}},
  \bibinfo {author} {\bibfnamefont {N.~A.}\ \bibnamefont {Peppas}},\ and\
  \bibinfo {author} {\bibfnamefont {R.}~\bibnamefont {Langer}},\ }\href@noop {}
  {\bibfield  {journal} {\bibinfo  {journal} {Nat. Rev. Drug Discov.}\ }\textbf
  {\bibinfo {volume} {20}},\ \bibinfo {pages} {101} (\bibinfo {year}
  {2021})}\BibitemShut {NoStop}%
\bibitem [{\citenamefont {Liu}\ and\ \citenamefont
  {Fr{\'e}chet}(1999)}]{liu1999designing}%
  \BibitemOpen
  \bibfield  {author} {\bibinfo {author} {\bibfnamefont {M.}~\bibnamefont
  {Liu}}\ and\ \bibinfo {author} {\bibfnamefont {J.~M.}\ \bibnamefont
  {Fr{\'e}chet}},\ }\href@noop {} {\bibfield  {journal} {\bibinfo  {journal}
  {Pharm. Sci. Technol. To.}\ }\textbf {\bibinfo {volume} {2}},\ \bibinfo
  {pages} {393} (\bibinfo {year} {1999})}\BibitemShut {NoStop}%
\bibitem [{\citenamefont {Gillies}\ and\ \citenamefont
  {Frechet}(2005)}]{gillies2005dendrimers}%
  \BibitemOpen
  \bibfield  {author} {\bibinfo {author} {\bibfnamefont {E.~R.}\ \bibnamefont
  {Gillies}}\ and\ \bibinfo {author} {\bibfnamefont {J.~M.}\ \bibnamefont
  {Frechet}},\ }\href@noop {} {\bibfield  {journal} {\bibinfo  {journal} {Drug
  Discov. Today}\ }\textbf {\bibinfo {volume} {10}},\ \bibinfo {pages} {35}
  (\bibinfo {year} {2005})}\BibitemShut {NoStop}%
\bibitem [{\citenamefont {Baghbanbashi}\ and\ \citenamefont
  {Kakkar}(2022)}]{baghbanbashi2022polymersomes}%
  \BibitemOpen
  \bibfield  {author} {\bibinfo {author} {\bibfnamefont {M.}~\bibnamefont
  {Baghbanbashi}}\ and\ \bibinfo {author} {\bibfnamefont {A.}~\bibnamefont
  {Kakkar}},\ }\href@noop {} {\bibfield  {journal} {\bibinfo  {journal} {Mol.
  Pharm.}\ }\textbf {\bibinfo {volume} {19}},\ \bibinfo {pages} {1687}
  (\bibinfo {year} {2022})}\BibitemShut {NoStop}%
\bibitem [{\citenamefont {Lee}\ \emph {et~al.}(2005)\citenamefont {Lee},
  \citenamefont {MacKay}, \citenamefont {Fr{\'e}chet},\ and\ \citenamefont
  {Szoka}}]{lee2005designing}%
  \BibitemOpen
  \bibfield  {author} {\bibinfo {author} {\bibfnamefont {C.~C.}\ \bibnamefont
  {Lee}}, \bibinfo {author} {\bibfnamefont {J.~A.}\ \bibnamefont {MacKay}},
  \bibinfo {author} {\bibfnamefont {J.~M.}\ \bibnamefont {Fr{\'e}chet}},\ and\
  \bibinfo {author} {\bibfnamefont {F.~C.}\ \bibnamefont {Szoka}},\ }\href@noop
  {} {\bibfield  {journal} {\bibinfo  {journal} {Nat. Biotechnol.}\ }\textbf
  {\bibinfo {volume} {23}},\ \bibinfo {pages} {1517} (\bibinfo {year}
  {2005})}\BibitemShut {NoStop}%
\bibitem [{\citenamefont {Harreis}, \citenamefont {Likos},\ and\ \citenamefont
  {Ballauff}(2003)}]{harreis2003can}%
  \BibitemOpen
  \bibfield  {author} {\bibinfo {author} {\bibfnamefont {H.}~\bibnamefont
  {Harreis}}, \bibinfo {author} {\bibfnamefont {C.}~\bibnamefont {Likos}},\
  and\ \bibinfo {author} {\bibfnamefont {M.}~\bibnamefont {Ballauff}},\
  }\href@noop {} {\bibfield  {journal} {\bibinfo  {journal} {J. Chem. Phys.}\
  }\textbf {\bibinfo {volume} {118}},\ \bibinfo {pages} {1979} (\bibinfo {year}
  {2003})}\BibitemShut {NoStop}%
\bibitem [{\citenamefont {Vlassopoulos}(2004)}]{vlassopoulos2004colloidal}%
  \BibitemOpen
  \bibfield  {author} {\bibinfo {author} {\bibfnamefont {D.}~\bibnamefont
  {Vlassopoulos}},\ }\href@noop {} {\bibfield  {journal} {\bibinfo  {journal}
  {J. Polym. Sci., Part B: Polym. Phys.}\ }\textbf {\bibinfo {volume} {42}},\
  \bibinfo {pages} {2931} (\bibinfo {year} {2004})}\BibitemShut {NoStop}%
\bibitem [{\citenamefont {Likos}\ \emph {et~al.}(1998)\citenamefont {Likos},
  \citenamefont {L{\"o}wen}, \citenamefont {Watzlawek}, \citenamefont {Abbas},
  \citenamefont {Jucknischke}, \citenamefont {Allgaier},\ and\ \citenamefont
  {Richter}}]{likos1998star}%
  \BibitemOpen
  \bibfield  {author} {\bibinfo {author} {\bibfnamefont {C.}~\bibnamefont
  {Likos}}, \bibinfo {author} {\bibfnamefont {H.}~\bibnamefont {L{\"o}wen}},
  \bibinfo {author} {\bibfnamefont {M.}~\bibnamefont {Watzlawek}}, \bibinfo
  {author} {\bibfnamefont {B.}~\bibnamefont {Abbas}}, \bibinfo {author}
  {\bibfnamefont {O.}~\bibnamefont {Jucknischke}}, \bibinfo {author}
  {\bibfnamefont {J.}~\bibnamefont {Allgaier}},\ and\ \bibinfo {author}
  {\bibfnamefont {D.}~\bibnamefont {Richter}},\ }\href@noop {} {\bibfield
  {journal} {\bibinfo  {journal} {Phys. Rev. Lett.}\ }\textbf {\bibinfo
  {volume} {80}},\ \bibinfo {pages} {4450} (\bibinfo {year}
  {1998})}\BibitemShut {NoStop}%
\bibitem [{\citenamefont {Bosko}\ and\ \citenamefont
  {Prakash}(2011)}]{bosko2011universal}%
  \BibitemOpen
  \bibfield  {author} {\bibinfo {author} {\bibfnamefont {J.~T.}\ \bibnamefont
  {Bosko}}\ and\ \bibinfo {author} {\bibfnamefont {J.~R.}\ \bibnamefont
  {Prakash}},\ }\href@noop {} {\bibfield  {journal} {\bibinfo  {journal}
  {Macromolecules}\ }\textbf {\bibinfo {volume} {44}},\ \bibinfo {pages} {660}
  (\bibinfo {year} {2011})}\BibitemShut {NoStop}%
\bibitem [{\citenamefont {Mason}\ and\ \citenamefont
  {Weitz}(1995)}]{mason1995optical}%
  \BibitemOpen
  \bibfield  {author} {\bibinfo {author} {\bibfnamefont {T.~G.}\ \bibnamefont
  {Mason}}\ and\ \bibinfo {author} {\bibfnamefont {D.~A.}\ \bibnamefont
  {Weitz}},\ }\href@noop {} {\bibfield  {journal} {\bibinfo  {journal} {Phys.
  Rev. Lett.}\ }\textbf {\bibinfo {volume} {74}},\ \bibinfo {pages} {1250}
  (\bibinfo {year} {1995})}\BibitemShut {NoStop}%
\bibitem [{\citenamefont {Squires}\ and\ \citenamefont
  {Mason}(2010)}]{squires2010fluid}%
  \BibitemOpen
  \bibfield  {author} {\bibinfo {author} {\bibfnamefont {T.~M.}\ \bibnamefont
  {Squires}}\ and\ \bibinfo {author} {\bibfnamefont {T.~G.}\ \bibnamefont
  {Mason}},\ }\href@noop {} {\bibfield  {journal} {\bibinfo  {journal} {Annu.
  Rev. Fluid Mech.}\ }\textbf {\bibinfo {volume} {42}},\ \bibinfo {pages} {413}
  (\bibinfo {year} {2010})}\BibitemShut {NoStop}%
\bibitem [{\citenamefont {Mackay}\ \emph {et~al.}(2003)\citenamefont {Mackay},
  \citenamefont {Dao}, \citenamefont {Tuteja}, \citenamefont {Ho},
  \citenamefont {Van~Horn}, \citenamefont {Kim},\ and\ \citenamefont
  {Hawker}}]{mackay2003nanoscale}%
  \BibitemOpen
  \bibfield  {author} {\bibinfo {author} {\bibfnamefont {M.~E.}\ \bibnamefont
  {Mackay}}, \bibinfo {author} {\bibfnamefont {T.~T.}\ \bibnamefont {Dao}},
  \bibinfo {author} {\bibfnamefont {A.}~\bibnamefont {Tuteja}}, \bibinfo
  {author} {\bibfnamefont {D.~L.}\ \bibnamefont {Ho}}, \bibinfo {author}
  {\bibfnamefont {B.}~\bibnamefont {Van~Horn}}, \bibinfo {author}
  {\bibfnamefont {H.-C.}\ \bibnamefont {Kim}},\ and\ \bibinfo {author}
  {\bibfnamefont {C.~J.}\ \bibnamefont {Hawker}},\ }\href@noop {} {\bibfield
  {journal} {\bibinfo  {journal} {Nat. Mater.}\ }\textbf {\bibinfo {volume}
  {2}},\ \bibinfo {pages} {762} (\bibinfo {year} {2003})}\BibitemShut {NoStop}%
\bibitem [{\citenamefont {Ye}, \citenamefont {Tong},\ and\ \citenamefont
  {Fetters}(1998)}]{ye1998transport}%
  \BibitemOpen
  \bibfield  {author} {\bibinfo {author} {\bibfnamefont {X.}~\bibnamefont
  {Ye}}, \bibinfo {author} {\bibfnamefont {P.}~\bibnamefont {Tong}},\ and\
  \bibinfo {author} {\bibfnamefont {L.}~\bibnamefont {Fetters}},\ }\href@noop
  {} {\bibfield  {journal} {\bibinfo  {journal} {Macromolecules}\ }\textbf
  {\bibinfo {volume} {31}},\ \bibinfo {pages} {5785} (\bibinfo {year}
  {1998})}\BibitemShut {NoStop}%
\bibitem [{\citenamefont {Maldonado-Camargo}\ and\ \citenamefont
  {Rinaldi}(2016)}]{maldonado2016breakdown}%
  \BibitemOpen
  \bibfield  {author} {\bibinfo {author} {\bibfnamefont {L.}~\bibnamefont
  {Maldonado-Camargo}}\ and\ \bibinfo {author} {\bibfnamefont {C.}~\bibnamefont
  {Rinaldi}},\ }\href@noop {} {\bibfield  {journal} {\bibinfo  {journal} {Nano
  Lett.}\ }\textbf {\bibinfo {volume} {16}},\ \bibinfo {pages} {6767} (\bibinfo
  {year} {2016})}\BibitemShut {NoStop}%
\bibitem [{\citenamefont {Ge}(2023)}]{ge2023scaling}%
  \BibitemOpen
  \bibfield  {author} {\bibinfo {author} {\bibfnamefont {T.}~\bibnamefont
  {Ge}},\ }\href@noop {} {\bibfield  {journal} {\bibinfo  {journal}
  {Macromolecules}\ }\textbf {\bibinfo {volume} {56}},\ \bibinfo {pages}
  {3809–3837} (\bibinfo {year} {2023})}\BibitemShut {NoStop}%
\bibitem [{\citenamefont {Langevin}\ and\ \citenamefont
  {Rondelez}(1978)}]{langevin1978sedimentation}%
  \BibitemOpen
  \bibfield  {author} {\bibinfo {author} {\bibfnamefont {D.}~\bibnamefont
  {Langevin}}\ and\ \bibinfo {author} {\bibfnamefont {F.}~\bibnamefont
  {Rondelez}},\ }\href@noop {} {\bibfield  {journal} {\bibinfo  {journal}
  {Polymer}\ }\textbf {\bibinfo {volume} {19}},\ \bibinfo {pages} {875}
  (\bibinfo {year} {1978})}\BibitemShut {NoStop}%
\bibitem [{\citenamefont {Cukier}(1984)}]{cukier1984diffusion}%
  \BibitemOpen
  \bibfield  {author} {\bibinfo {author} {\bibfnamefont {R.}~\bibnamefont
  {Cukier}},\ }\href@noop {} {\bibfield  {journal} {\bibinfo  {journal}
  {Macromolecules}\ }\textbf {\bibinfo {volume} {17}},\ \bibinfo {pages} {252}
  (\bibinfo {year} {1984})}\BibitemShut {NoStop}%
\bibitem [{\citenamefont {Altenberger}\ and\ \citenamefont
  {Tirrell}(1984)}]{altenberger1984theory}%
  \BibitemOpen
  \bibfield  {author} {\bibinfo {author} {\bibfnamefont {A.~R.}\ \bibnamefont
  {Altenberger}}\ and\ \bibinfo {author} {\bibfnamefont {M.}~\bibnamefont
  {Tirrell}},\ }\href@noop {} {\bibfield  {journal} {\bibinfo  {journal} {J.
  Chem. Phys.}\ }\textbf {\bibinfo {volume} {80}},\ \bibinfo {pages} {2208}
  (\bibinfo {year} {1984})}\BibitemShut {NoStop}%
\bibitem [{\citenamefont {Phillies}\ \emph {et~al.}(1989)\citenamefont
  {Phillies}, \citenamefont {Gong}, \citenamefont {Li}, \citenamefont {Rau},
  \citenamefont {Zhang}, \citenamefont {Yu},\ and\ \citenamefont
  {Rollings}}]{phillies1989macroparticle}%
  \BibitemOpen
  \bibfield  {author} {\bibinfo {author} {\bibfnamefont {G.~D.}\ \bibnamefont
  {Phillies}}, \bibinfo {author} {\bibfnamefont {J.}~\bibnamefont {Gong}},
  \bibinfo {author} {\bibfnamefont {L.}~\bibnamefont {Li}}, \bibinfo {author}
  {\bibfnamefont {A.}~\bibnamefont {Rau}}, \bibinfo {author} {\bibfnamefont
  {K.}~\bibnamefont {Zhang}}, \bibinfo {author} {\bibfnamefont {L.~P.}\
  \bibnamefont {Yu}},\ and\ \bibinfo {author} {\bibfnamefont {J.}~\bibnamefont
  {Rollings}},\ }\href@noop {} {\bibfield  {journal} {\bibinfo  {journal} {J.
  Phys. Chem. A}\ }\textbf {\bibinfo {volume} {93}},\ \bibinfo {pages} {6219}
  (\bibinfo {year} {1989})}\BibitemShut {NoStop}%
\bibitem [{\citenamefont {de~Gennes}(1979)}]{de1979scaling}%
  \BibitemOpen
  \bibfield  {author} {\bibinfo {author} {\bibfnamefont {P.-G.}\ \bibnamefont
  {de~Gennes}},\ }\href@noop {} {\emph {\bibinfo {title} {Scaling Concepts in
  Polymer Physics}}}\ (\bibinfo  {publisher} {Cornell University Press},\
  \bibinfo {year} {1979})\BibitemShut {NoStop}%
\bibitem [{\citenamefont {Tong}\ \emph {et~al.}(1997)\citenamefont {Tong},
  \citenamefont {Ye}, \citenamefont {Ackerson},\ and\ \citenamefont
  {Fetters}}]{tong1997sedimentation}%
  \BibitemOpen
  \bibfield  {author} {\bibinfo {author} {\bibfnamefont {P.}~\bibnamefont
  {Tong}}, \bibinfo {author} {\bibfnamefont {X.}~\bibnamefont {Ye}}, \bibinfo
  {author} {\bibfnamefont {B.~J.}\ \bibnamefont {Ackerson}},\ and\ \bibinfo
  {author} {\bibfnamefont {L.}~\bibnamefont {Fetters}},\ }\href@noop {}
  {\bibfield  {journal} {\bibinfo  {journal} {Phys. Rev. Lett.}\ }\textbf
  {\bibinfo {volume} {79}},\ \bibinfo {pages} {2363} (\bibinfo {year}
  {1997})}\BibitemShut {NoStop}%
\bibitem [{\citenamefont {Cheng}, \citenamefont {Prud'Homme},\ and\
  \citenamefont {Thomas}(2002)}]{cheng2002diffusion}%
  \BibitemOpen
  \bibfield  {author} {\bibinfo {author} {\bibfnamefont {Y.}~\bibnamefont
  {Cheng}}, \bibinfo {author} {\bibfnamefont {R.~K.}\ \bibnamefont
  {Prud'Homme}},\ and\ \bibinfo {author} {\bibfnamefont {J.~L.}\ \bibnamefont
  {Thomas}},\ }\href@noop {} {\bibfield  {journal} {\bibinfo  {journal}
  {Macromolecules}\ }\textbf {\bibinfo {volume} {35}},\ \bibinfo {pages} {8111}
  (\bibinfo {year} {2002})}\BibitemShut {NoStop}%
\bibitem [{\citenamefont {Holyst}\ \emph {et~al.}(2009)\citenamefont {Holyst},
  \citenamefont {Bielejewska}, \citenamefont {Szyma{\'n}ski}, \citenamefont
  {Wilk}, \citenamefont {Patkowski}, \citenamefont {Gapi{\'n}ski},
  \citenamefont {{\.Z}ywoci{\'n}ski}, \citenamefont {Kalwarczyk}, \citenamefont
  {Kalwarczyk}, \citenamefont {Tabaka} \emph {et~al.}}]{holyst2009scaling}%
  \BibitemOpen
  \bibfield  {author} {\bibinfo {author} {\bibfnamefont {R.}~\bibnamefont
  {Holyst}}, \bibinfo {author} {\bibfnamefont {A.}~\bibnamefont {Bielejewska}},
  \bibinfo {author} {\bibfnamefont {J.}~\bibnamefont {Szyma{\'n}ski}}, \bibinfo
  {author} {\bibfnamefont {A.}~\bibnamefont {Wilk}}, \bibinfo {author}
  {\bibfnamefont {A.}~\bibnamefont {Patkowski}}, \bibinfo {author}
  {\bibfnamefont {J.}~\bibnamefont {Gapi{\'n}ski}}, \bibinfo {author}
  {\bibfnamefont {A.}~\bibnamefont {{\.Z}ywoci{\'n}ski}}, \bibinfo {author}
  {\bibfnamefont {T.}~\bibnamefont {Kalwarczyk}}, \bibinfo {author}
  {\bibfnamefont {E.}~\bibnamefont {Kalwarczyk}}, \bibinfo {author}
  {\bibfnamefont {M.}~\bibnamefont {Tabaka}}, \emph {et~al.},\ }\href@noop {}
  {\bibfield  {journal} {\bibinfo  {journal} {Phys. Chem. Chem. Phys.}\
  }\textbf {\bibinfo {volume} {11}},\ \bibinfo {pages} {9025} (\bibinfo {year}
  {2009})}\BibitemShut {NoStop}%
\bibitem [{\citenamefont {Kalwarczyk}\ \emph {et~al.}(2015)\citenamefont
  {Kalwarczyk}, \citenamefont {Sozanski}, \citenamefont {Ochab-Marcinek},
  \citenamefont {Szymanski}, \citenamefont {Tabaka}, \citenamefont {Hou},\ and\
  \citenamefont {Holyst}}]{kalwarczyk2015motion}%
  \BibitemOpen
  \bibfield  {author} {\bibinfo {author} {\bibfnamefont {T.}~\bibnamefont
  {Kalwarczyk}}, \bibinfo {author} {\bibfnamefont {K.}~\bibnamefont
  {Sozanski}}, \bibinfo {author} {\bibfnamefont {A.}~\bibnamefont
  {Ochab-Marcinek}}, \bibinfo {author} {\bibfnamefont {J.}~\bibnamefont
  {Szymanski}}, \bibinfo {author} {\bibfnamefont {M.}~\bibnamefont {Tabaka}},
  \bibinfo {author} {\bibfnamefont {S.}~\bibnamefont {Hou}},\ and\ \bibinfo
  {author} {\bibfnamefont {R.}~\bibnamefont {Holyst}},\ }\href@noop {}
  {\bibfield  {journal} {\bibinfo  {journal} {Adv. Colloid Interface Sci.}\
  }\textbf {\bibinfo {volume} {223}},\ \bibinfo {pages} {55} (\bibinfo {year}
  {2015})}\BibitemShut {NoStop}%
\bibitem [{\citenamefont {Kalwarczyk}\ \emph {et~al.}(2011)\citenamefont
  {Kalwarczyk}, \citenamefont {Ziebacz}, \citenamefont {Bielejewska},
  \citenamefont {Zaboklicka}, \citenamefont {Koynov}, \citenamefont
  {Szymanski}, \citenamefont {Wilk}, \citenamefont {Patkowski}, \citenamefont
  {Gapinski}, \citenamefont {Butt},\ and\ \citenamefont
  {Ho{\l}yst}}]{kalwarczyk2011comparative}%
  \BibitemOpen
  \bibfield  {author} {\bibinfo {author} {\bibfnamefont {T.}~\bibnamefont
  {Kalwarczyk}}, \bibinfo {author} {\bibfnamefont {N.}~\bibnamefont {Ziebacz}},
  \bibinfo {author} {\bibfnamefont {A.}~\bibnamefont {Bielejewska}}, \bibinfo
  {author} {\bibfnamefont {E.}~\bibnamefont {Zaboklicka}}, \bibinfo {author}
  {\bibfnamefont {K.}~\bibnamefont {Koynov}}, \bibinfo {author} {\bibfnamefont
  {J.}~\bibnamefont {Szymanski}}, \bibinfo {author} {\bibfnamefont
  {A.}~\bibnamefont {Wilk}}, \bibinfo {author} {\bibfnamefont {A.}~\bibnamefont
  {Patkowski}}, \bibinfo {author} {\bibfnamefont {J.}~\bibnamefont {Gapinski}},
  \bibinfo {author} {\bibfnamefont {H.-J.}\ \bibnamefont {Butt}},\ and\
  \bibinfo {author} {\bibfnamefont {R.}~\bibnamefont {Ho{\l}yst}},\ }\href@noop
  {} {\bibfield  {journal} {\bibinfo  {journal} {Nano Lett.}\ }\textbf
  {\bibinfo {volume} {11}},\ \bibinfo {pages} {2157} (\bibinfo {year}
  {2011})}\BibitemShut {NoStop}%
\bibitem [{\citenamefont {Soza{\'n}ski}\ \emph {et~al.}(2013)\citenamefont
  {Soza{\'n}ski}, \citenamefont {Wi{\'s}niewska}, \citenamefont {Kalwarczyk},\
  and\ \citenamefont {Ho{\l}yst}}]{sozanski2013activation}%
  \BibitemOpen
  \bibfield  {author} {\bibinfo {author} {\bibfnamefont {K.}~\bibnamefont
  {Soza{\'n}ski}}, \bibinfo {author} {\bibfnamefont {A.}~\bibnamefont
  {Wi{\'s}niewska}}, \bibinfo {author} {\bibfnamefont {T.}~\bibnamefont
  {Kalwarczyk}},\ and\ \bibinfo {author} {\bibfnamefont {R.}~\bibnamefont
  {Ho{\l}yst}},\ }\href@noop {} {\bibfield  {journal} {\bibinfo  {journal}
  {Phys. Rev. Lett.}\ }\textbf {\bibinfo {volume} {111}},\ \bibinfo {pages}
  {228301} (\bibinfo {year} {2013})}\BibitemShut {NoStop}%
\bibitem [{\citenamefont {Chen}, \citenamefont {Zhao},\ and\ \citenamefont
  {Hou}(2017)}]{chen2017effect}%
  \BibitemOpen
  \bibfield  {author} {\bibinfo {author} {\bibfnamefont {A.}~\bibnamefont
  {Chen}}, \bibinfo {author} {\bibfnamefont {N.}~\bibnamefont {Zhao}},\ and\
  \bibinfo {author} {\bibfnamefont {Z.}~\bibnamefont {Hou}},\ }\href@noop {}
  {\bibfield  {journal} {\bibinfo  {journal} {Soft Matter}\ }\textbf {\bibinfo
  {volume} {13}},\ \bibinfo {pages} {8625} (\bibinfo {year}
  {2017})}\BibitemShut {NoStop}%
\bibitem [{\citenamefont {Cai}, \citenamefont {Panyukov},\ and\ \citenamefont
  {Rubinstein}(2011)}]{cai2011mobility}%
  \BibitemOpen
  \bibfield  {author} {\bibinfo {author} {\bibfnamefont {L.-H.}\ \bibnamefont
  {Cai}}, \bibinfo {author} {\bibfnamefont {S.}~\bibnamefont {Panyukov}},\ and\
  \bibinfo {author} {\bibfnamefont {M.}~\bibnamefont {Rubinstein}},\
  }\href@noop {} {\bibfield  {journal} {\bibinfo  {journal} {Macromolecules}\
  }\textbf {\bibinfo {volume} {44}},\ \bibinfo {pages} {7853} (\bibinfo {year}
  {2011})}\BibitemShut {NoStop}%
\bibitem [{\citenamefont {Poling-Skutvik}, \citenamefont {Krishnamoorti},\ and\
  \citenamefont {Conrad}(2015)}]{poling2015size}%
  \BibitemOpen
  \bibfield  {author} {\bibinfo {author} {\bibfnamefont {R.}~\bibnamefont
  {Poling-Skutvik}}, \bibinfo {author} {\bibfnamefont {R.}~\bibnamefont
  {Krishnamoorti}},\ and\ \bibinfo {author} {\bibfnamefont {J.~C.}\
  \bibnamefont {Conrad}},\ }\href@noop {} {\bibfield  {journal} {\bibinfo
  {journal} {ACS Macro Lett.}\ }\textbf {\bibinfo {volume} {4}},\ \bibinfo
  {pages} {1169} (\bibinfo {year} {2015})}\BibitemShut {NoStop}%
\bibitem [{\citenamefont {Chen}\ \emph {et~al.}(2018)\citenamefont {Chen},
  \citenamefont {Poling-Skutvik}, \citenamefont {Nikoubashman}, \citenamefont
  {Howard}, \citenamefont {Conrad},\ and\ \citenamefont
  {Palmer}}]{chen2018coupling}%
  \BibitemOpen
  \bibfield  {author} {\bibinfo {author} {\bibfnamefont {R.}~\bibnamefont
  {Chen}}, \bibinfo {author} {\bibfnamefont {R.}~\bibnamefont
  {Poling-Skutvik}}, \bibinfo {author} {\bibfnamefont {A.}~\bibnamefont
  {Nikoubashman}}, \bibinfo {author} {\bibfnamefont {M.~P.}\ \bibnamefont
  {Howard}}, \bibinfo {author} {\bibfnamefont {J.~C.}\ \bibnamefont {Conrad}},\
  and\ \bibinfo {author} {\bibfnamefont {J.~C.}\ \bibnamefont {Palmer}},\
  }\href@noop {} {\bibfield  {journal} {\bibinfo  {journal} {Macromolecules}\
  }\textbf {\bibinfo {volume} {51}},\ \bibinfo {pages} {1865} (\bibinfo {year}
  {2018})}\BibitemShut {NoStop}%
\bibitem [{\citenamefont {de~Kort}\ \emph {et~al.}(2015)\citenamefont
  {de~Kort}, \citenamefont {Rombouts}, \citenamefont {Hoeben}, \citenamefont
  {Janssen}, \citenamefont {Van~As},\ and\ \citenamefont {van
  Duynhoven}}]{de2015scaling}%
  \BibitemOpen
  \bibfield  {author} {\bibinfo {author} {\bibfnamefont {D.~W.}\ \bibnamefont
  {de~Kort}}, \bibinfo {author} {\bibfnamefont {W.~H.}\ \bibnamefont
  {Rombouts}}, \bibinfo {author} {\bibfnamefont {F.~J.}\ \bibnamefont
  {Hoeben}}, \bibinfo {author} {\bibfnamefont {H.~M.}\ \bibnamefont {Janssen}},
  \bibinfo {author} {\bibfnamefont {H.}~\bibnamefont {Van~As}},\ and\ \bibinfo
  {author} {\bibfnamefont {J.~P.}\ \bibnamefont {van Duynhoven}},\ }\href@noop
  {} {\bibfield  {journal} {\bibinfo  {journal} {Macromolecules}\ }\textbf
  {\bibinfo {volume} {48}},\ \bibinfo {pages} {7585} (\bibinfo {year}
  {2015})}\BibitemShut {NoStop}%
\bibitem [{\citenamefont {Cai}, \citenamefont {Panyukov},\ and\ \citenamefont
  {Rubinstein}(2015)}]{cai2015hopping}%
  \BibitemOpen
  \bibfield  {author} {\bibinfo {author} {\bibfnamefont {L.-H.}\ \bibnamefont
  {Cai}}, \bibinfo {author} {\bibfnamefont {S.}~\bibnamefont {Panyukov}},\ and\
  \bibinfo {author} {\bibfnamefont {M.}~\bibnamefont {Rubinstein}},\
  }\href@noop {} {\bibfield  {journal} {\bibinfo  {journal} {Macromolecules}\
  }\textbf {\bibinfo {volume} {48}},\ \bibinfo {pages} {847} (\bibinfo {year}
  {2015})}\BibitemShut {NoStop}%
\bibitem [{\citenamefont {Bird}\ \emph {et~al.}(1987)\citenamefont {Bird},
  \citenamefont {Curtiss}, \citenamefont {Armstrong},\ and\ \citenamefont
  {Hassager}}]{Bird}%
  \BibitemOpen
  \bibfield  {author} {\bibinfo {author} {\bibfnamefont {R.~B.}\ \bibnamefont
  {Bird}}, \bibinfo {author} {\bibfnamefont {C.~F.}\ \bibnamefont {Curtiss}},
  \bibinfo {author} {\bibfnamefont {R.~C.}\ \bibnamefont {Armstrong}},\ and\
  \bibinfo {author} {\bibfnamefont {O.}~\bibnamefont {Hassager}},\ }\href@noop
  {} {\emph {\bibinfo {title} {Dynamics of Polymeric Liquids}}},\ Vol.~\bibinfo
  {volume} {2}\ (\bibinfo  {publisher} {John Wiley and Sons, New York},\
  \bibinfo {year} {1987})\BibitemShut {NoStop}%
\bibitem [{\citenamefont {Soddemann}, \citenamefont {D{\"u}nweg},\ and\
  \citenamefont {Kremer}(2001)}]{SDK}%
  \BibitemOpen
  \bibfield  {author} {\bibinfo {author} {\bibfnamefont {T.}~\bibnamefont
  {Soddemann}}, \bibinfo {author} {\bibfnamefont {B.}~\bibnamefont
  {D{\"u}nweg}},\ and\ \bibinfo {author} {\bibfnamefont {K.}~\bibnamefont
  {Kremer}},\ }\href@noop {} {\bibfield  {journal} {\bibinfo  {journal} {Eur.
  Phys. J. E}\ }\textbf {\bibinfo {volume} {6}},\ \bibinfo {pages} {409}
  (\bibinfo {year} {2001})}\BibitemShut {NoStop}%
\bibitem [{\citenamefont {Santra}\ \emph {et~al.}(2019)\citenamefont {Santra},
  \citenamefont {Kumari}, \citenamefont {Padinhateeri}, \citenamefont
  {D{\"u}nweg},\ and\ \citenamefont {Prakash}}]{santra2019universality}%
  \BibitemOpen
  \bibfield  {author} {\bibinfo {author} {\bibfnamefont {A.}~\bibnamefont
  {Santra}}, \bibinfo {author} {\bibfnamefont {K.}~\bibnamefont {Kumari}},
  \bibinfo {author} {\bibfnamefont {R.}~\bibnamefont {Padinhateeri}}, \bibinfo
  {author} {\bibfnamefont {B.}~\bibnamefont {D{\"u}nweg}},\ and\ \bibinfo
  {author} {\bibfnamefont {J.~R.}\ \bibnamefont {Prakash}},\ }\href@noop {}
  {\bibfield  {journal} {\bibinfo  {journal} {Soft Matter}\ }\textbf {\bibinfo
  {volume} {15}},\ \bibinfo {pages} {7876} (\bibinfo {year}
  {2019})}\BibitemShut {NoStop}%
\bibitem [{\citenamefont {Anderson}, \citenamefont {Glaser},\ and\
  \citenamefont {Glotzer}(2020)}]{anderson2020hoomd}%
  \BibitemOpen
  \bibfield  {author} {\bibinfo {author} {\bibfnamefont {J.~A.}\ \bibnamefont
  {Anderson}}, \bibinfo {author} {\bibfnamefont {J.}~\bibnamefont {Glaser}},\
  and\ \bibinfo {author} {\bibfnamefont {S.~C.}\ \bibnamefont {Glotzer}},\
  }\href@noop {} {\bibfield  {journal} {\bibinfo  {journal} {Comput. Mater.
  Sci.}\ }\textbf {\bibinfo {volume} {173}},\ \bibinfo {pages} {109363}
  (\bibinfo {year} {2020})}\BibitemShut {NoStop}%
\bibitem [{\citenamefont {Robe}\ \emph {et~al.}(2023)\citenamefont {Robe},
  \citenamefont {Santra}, \citenamefont {McKinley},\ and\ \citenamefont
  {Prakash}}]{robe2023evanescent}%
  \BibitemOpen
  \bibfield  {author} {\bibinfo {author} {\bibfnamefont {D.}~\bibnamefont
  {Robe}}, \bibinfo {author} {\bibfnamefont {A.}~\bibnamefont {Santra}},
  \bibinfo {author} {\bibfnamefont {G.}~\bibnamefont {McKinley}},\ and\
  \bibinfo {author} {\bibfnamefont {J.~R.}\ \bibnamefont {Prakash}},\
  }\href@noop {} {\bibfield  {journal} {\bibinfo  {journal} {arXiv:2302.13623;
  cond-mat.soft}\ } (\bibinfo {year} {2023})},\ \Eprint
  {https://arxiv.org/abs/2302.13623} {arXiv:2302.13623 [cond-mat.soft]}
  \BibitemShut {NoStop}%
\bibitem [{\citenamefont {Jain}\ \emph {et~al.}(2012)\citenamefont {Jain},
  \citenamefont {Sunthar}, \citenamefont {D{\"u}nweg},\ and\ \citenamefont
  {Prakash}}]{jain2012optimization}%
  \BibitemOpen
  \bibfield  {author} {\bibinfo {author} {\bibfnamefont {A.}~\bibnamefont
  {Jain}}, \bibinfo {author} {\bibfnamefont {P.}~\bibnamefont {Sunthar}},
  \bibinfo {author} {\bibfnamefont {B.}~\bibnamefont {D{\"u}nweg}},\ and\
  \bibinfo {author} {\bibfnamefont {J.~R.}\ \bibnamefont {Prakash}},\
  }\href@noop {} {\bibfield  {journal} {\bibinfo  {journal} {Phys. Rev. E}\
  }\textbf {\bibinfo {volume} {85}},\ \bibinfo {pages} {066703} (\bibinfo
  {year} {2012})}\BibitemShut {NoStop}%
\bibitem [{\citenamefont {Jain}\ \emph {et~al.}(2015)\citenamefont {Jain},
  \citenamefont {Sasmal}, \citenamefont {Hartkamp}, \citenamefont {Todd},\ and\
  \citenamefont {Prakash}}]{jain2015brownian}%
  \BibitemOpen
  \bibfield  {author} {\bibinfo {author} {\bibfnamefont {A.}~\bibnamefont
  {Jain}}, \bibinfo {author} {\bibfnamefont {C.}~\bibnamefont {Sasmal}},
  \bibinfo {author} {\bibfnamefont {R.}~\bibnamefont {Hartkamp}}, \bibinfo
  {author} {\bibfnamefont {B.~D.}\ \bibnamefont {Todd}},\ and\ \bibinfo
  {author} {\bibfnamefont {J.~R.}\ \bibnamefont {Prakash}},\ }\href@noop {}
  {\bibfield  {journal} {\bibinfo  {journal} {Chem. Eng. Sci.}\ }\textbf
  {\bibinfo {volume} {121}},\ \bibinfo {pages} {245} (\bibinfo {year}
  {2015})}\BibitemShut {NoStop}%
\bibitem [{\citenamefont {Santra}, \citenamefont {D{\"u}nweg},\ and\
  \citenamefont {Ravi~Prakash}(2021)}]{santra2021universal}%
  \BibitemOpen
  \bibfield  {author} {\bibinfo {author} {\bibfnamefont {A.}~\bibnamefont
  {Santra}}, \bibinfo {author} {\bibfnamefont {B.}~\bibnamefont {D{\"u}nweg}},\
  and\ \bibinfo {author} {\bibfnamefont {J.}~\bibnamefont {Ravi~Prakash}},\
  }\href@noop {} {\bibfield  {journal} {\bibinfo  {journal} {J. Rheol.}\
  }\textbf {\bibinfo {volume} {65}},\ \bibinfo {pages} {549} (\bibinfo {year}
  {2021})}\BibitemShut {NoStop}%
\bibitem [{\citenamefont {Fiore}\ \emph {et~al.}(2017)\citenamefont {Fiore},
  \citenamefont {Balboa~Usabiaga}, \citenamefont {Donev},\ and\ \citenamefont
  {Swan}}]{fiore2017rapid}%
  \BibitemOpen
  \bibfield  {author} {\bibinfo {author} {\bibfnamefont {A.~M.}\ \bibnamefont
  {Fiore}}, \bibinfo {author} {\bibfnamefont {F.}~\bibnamefont
  {Balboa~Usabiaga}}, \bibinfo {author} {\bibfnamefont {A.}~\bibnamefont
  {Donev}},\ and\ \bibinfo {author} {\bibfnamefont {J.~W.}\ \bibnamefont
  {Swan}},\ }\href@noop {} {\bibfield  {journal} {\bibinfo  {journal} {J. Chem.
  Phys.}\ }\textbf {\bibinfo {volume} {146}},\ \bibinfo {pages} {124116}
  (\bibinfo {year} {2017})}\BibitemShut {NoStop}%
\bibitem [{\citenamefont {Colby}\ and\ \citenamefont
  {Rubinstein}(2003)}]{colby2003polymer}%
  \BibitemOpen
  \bibfield  {author} {\bibinfo {author} {\bibfnamefont {R.~H.}\ \bibnamefont
  {Colby}}\ and\ \bibinfo {author} {\bibfnamefont {M.}~\bibnamefont
  {Rubinstein}},\ }\href@noop {} {\bibfield  {journal} {\bibinfo  {journal}
  {New-York: Oxford University}\ }\textbf {\bibinfo {volume} {100}},\ \bibinfo
  {pages} {274} (\bibinfo {year} {2003})}\BibitemShut {NoStop}%
\bibitem [{\citenamefont {Theodorou}\ and\ \citenamefont
  {Suter}(1985)}]{theodorou1985shape}%
  \BibitemOpen
  \bibfield  {author} {\bibinfo {author} {\bibfnamefont {D.~N.}\ \bibnamefont
  {Theodorou}}\ and\ \bibinfo {author} {\bibfnamefont {U.~W.}\ \bibnamefont
  {Suter}},\ }\href@noop {} {\bibfield  {journal} {\bibinfo  {journal}
  {Macromolecules}\ }\textbf {\bibinfo {volume} {18}},\ \bibinfo {pages} {1206}
  (\bibinfo {year} {1985})}\BibitemShut {NoStop}%
\bibitem [{\citenamefont {Bishop}\ and\ \citenamefont
  {Michels}(1986)}]{bishop1986polymer}%
  \BibitemOpen
  \bibfield  {author} {\bibinfo {author} {\bibfnamefont {M.}~\bibnamefont
  {Bishop}}\ and\ \bibinfo {author} {\bibfnamefont {J.}~\bibnamefont
  {Michels}},\ }\href@noop {} {\bibfield  {journal} {\bibinfo  {journal} {J.
  Chem. Phys.}\ }\textbf {\bibinfo {volume} {85}},\ \bibinfo {pages} {5961}
  (\bibinfo {year} {1986})}\BibitemShut {NoStop}%
\bibitem [{\citenamefont {Kumari}\ \emph {et~al.}(2020)\citenamefont {Kumari},
  \citenamefont {Duenweg}, \citenamefont {Padinhateeri},\ and\ \citenamefont
  {Prakash}}]{kumari2020computing}%
  \BibitemOpen
  \bibfield  {author} {\bibinfo {author} {\bibfnamefont {K.}~\bibnamefont
  {Kumari}}, \bibinfo {author} {\bibfnamefont {B.}~\bibnamefont {Duenweg}},
  \bibinfo {author} {\bibfnamefont {R.}~\bibnamefont {Padinhateeri}},\ and\
  \bibinfo {author} {\bibfnamefont {J.~R.}\ \bibnamefont {Prakash}},\
  }\href@noop {} {\bibfield  {journal} {\bibinfo  {journal} {Biophys. J.}\
  }\textbf {\bibinfo {volume} {118}},\ \bibinfo {pages} {2193} (\bibinfo {year}
  {2020})}\BibitemShut {NoStop}%
\bibitem [{\citenamefont {Kr{\"o}ger}\ \emph {et~al.}(2000)\citenamefont
  {Kr{\"o}ger}, \citenamefont {Alba-P{\'e}rez}, \citenamefont {Laso},\ and\
  \citenamefont {{\"O}ttinger}}]{kroger2000variance}%
  \BibitemOpen
  \bibfield  {author} {\bibinfo {author} {\bibfnamefont {M.}~\bibnamefont
  {Kr{\"o}ger}}, \bibinfo {author} {\bibfnamefont {A.}~\bibnamefont
  {Alba-P{\'e}rez}}, \bibinfo {author} {\bibfnamefont {M.}~\bibnamefont
  {Laso}},\ and\ \bibinfo {author} {\bibfnamefont {H.~C.}\ \bibnamefont
  {{\"O}ttinger}},\ }\href@noop {} {\bibfield  {journal} {\bibinfo  {journal}
  {J. Chem. Phys.}\ }\textbf {\bibinfo {volume} {113}},\ \bibinfo {pages}
  {4767} (\bibinfo {year} {2000})}\BibitemShut {NoStop}%
\bibitem [{\citenamefont {Sunthar}\ and\ \citenamefont
  {Prakash}(2006)}]{sunthar2006dynamic}%
  \BibitemOpen
  \bibfield  {author} {\bibinfo {author} {\bibfnamefont {P.}~\bibnamefont
  {Sunthar}}\ and\ \bibinfo {author} {\bibfnamefont {J.~R.}\ \bibnamefont
  {Prakash}},\ }\href@noop {} {\bibfield  {journal} {\bibinfo  {journal} {EPL}\
  }\textbf {\bibinfo {volume} {75}},\ \bibinfo {pages} {77} (\bibinfo {year}
  {2006})}\BibitemShut {NoStop}%
\bibitem [{\citenamefont {Pan}\ \emph {et~al.}(2014)\citenamefont {Pan},
  \citenamefont {Ahirwal}, \citenamefont {Nguyen}, \citenamefont {Sridhar},
  \citenamefont {Sunthar},\ and\ \citenamefont {Prakash}}]{pan2014viscosity}%
  \BibitemOpen
  \bibfield  {author} {\bibinfo {author} {\bibfnamefont {S.}~\bibnamefont
  {Pan}}, \bibinfo {author} {\bibfnamefont {D.}~\bibnamefont {Ahirwal}},
  \bibinfo {author} {\bibfnamefont {D.~A.}\ \bibnamefont {Nguyen}}, \bibinfo
  {author} {\bibfnamefont {T.}~\bibnamefont {Sridhar}}, \bibinfo {author}
  {\bibfnamefont {P.}~\bibnamefont {Sunthar}},\ and\ \bibinfo {author}
  {\bibfnamefont {J.~R.}\ \bibnamefont {Prakash}},\ }\href@noop {} {\bibfield
  {journal} {\bibinfo  {journal} {Macromolecules}\ }\textbf {\bibinfo {volume}
  {47}},\ \bibinfo {pages} {7548} (\bibinfo {year} {2014})}\BibitemShut
  {NoStop}%
\bibitem [{\citenamefont {Guinier}\ \emph {et~al.}(1955)\citenamefont
  {Guinier}, \citenamefont {Fournet}, \citenamefont {Walker},\ and\
  \citenamefont {Yudowitch}}]{guinier1955small}%
  \BibitemOpen
  \bibfield  {author} {\bibinfo {author} {\bibfnamefont {A.}~\bibnamefont
  {Guinier}}, \bibinfo {author} {\bibfnamefont {G.}~\bibnamefont {Fournet}},
  \bibinfo {author} {\bibfnamefont {C.~B.}\ \bibnamefont {Walker}},\ and\
  \bibinfo {author} {\bibfnamefont {K.~L.}\ \bibnamefont {Yudowitch}},\
  }\href@noop {} {\emph {\bibinfo {title} {Small-angle Scattering of X-rays}}}\
  (\bibinfo  {publisher} {Wiley New York},\ \bibinfo {year} {1955})\BibitemShut
  {NoStop}%
\bibitem [{\citenamefont {Huang}\ \emph {et~al.}(2010)\citenamefont {Huang},
  \citenamefont {Winkler}, \citenamefont {Sutmann},\ and\ \citenamefont
  {Gompper}}]{huang2010semidilute}%
  \BibitemOpen
  \bibfield  {author} {\bibinfo {author} {\bibfnamefont {C.-C.}\ \bibnamefont
  {Huang}}, \bibinfo {author} {\bibfnamefont {R.~G.}\ \bibnamefont {Winkler}},
  \bibinfo {author} {\bibfnamefont {G.}~\bibnamefont {Sutmann}},\ and\ \bibinfo
  {author} {\bibfnamefont {G.}~\bibnamefont {Gompper}},\ }\href@noop {}
  {\bibfield  {journal} {\bibinfo  {journal} {Macromolecules}\ }\textbf
  {\bibinfo {volume} {43}},\ \bibinfo {pages} {10107} (\bibinfo {year}
  {2010})}\BibitemShut {NoStop}%
\bibitem [{\citenamefont {Daoud}\ \emph {et~al.}(1975)\citenamefont {Daoud},
  \citenamefont {Cotton}, \citenamefont {Farnoux}, \citenamefont {Jannink},
  \citenamefont {Sarma}, \citenamefont {Benoit}, \citenamefont {Duplessix},
  \citenamefont {Picot},\ and\ \citenamefont {de~Gennes}}]{daoud1975solutions}%
  \BibitemOpen
  \bibfield  {author} {\bibinfo {author} {\bibfnamefont {M.}~\bibnamefont
  {Daoud}}, \bibinfo {author} {\bibfnamefont {J.}~\bibnamefont {Cotton}},
  \bibinfo {author} {\bibfnamefont {B.}~\bibnamefont {Farnoux}}, \bibinfo
  {author} {\bibfnamefont {G.}~\bibnamefont {Jannink}}, \bibinfo {author}
  {\bibfnamefont {G.}~\bibnamefont {Sarma}}, \bibinfo {author} {\bibfnamefont
  {H.}~\bibnamefont {Benoit}}, \bibinfo {author} {\bibfnamefont
  {C.}~\bibnamefont {Duplessix}}, \bibinfo {author} {\bibfnamefont
  {C.}~\bibnamefont {Picot}},\ and\ \bibinfo {author} {\bibfnamefont
  {P.}~\bibnamefont {de~Gennes}},\ }\href@noop {} {\bibfield  {journal}
  {\bibinfo  {journal} {Macromolecules}\ }\textbf {\bibinfo {volume} {8}},\
  \bibinfo {pages} {804} (\bibinfo {year} {1975})}\BibitemShut {NoStop}%
\bibitem [{\citenamefont {Doi}\ and\ \citenamefont
  {Edwards}(1988)}]{doi1988theory}%
  \BibitemOpen
  \bibfield  {author} {\bibinfo {author} {\bibfnamefont {M.}~\bibnamefont
  {Doi}}\ and\ \bibinfo {author} {\bibfnamefont {S.~F.}\ \bibnamefont
  {Edwards}},\ }\href@noop {} {\emph {\bibinfo {title} {The Theory of Polymer
  Dynamics}}},\ Vol.~\bibinfo {volume} {73}\ (\bibinfo  {publisher} {Oxford
  University Press},\ \bibinfo {year} {1988})\BibitemShut {NoStop}%
\bibitem [{\citenamefont {Cannon}, \citenamefont {Aronovitz},\ and\
  \citenamefont {Goldbart}(1991)}]{cannon1991equilibrium}%
  \BibitemOpen
  \bibfield  {author} {\bibinfo {author} {\bibfnamefont {J.~W.}\ \bibnamefont
  {Cannon}}, \bibinfo {author} {\bibfnamefont {J.~A.}\ \bibnamefont
  {Aronovitz}},\ and\ \bibinfo {author} {\bibfnamefont {P.}~\bibnamefont
  {Goldbart}},\ }\href@noop {} {\bibfield  {journal} {\bibinfo  {journal} {J.
  Phys. I}\ }\textbf {\bibinfo {volume} {1}},\ \bibinfo {pages} {629} (\bibinfo
  {year} {1991})}\BibitemShut {NoStop}%
\bibitem [{\citenamefont {Tande}\ \emph {et~al.}(2001)\citenamefont {Tande},
  \citenamefont {Wagner}, \citenamefont {Mackay}, \citenamefont {Hawker},\ and\
  \citenamefont {Jeong}}]{tande2001viscosimetric}%
  \BibitemOpen
  \bibfield  {author} {\bibinfo {author} {\bibfnamefont {B.~M.}\ \bibnamefont
  {Tande}}, \bibinfo {author} {\bibfnamefont {N.~J.}\ \bibnamefont {Wagner}},
  \bibinfo {author} {\bibfnamefont {M.~E.}\ \bibnamefont {Mackay}}, \bibinfo
  {author} {\bibfnamefont {C.~J.}\ \bibnamefont {Hawker}},\ and\ \bibinfo
  {author} {\bibfnamefont {M.}~\bibnamefont {Jeong}},\ }\href@noop {}
  {\bibfield  {journal} {\bibinfo  {journal} {Macromolecules}\ }\textbf
  {\bibinfo {volume} {34}},\ \bibinfo {pages} {8580} (\bibinfo {year}
  {2001})}\BibitemShut {NoStop}%
\bibitem [{\citenamefont {Phillies}(2015)}]{phillies2015complex}%
  \BibitemOpen
  \bibfield  {author} {\bibinfo {author} {\bibfnamefont {G.~D.~J.}\
  \bibnamefont {Phillies}},\ }\href@noop {} {\bibfield  {journal} {\bibinfo
  {journal} {Soft Matter}\ }\textbf {\bibinfo {volume} {11}},\ \bibinfo {pages}
  {580} (\bibinfo {year} {2015})}\BibitemShut {NoStop}%
\bibitem [{\citenamefont {Kumar}\ \emph {et~al.}(2019)\citenamefont {Kumar},
  \citenamefont {Theeyancheri}, \citenamefont {Chaki},\ and\ \citenamefont
  {Chakrabarti}}]{kumar2019transport}%
  \BibitemOpen
  \bibfield  {author} {\bibinfo {author} {\bibfnamefont {P.}~\bibnamefont
  {Kumar}}, \bibinfo {author} {\bibfnamefont {L.}~\bibnamefont {Theeyancheri}},
  \bibinfo {author} {\bibfnamefont {S.}~\bibnamefont {Chaki}},\ and\ \bibinfo
  {author} {\bibfnamefont {R.}~\bibnamefont {Chakrabarti}},\ }\href@noop {}
  {\bibfield  {journal} {\bibinfo  {journal} {Soft Matter}\ }\textbf {\bibinfo
  {volume} {15}},\ \bibinfo {pages} {8992} (\bibinfo {year}
  {2019})}\BibitemShut {NoStop}%
\bibitem [{\citenamefont {Wang}\ \emph {et~al.}(2012)\citenamefont {Wang},
  \citenamefont {Kuo}, \citenamefont {Bae},\ and\ \citenamefont
  {Granick}}]{wang2012brownian}%
  \BibitemOpen
  \bibfield  {author} {\bibinfo {author} {\bibfnamefont {B.}~\bibnamefont
  {Wang}}, \bibinfo {author} {\bibfnamefont {J.}~\bibnamefont {Kuo}}, \bibinfo
  {author} {\bibfnamefont {S.~C.}\ \bibnamefont {Bae}},\ and\ \bibinfo {author}
  {\bibfnamefont {S.}~\bibnamefont {Granick}},\ }\href@noop {} {\bibfield
  {journal} {\bibinfo  {journal} {Nat. Mater.}\ }\textbf {\bibinfo {volume}
  {11}},\ \bibinfo {pages} {481} (\bibinfo {year} {2012})}\BibitemShut
  {NoStop}%
\bibitem [{\citenamefont {Wang}\ \emph {et~al.}(2009)\citenamefont {Wang},
  \citenamefont {Anthony}, \citenamefont {Bae},\ and\ \citenamefont
  {Granick}}]{wang2009anomalous}%
  \BibitemOpen
  \bibfield  {author} {\bibinfo {author} {\bibfnamefont {B.}~\bibnamefont
  {Wang}}, \bibinfo {author} {\bibfnamefont {S.~M.}\ \bibnamefont {Anthony}},
  \bibinfo {author} {\bibfnamefont {S.~C.}\ \bibnamefont {Bae}},\ and\ \bibinfo
  {author} {\bibfnamefont {S.}~\bibnamefont {Granick}},\ }\href@noop {}
  {\bibfield  {journal} {\bibinfo  {journal} {Proc. Natl. Acad. Sci. U.S.A.}\
  }\textbf {\bibinfo {volume} {106}},\ \bibinfo {pages} {15160} (\bibinfo
  {year} {2009})}\BibitemShut {NoStop}%
\bibitem [{\citenamefont {Slim}, \citenamefont {Poling-Skutvik},\ and\
  \citenamefont {Conrad}(2020)}]{slim2020local}%
  \BibitemOpen
  \bibfield  {author} {\bibinfo {author} {\bibfnamefont {A.~H.}\ \bibnamefont
  {Slim}}, \bibinfo {author} {\bibfnamefont {R.}~\bibnamefont
  {Poling-Skutvik}},\ and\ \bibinfo {author} {\bibfnamefont {J.~C.}\
  \bibnamefont {Conrad}},\ }\href@noop {} {\bibfield  {journal} {\bibinfo
  {journal} {Langmuir}\ }\textbf {\bibinfo {volume} {36}},\ \bibinfo {pages}
  {9153} (\bibinfo {year} {2020})}\BibitemShut {NoStop}%
\bibitem [{\citenamefont {Smith}\ \emph {et~al.}(2021)\citenamefont {Smith},
  \citenamefont {Poling-Skutvik}, \citenamefont {Slim}, \citenamefont
  {Willson},\ and\ \citenamefont {Conrad}}]{smith2021dynamics}%
  \BibitemOpen
  \bibfield  {author} {\bibinfo {author} {\bibfnamefont {M.}~\bibnamefont
  {Smith}}, \bibinfo {author} {\bibfnamefont {R.}~\bibnamefont
  {Poling-Skutvik}}, \bibinfo {author} {\bibfnamefont {A.~H.}\ \bibnamefont
  {Slim}}, \bibinfo {author} {\bibfnamefont {R.~C.}\ \bibnamefont {Willson}},\
  and\ \bibinfo {author} {\bibfnamefont {J.~C.}\ \bibnamefont {Conrad}},\
  }\href@noop {} {\bibfield  {journal} {\bibinfo  {journal} {Macromolecules}\
  }\textbf {\bibinfo {volume} {54}},\ \bibinfo {pages} {4557} (\bibinfo {year}
  {2021})}\BibitemShut {NoStop}%
\bibitem [{\citenamefont {Lampo}\ \emph {et~al.}(2017)\citenamefont {Lampo},
  \citenamefont {Stylianidou}, \citenamefont {Backlund}, \citenamefont
  {Wiggins},\ and\ \citenamefont {Spakowitz}}]{lampo2017cytoplasmic}%
  \BibitemOpen
  \bibfield  {author} {\bibinfo {author} {\bibfnamefont {T.~J.}\ \bibnamefont
  {Lampo}}, \bibinfo {author} {\bibfnamefont {S.}~\bibnamefont {Stylianidou}},
  \bibinfo {author} {\bibfnamefont {M.~P.}\ \bibnamefont {Backlund}}, \bibinfo
  {author} {\bibfnamefont {P.~A.}\ \bibnamefont {Wiggins}},\ and\ \bibinfo
  {author} {\bibfnamefont {A.~J.}\ \bibnamefont {Spakowitz}},\ }\href@noop {}
  {\bibfield  {journal} {\bibinfo  {journal} {Biophys. J.}\ }\textbf {\bibinfo
  {volume} {112}},\ \bibinfo {pages} {532} (\bibinfo {year}
  {2017})}\BibitemShut {NoStop}%
\bibitem [{\citenamefont {He}\ \emph {et~al.}(2016)\citenamefont {He},
  \citenamefont {Song}, \citenamefont {Su}, \citenamefont {Geng}, \citenamefont
  {Ackerson}, \citenamefont {Peng},\ and\ \citenamefont
  {Tong}}]{he2016dynamic}%
  \BibitemOpen
  \bibfield  {author} {\bibinfo {author} {\bibfnamefont {W.}~\bibnamefont
  {He}}, \bibinfo {author} {\bibfnamefont {H.}~\bibnamefont {Song}}, \bibinfo
  {author} {\bibfnamefont {Y.}~\bibnamefont {Su}}, \bibinfo {author}
  {\bibfnamefont {L.}~\bibnamefont {Geng}}, \bibinfo {author} {\bibfnamefont
  {B.~J.}\ \bibnamefont {Ackerson}}, \bibinfo {author} {\bibfnamefont
  {H.}~\bibnamefont {Peng}},\ and\ \bibinfo {author} {\bibfnamefont
  {P.}~\bibnamefont {Tong}},\ }\href@noop {} {\bibfield  {journal} {\bibinfo
  {journal} {Nat. Commun.}\ }\textbf {\bibinfo {volume} {7}},\ \bibinfo {pages}
  {11701} (\bibinfo {year} {2016})}\BibitemShut {NoStop}%
\bibitem [{\citenamefont {Jee}, \citenamefont {Curtis-Fisk},\ and\
  \citenamefont {Granick}(2014)}]{jee2014nanoparticle}%
  \BibitemOpen
  \bibfield  {author} {\bibinfo {author} {\bibfnamefont {A.-Y.}\ \bibnamefont
  {Jee}}, \bibinfo {author} {\bibfnamefont {J.~L.}\ \bibnamefont
  {Curtis-Fisk}},\ and\ \bibinfo {author} {\bibfnamefont {S.}~\bibnamefont
  {Granick}},\ }\href@noop {} {\bibfield  {journal} {\bibinfo  {journal}
  {Macromolecules}\ }\textbf {\bibinfo {volume} {47}},\ \bibinfo {pages} {5793}
  (\bibinfo {year} {2014})}\BibitemShut {NoStop}%
\bibitem [{\citenamefont {Guan}, \citenamefont {Wang},\ and\ \citenamefont
  {Granick}(2014)}]{guan2014even}%
  \BibitemOpen
  \bibfield  {author} {\bibinfo {author} {\bibfnamefont {J.}~\bibnamefont
  {Guan}}, \bibinfo {author} {\bibfnamefont {B.}~\bibnamefont {Wang}},\ and\
  \bibinfo {author} {\bibfnamefont {S.}~\bibnamefont {Granick}},\ }\href@noop
  {} {\bibfield  {journal} {\bibinfo  {journal} {ACS Nano}\ }\textbf {\bibinfo
  {volume} {8}},\ \bibinfo {pages} {3331} (\bibinfo {year} {2014})}\BibitemShut
  {NoStop}%
\bibitem [{\citenamefont {Xue}\ \emph {et~al.}(2016)\citenamefont {Xue},
  \citenamefont {Zheng}, \citenamefont {Chen}, \citenamefont {Tian},\ and\
  \citenamefont {Hu}}]{xue2016probing}%
  \BibitemOpen
  \bibfield  {author} {\bibinfo {author} {\bibfnamefont {C.}~\bibnamefont
  {Xue}}, \bibinfo {author} {\bibfnamefont {X.}~\bibnamefont {Zheng}}, \bibinfo
  {author} {\bibfnamefont {K.}~\bibnamefont {Chen}}, \bibinfo {author}
  {\bibfnamefont {Y.}~\bibnamefont {Tian}},\ and\ \bibinfo {author}
  {\bibfnamefont {G.}~\bibnamefont {Hu}},\ }\href@noop {} {\bibfield  {journal}
  {\bibinfo  {journal} {J. Phys. Chem. Lett.}\ }\textbf {\bibinfo {volume}
  {7}},\ \bibinfo {pages} {514} (\bibinfo {year} {2016})}\BibitemShut {NoStop}%
\bibitem [{\citenamefont {Montroll}\ and\ \citenamefont
  {Weiss}(1965)}]{montroll1965random}%
  \BibitemOpen
  \bibfield  {author} {\bibinfo {author} {\bibfnamefont {E.~W.}\ \bibnamefont
  {Montroll}}\ and\ \bibinfo {author} {\bibfnamefont {G.~H.}\ \bibnamefont
  {Weiss}},\ }\href@noop {} {\bibfield  {journal} {\bibinfo  {journal} {Journal
  of Mathematical Physics}\ }\textbf {\bibinfo {volume} {6}},\ \bibinfo {pages}
  {167} (\bibinfo {year} {1965})}\BibitemShut {NoStop}%
\bibitem [{\citenamefont {Mandelbrot}\ and\ \citenamefont
  {Van~Ness}(1968)}]{mandelbrot1968fractional}%
  \BibitemOpen
  \bibfield  {author} {\bibinfo {author} {\bibfnamefont {B.~B.}\ \bibnamefont
  {Mandelbrot}}\ and\ \bibinfo {author} {\bibfnamefont {J.~W.}\ \bibnamefont
  {Van~Ness}},\ }\href@noop {} {\bibfield  {journal} {\bibinfo  {journal} {SIAM
  Rev.}\ }\textbf {\bibinfo {volume} {10}},\ \bibinfo {pages} {422} (\bibinfo
  {year} {1968})}\BibitemShut {NoStop}%
\bibitem [{\citenamefont {Chen}\ \emph {et~al.}(2022)\citenamefont {Chen},
  \citenamefont {Xu}, \citenamefont {Ma}, \citenamefont {Liu},\ and\
  \citenamefont {Zhang}}]{chen2022diffusion}%
  \BibitemOpen
  \bibfield  {author} {\bibinfo {author} {\bibfnamefont {Y.}~\bibnamefont
  {Chen}}, \bibinfo {author} {\bibfnamefont {H.}~\bibnamefont {Xu}}, \bibinfo
  {author} {\bibfnamefont {Y.}~\bibnamefont {Ma}}, \bibinfo {author}
  {\bibfnamefont {J.}~\bibnamefont {Liu}},\ and\ \bibinfo {author}
  {\bibfnamefont {L.}~\bibnamefont {Zhang}},\ }\href@noop {} {\bibfield
  {journal} {\bibinfo  {journal} {Phys. Chem. Chem. Phys.}\ }\textbf {\bibinfo
  {volume} {24}},\ \bibinfo {pages} {11322} (\bibinfo {year}
  {2022})}\BibitemShut {NoStop}%
\bibitem [{\citenamefont {Kohli}\ and\ \citenamefont
  {Mukhopadhyay}(2012)}]{kohli2012diffusion}%
  \BibitemOpen
  \bibfield  {author} {\bibinfo {author} {\bibfnamefont {I.}~\bibnamefont
  {Kohli}}\ and\ \bibinfo {author} {\bibfnamefont {A.}~\bibnamefont
  {Mukhopadhyay}},\ }\href@noop {} {\bibfield  {journal} {\bibinfo  {journal}
  {Macromolecules}\ }\textbf {\bibinfo {volume} {45}},\ \bibinfo {pages} {6143}
  (\bibinfo {year} {2012})}\BibitemShut {NoStop}%
\bibitem [{\citenamefont {Poling-Skutvik}\ \emph {et~al.}(2019)\citenamefont
  {Poling-Skutvik}, \citenamefont {Slim}, \citenamefont {Narayanan},
  \citenamefont {Conrad},\ and\ \citenamefont
  {Krishnamoorti}}]{poling2019soft}%
  \BibitemOpen
  \bibfield  {author} {\bibinfo {author} {\bibfnamefont {R.}~\bibnamefont
  {Poling-Skutvik}}, \bibinfo {author} {\bibfnamefont {A.~H.}\ \bibnamefont
  {Slim}}, \bibinfo {author} {\bibfnamefont {S.}~\bibnamefont {Narayanan}},
  \bibinfo {author} {\bibfnamefont {J.~C.}\ \bibnamefont {Conrad}},\ and\
  \bibinfo {author} {\bibfnamefont {R.}~\bibnamefont {Krishnamoorti}},\
  }\href@noop {} {\bibfield  {journal} {\bibinfo  {journal} {ACS Macro
  Letters}\ }\textbf {\bibinfo {volume} {8}},\ \bibinfo {pages} {917} (\bibinfo
  {year} {2019})}\BibitemShut {NoStop}%
\bibitem [{\citenamefont {Mainardi}(2020)}]{mainardi2020mittag}%
  \BibitemOpen
  \bibfield  {author} {\bibinfo {author} {\bibfnamefont {F.}~\bibnamefont
  {Mainardi}},\ }\href@noop {} {\bibfield  {journal} {\bibinfo  {journal}
  {Entropy}\ }\textbf {\bibinfo {volume} {22}},\ \bibinfo {pages} {1359}
  (\bibinfo {year} {2020})}\BibitemShut {NoStop}%
\bibitem [{\citenamefont {Jaishankar}\ and\ \citenamefont
  {McKinley}(2013)}]{jaishankar2013power}%
  \BibitemOpen
  \bibfield  {author} {\bibinfo {author} {\bibfnamefont {A.}~\bibnamefont
  {Jaishankar}}\ and\ \bibinfo {author} {\bibfnamefont {G.~H.}\ \bibnamefont
  {McKinley}},\ }\href@noop {} {\bibfield  {journal} {\bibinfo  {journal}
  {Proc. R. Soc. A}\ }\textbf {\bibinfo {volume} {469}},\ \bibinfo {pages}
  {20120284} (\bibinfo {year} {2013})}\BibitemShut {NoStop}%
\end{thebibliography}%


\begin{thebibliography}{6}%
\makeatletter
\providecommand \@ifxundefined [1]{%
 \@ifx{#1\undefined}
}%
\providecommand \@ifnum [1]{%
 \ifnum #1\expandafter \@firstoftwo
 \else \expandafter \@secondoftwo
 \fi
}%
\providecommand \@ifx [1]{%
 \ifx #1\expandafter \@firstoftwo
 \else \expandafter \@secondoftwo
 \fi
}%
\providecommand \natexlab [1]{#1}%
\providecommand \enquote  [1]{``#1''}%
\providecommand \bibnamefont  [1]{#1}%
\providecommand \bibfnamefont [1]{#1}%
\providecommand \citenamefont [1]{#1}%
\providecommand \href@noop [0]{\@secondoftwo}%
\providecommand \href [0]{\begingroup \@sanitize@url \@href}%
\providecommand \@href[1]{\@@startlink{#1}\@@href}%
\providecommand \@@href[1]{\endgroup#1\@@endlink}%
\providecommand \@sanitize@url [0]{\catcode `\\12\catcode `\$12\catcode
  `\&12\catcode `\#12\catcode `\^12\catcode `\_12\catcode `\%12\relax}%
\providecommand \@@startlink[1]{}%
\providecommand \@@endlink[0]{}%
\providecommand \url  [0]{\begingroup\@sanitize@url \@url }%
\providecommand \@url [1]{\endgroup\@href {#1}{\urlprefix }}%
\providecommand \urlprefix  [0]{URL }%
\providecommand \Eprint [0]{\href }%
\providecommand \doibase [0]{https://doi.org/}%
\providecommand \selectlanguage [0]{\@gobble}%
\providecommand \bibinfo  [0]{\@secondoftwo}%
\providecommand \bibfield  [0]{\@secondoftwo}%
\providecommand \translation [1]{[#1]}%
\providecommand \BibitemOpen [0]{}%
\providecommand \bibitemStop [0]{}%
\providecommand \bibitemNoStop [0]{.\EOS\space}%
\providecommand \EOS [0]{\spacefactor3000\relax}%
\providecommand \BibitemShut  [1]{\csname bibitem#1\endcsname}%
\let\auto@bib@innerbib\@empty
\bibitem [{\citenamefont {Colby}\ and\ \citenamefont
  {Rubinstein}(2003)}]{colby2003polymer}%
  \BibitemOpen
  \bibfield  {author} {\bibinfo {author} {\bibfnamefont {R.~H.}\ \bibnamefont
  {Colby}}\ and\ \bibinfo {author} {\bibfnamefont {M.}~\bibnamefont
  {Rubinstein}},\ }\href@noop {} {\bibfield  {journal} {\bibinfo  {journal}
  {New-York: Oxford University}\ }\textbf {\bibinfo {volume} {100}},\ \bibinfo
  {pages} {274} (\bibinfo {year} {2003})}\BibitemShut {NoStop}%
\bibitem [{\citenamefont {Theodorou}\ and\ \citenamefont
  {Suter}(1985)}]{theodorou1985shape}%
  \BibitemOpen
  \bibfield  {author} {\bibinfo {author} {\bibfnamefont {D.~N.}\ \bibnamefont
  {Theodorou}}\ and\ \bibinfo {author} {\bibfnamefont {U.~W.}\ \bibnamefont
  {Suter}},\ }\href@noop {} {\bibfield  {journal} {\bibinfo  {journal}
  {Macromolecules}\ }\textbf {\bibinfo {volume} {18}},\ \bibinfo {pages} {1206}
  (\bibinfo {year} {1985})}\BibitemShut {NoStop}%
\bibitem [{\citenamefont {Bishop}\ and\ \citenamefont
  {Michels}(1986)}]{bishop1986polymer}%
  \BibitemOpen
  \bibfield  {author} {\bibinfo {author} {\bibfnamefont {M.}~\bibnamefont
  {Bishop}}\ and\ \bibinfo {author} {\bibfnamefont {J.}~\bibnamefont
  {Michels}},\ }\href@noop {} {\bibfield  {journal} {\bibinfo  {journal} {J.
  Chem. Phys.}\ }\textbf {\bibinfo {volume} {85}},\ \bibinfo {pages} {5961}
  (\bibinfo {year} {1986})}\BibitemShut {NoStop}%
\bibitem [{\citenamefont {Kumari}\ \emph {et~al.}(2020)\citenamefont {Kumari},
  \citenamefont {Duenweg}, \citenamefont {Padinhateeri},\ and\ \citenamefont
  {Prakash}}]{kumari2020computing}%
  \BibitemOpen
  \bibfield  {author} {\bibinfo {author} {\bibfnamefont {K.}~\bibnamefont
  {Kumari}}, \bibinfo {author} {\bibfnamefont {B.}~\bibnamefont {Duenweg}},
  \bibinfo {author} {\bibfnamefont {R.}~\bibnamefont {Padinhateeri}},\ and\
  \bibinfo {author} {\bibfnamefont {J.~R.}\ \bibnamefont {Prakash}},\
  }\href@noop {} {\bibfield  {journal} {\bibinfo  {journal} {Biophys. J.}\
  }\textbf {\bibinfo {volume} {118}},\ \bibinfo {pages} {2193} (\bibinfo {year}
  {2020})}\BibitemShut {NoStop}%
\bibitem [{\citenamefont {Poling-Skutvik}\ \emph {et~al.}(2019)\citenamefont
  {Poling-Skutvik}, \citenamefont {Slim}, \citenamefont {Narayanan},
  \citenamefont {Conrad},\ and\ \citenamefont
  {Krishnamoorti}}]{poling2019soft}%
  \BibitemOpen
  \bibfield  {author} {\bibinfo {author} {\bibfnamefont {R.}~\bibnamefont
  {Poling-Skutvik}}, \bibinfo {author} {\bibfnamefont {A.~H.}\ \bibnamefont
  {Slim}}, \bibinfo {author} {\bibfnamefont {S.}~\bibnamefont {Narayanan}},
  \bibinfo {author} {\bibfnamefont {J.~C.}\ \bibnamefont {Conrad}},\ and\
  \bibinfo {author} {\bibfnamefont {R.}~\bibnamefont {Krishnamoorti}},\
  }\href@noop {} {\bibfield  {journal} {\bibinfo  {journal} {ACS Macro
  Letters}\ }\textbf {\bibinfo {volume} {8}},\ \bibinfo {pages} {917} (\bibinfo
  {year} {2019})}\BibitemShut {NoStop}%
\bibitem [{\citenamefont {Cai}, \citenamefont {Panyukov},\ and\ \citenamefont
  {Rubinstein}(2011)}]{cai2011mobility}%
  \BibitemOpen
  \bibfield  {author} {\bibinfo {author} {\bibfnamefont {L.-H.}\ \bibnamefont
  {Cai}}, \bibinfo {author} {\bibfnamefont {S.}~\bibnamefont {Panyukov}},\ and\
  \bibinfo {author} {\bibfnamefont {M.}~\bibnamefont {Rubinstein}},\
  }\href@noop {} {\bibfield  {journal} {\bibinfo  {journal} {Macromolecules}\
  }\textbf {\bibinfo {volume} {44}},\ \bibinfo {pages} {7853} (\bibinfo {year}
  {2011})}\BibitemShut {NoStop}%
\end{thebibliography}%
\bibliographystyle{aapmrev4-2} 

\end{document}


\beginsupplement
\title{Supplementary Material for: Universal scaling of the diffusivity of dendrimers in a semidilute solution of linear polymers}
\date{\today}
\author{Silpa Mariya}
\email{silpa.mariya@monash.edu}
\affiliation{IITB-Monash Research Academy, Indian Institute of Technology Bombay, Mumbai, 400076, India.}
\affiliation{Department of Chemical Engineering, 
Indian Institute of Technology Bombay, Mumbai, 400076, India.}
\affiliation{Department of Chemical Engineering, Monash University,
Melbourne, VIC 3800, Australia}
\author{Jeremy J. Barr}
\email{jeremy.barr@monash.edu}
\affiliation{School of Biological Sciences, Monash University, Clayton, VIC 3800, Australia.}
\author{P. Sunthar}
\email{p.sunthar@iitb.ac.in}
\affiliation{Department of Chemical Engineering,
  Indian Institute of Technology Bombay, Mumbai, 400076, India.}
\author{J. Ravi Prakash}
\email{ravi.jagadeeshan@monash.edu}
\affiliation{Department of Chemical Engineering, Monash University,
Melbourne, VIC 3800, Australia}

\maketitle

\section{Radius of gyration versus number of beads for various topologies}

\begin{figure*}[ptbh]
	\begin{center}
		\begin{tabular}{cc}
			\resizebox{7cm}{!} {\includegraphics[width=4cm]{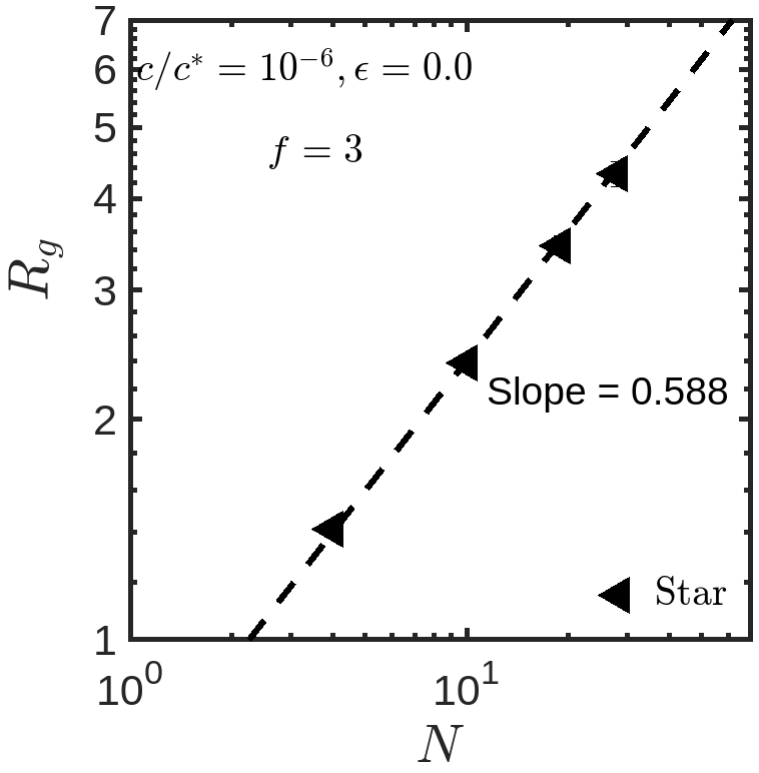}} &
			\resizebox{7cm}{!} {\includegraphics[width=4cm]{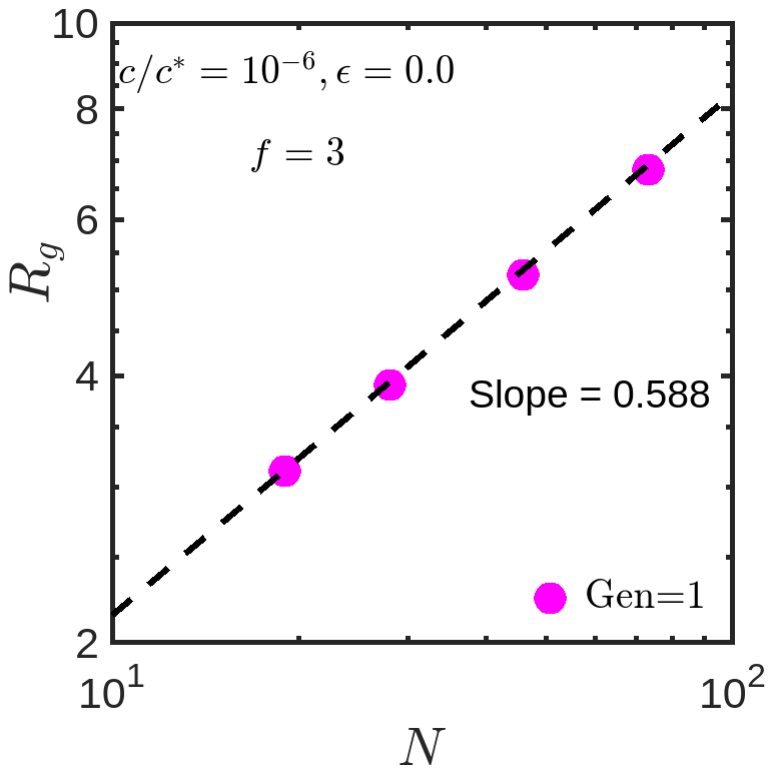}}\\
			(a) & (b)  \\
			\resizebox{7cm}{!} {\includegraphics[width=4cm]{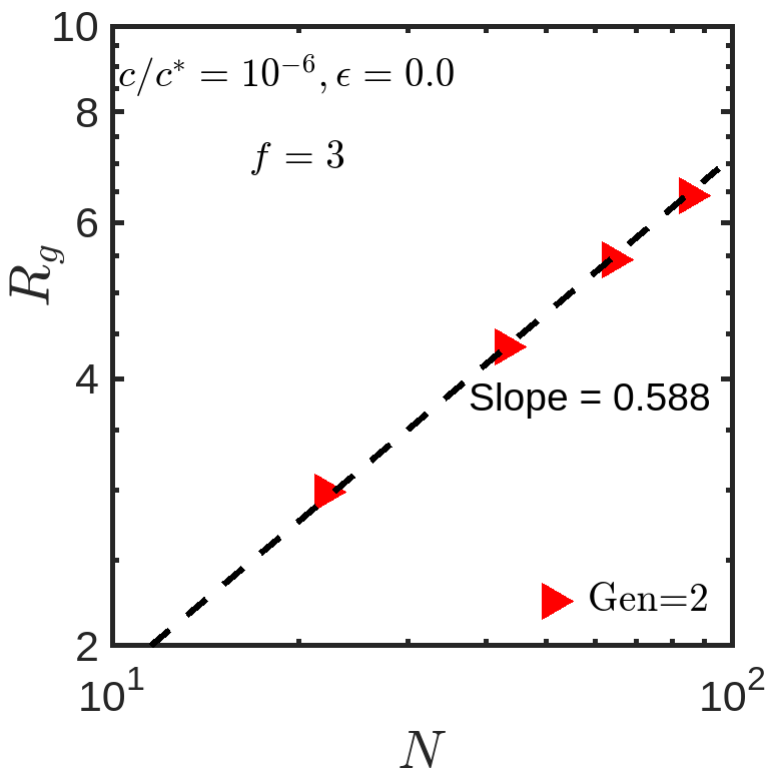}} &
			\resizebox{7cm}{!} {\includegraphics[width=4cm]{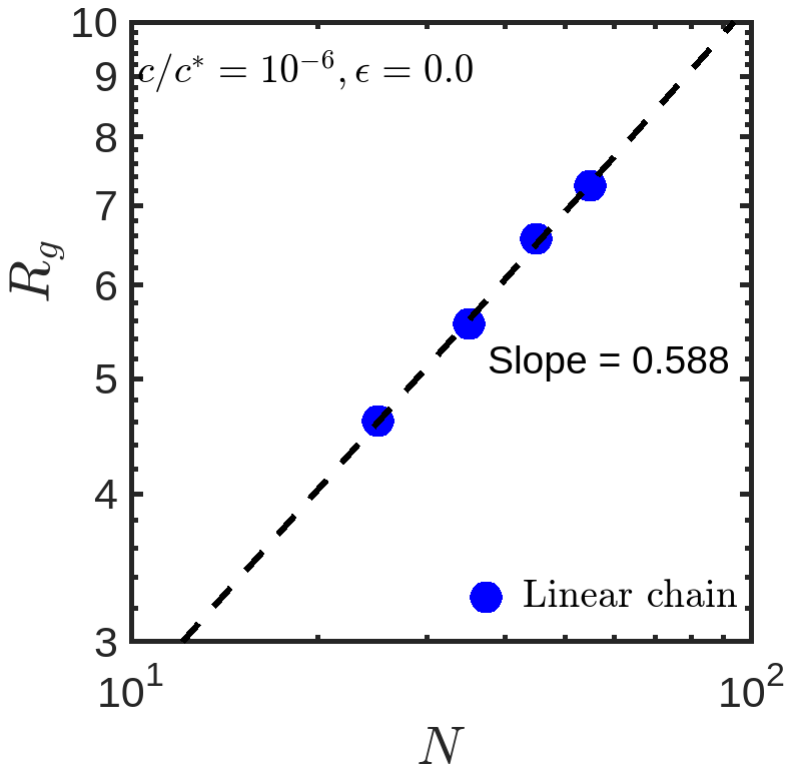}}\\
			(c) & (d)  \\
		\end{tabular}
	\end{center}
	\vspace{-15pt}
	\caption{ (Color online) Radius of gyration versus number of beads per molecule (a) Star polymers with functionality 3.  (b) Generation 1 dendrimer with functionality 3. (c) Generation 2 dendrimer with functionality 3. (d) Linear chains. The dashed lines are the scaling law given by eqns~\eqref{eq:Rg0_d} and ~\eqref{eq:Rg0_lc}}.
    \label{fig:Rg_plots}
\end{figure*}

In order to determine the dendrimer parameters and length of linear chains in the solution, we used their respective radius of gyration versus number of beads per molecule plots. In Fig ~\ref{fig:Rg_plots} (a),(b) and (c), the functionality ($f$) and generation number ($g$) of the dendrimer is fixed and number of beads is varied by changing the spacer length ($s$). Linear chains and dendrimers follow the scaling law given below:

\begin{align}
    R_{\textrm{g0}}^{{\textrm{lc}}} &= a^{\textrm{lc}} \left(N_b^{\textrm{lc}}\right)^{\nu} \label{eq:Rg0_d}\\
    R_{\textrm{g0}}^{{\textrm{d}}} &= a^{\textrm{d}} \left(N_b^{\textrm{d}}\right)^{\nu} \label{eq:Rg0_lc}
\end{align}
Here $\nu$ is the Flory exponent, $\nu=0.588$. This information is used to construct Table 1 in the main text, for the various systems we have studied.

\section{Size regimes}

\begin{figure}[ptbh]
	\begin{center}
			\resizebox{9.0cm}{!} {\includegraphics[width=4cm]{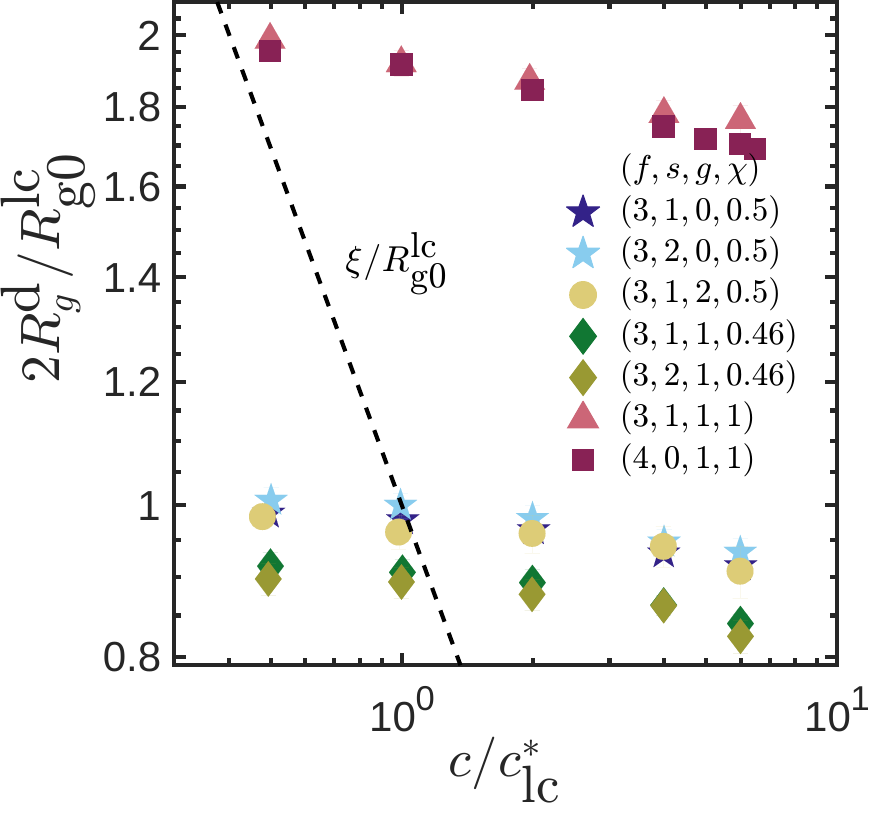}} 		
	\end{center}
	\vspace{-15pt}
	\caption{(Color online) Size regimes based on relative sizes of polymers and correlation length of the solution. The dendrimer architecture is included in the order ($f,s,g,\chi$). The \textit{y}-axis is the ratio between the size of the dendrimer and the radius of gyration of the linear chain in the dilute limit in each simulation system. The dashed line represents the scaling of $\xi$ with concentration given by eqn~\eqref{eq:xi}.}
\label{fig:size_regimes}
\end{figure}

The correlation length of the solution is given by \cite{colby2003polymer}, 

\begin{align}\label{eq:xi}
     \xi=R_{\textrm{g0}}^{\textrm{lc}} \left(\frac{c}{c^{\ast}_{\textrm{lc}}}\right)^{\frac{-\nu}{\left(3\nu -1 \right)}}
\end{align}

\noindent where $\nu$ is the Flory exponent, assumed here to be equal to $0.588$. At small concentrations, the size of the dendrimer is smaller than $\xi$ (Regime-1). The correlation length decreases with concentration and beyond a certain concentration, the dendrimer is larger than $\xi$ (Regime-2). Fig.~\ref{fig:size_regimes} shows these regimes based on the relative size of dendrimers and the correlation length of the solution.  The lengthscales are normalized with the radius of gyration of linear chains in the dilute limit to show the collapse of dendrimers with the same $\chi$ value. 

\begin{figure*}[t]
	\begin{center}
		\begin{tabular}{c}
			\resizebox{7.5cm}{!} {\includegraphics[width=5cm,height=!]{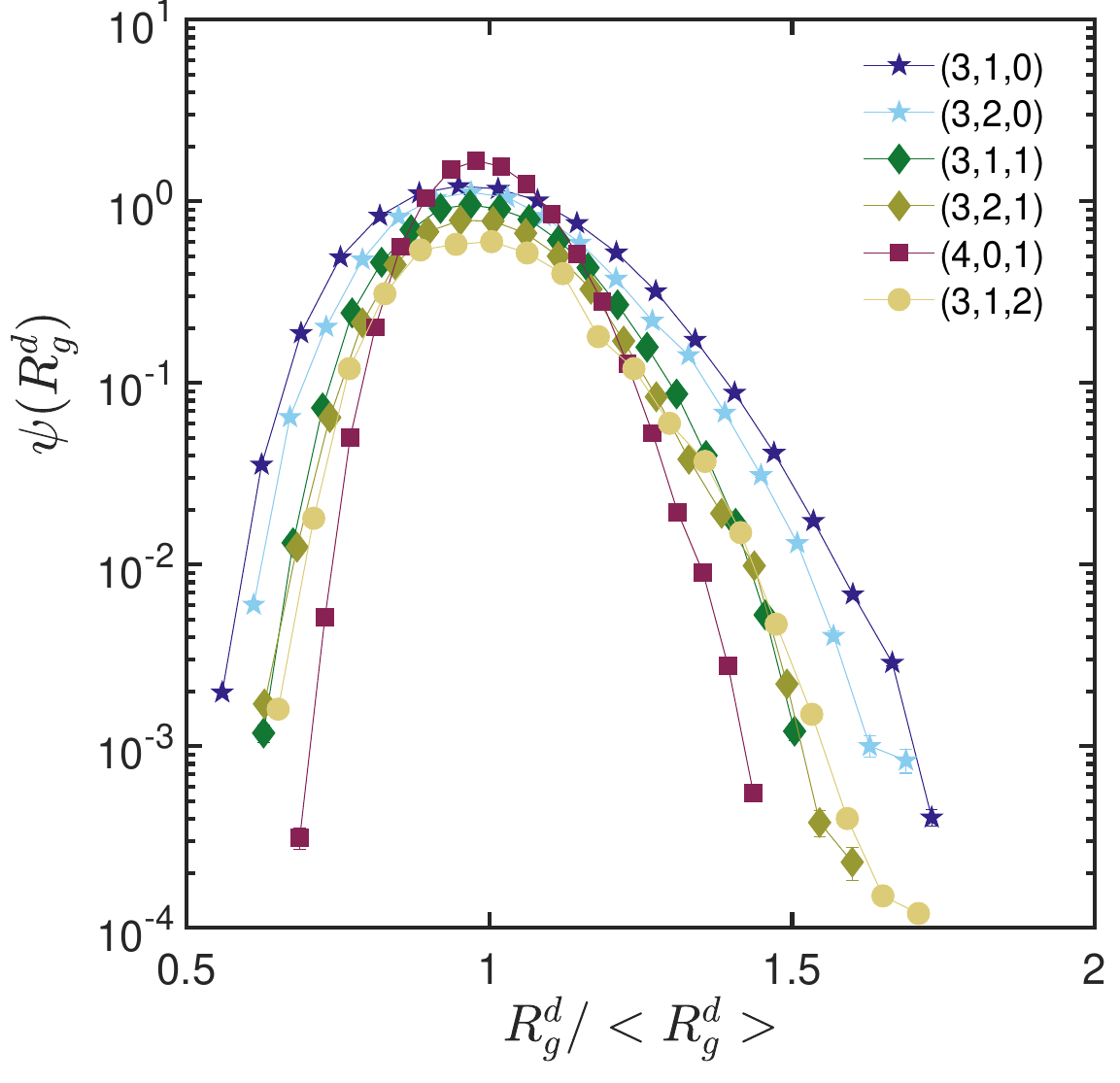}}  \\
                    	(a)   \\			
   			\resizebox{7.5cm}{!} {\includegraphics[width=5cm,height=!]{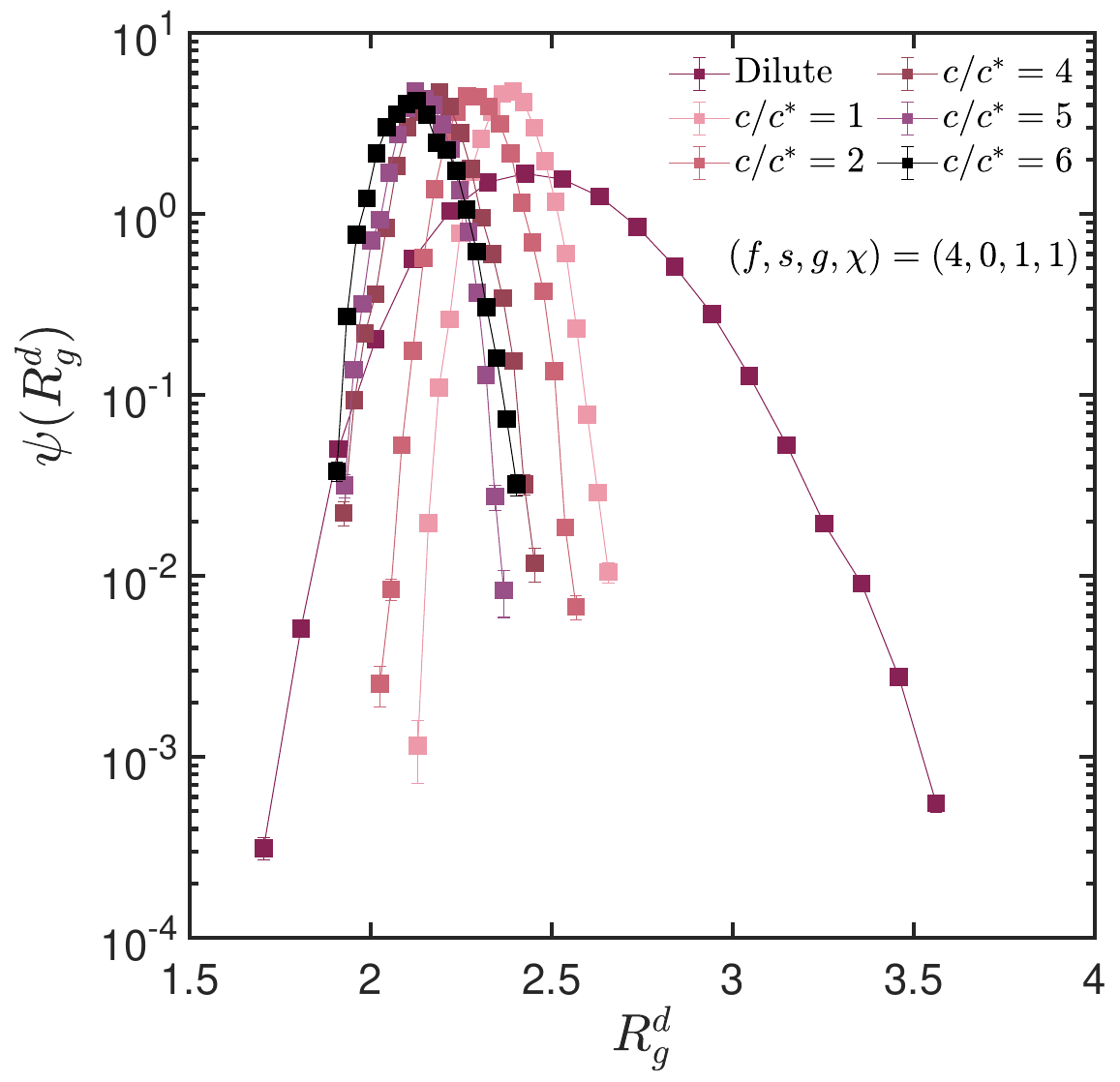}} \\	
			  (b) \\		 
			\resizebox{7.5cm}{!} {\includegraphics[width=5cm,height=!]{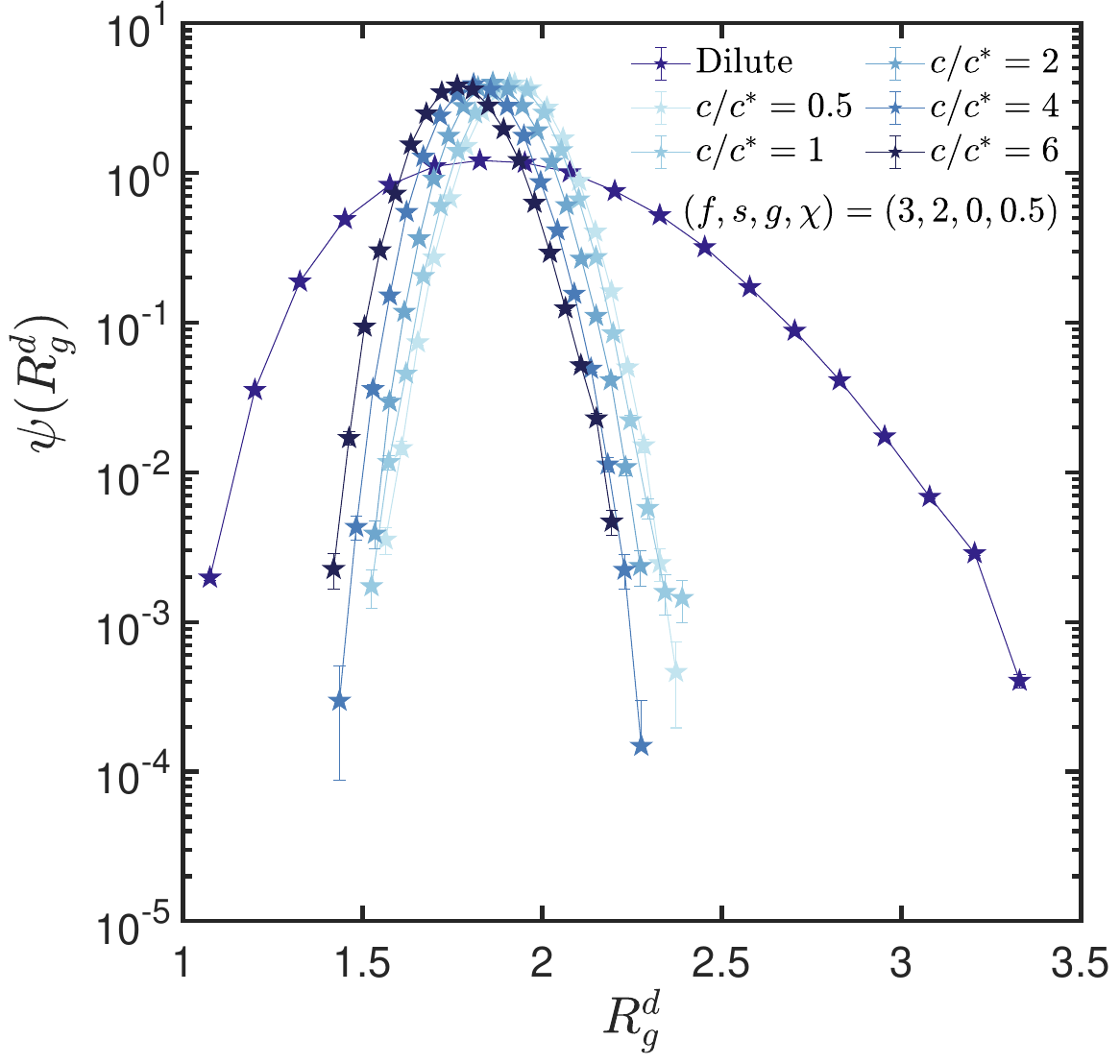}} \\
                   (c) \\
                \end{tabular}	
	\end{center}
	\vspace{-15pt}
	\caption{ (Color online) Distribution function of the radius of gyration of dendrimers (a) Dendrimers of different architectures in dilute solution. The dendrimer architecture is included in the order ($f,s,g)$. Note that the $x$-axis is the normalised radius of gyration. (b) Effect of concentration on the size fluctuation of $(f,s,g,\chi)=(4,0,1,1)$ dendrimers. (c) Effect of concentration on the size fluctuation of $(f,s,g,\chi)=(3,1,0,0.5)$ dendrimers.}
    \label{fig:rg_dist}
\end{figure*}

\section{Distribution of radius of gyration}

\begin{figure*}[t]
	\begin{center}
		\begin{tabular}{c}
			\resizebox{7cm}{!} {\includegraphics[width=7cm]{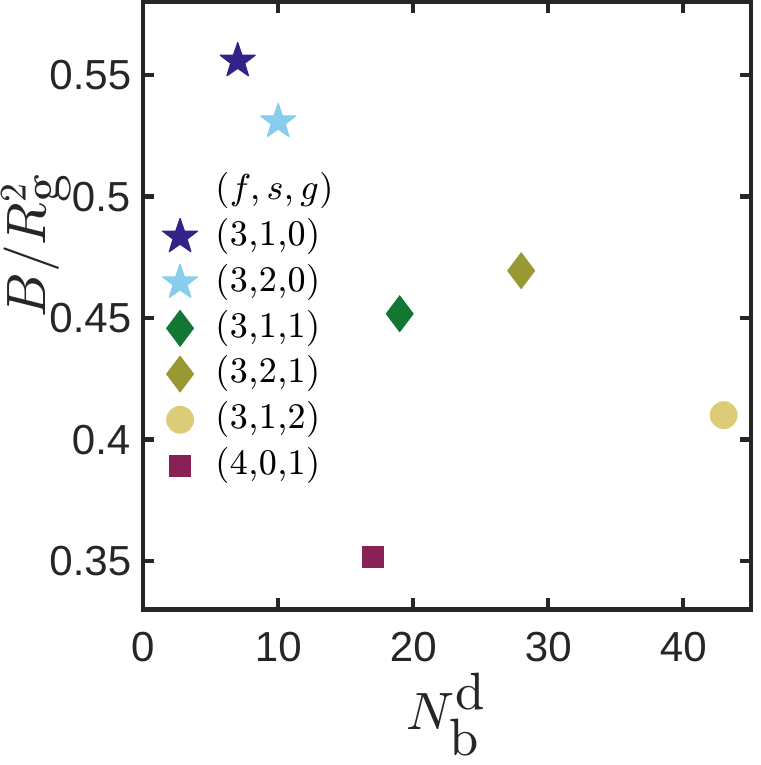}} \\
			(a) \\
			\resizebox{7cm}{!} {\includegraphics[width=7cm]{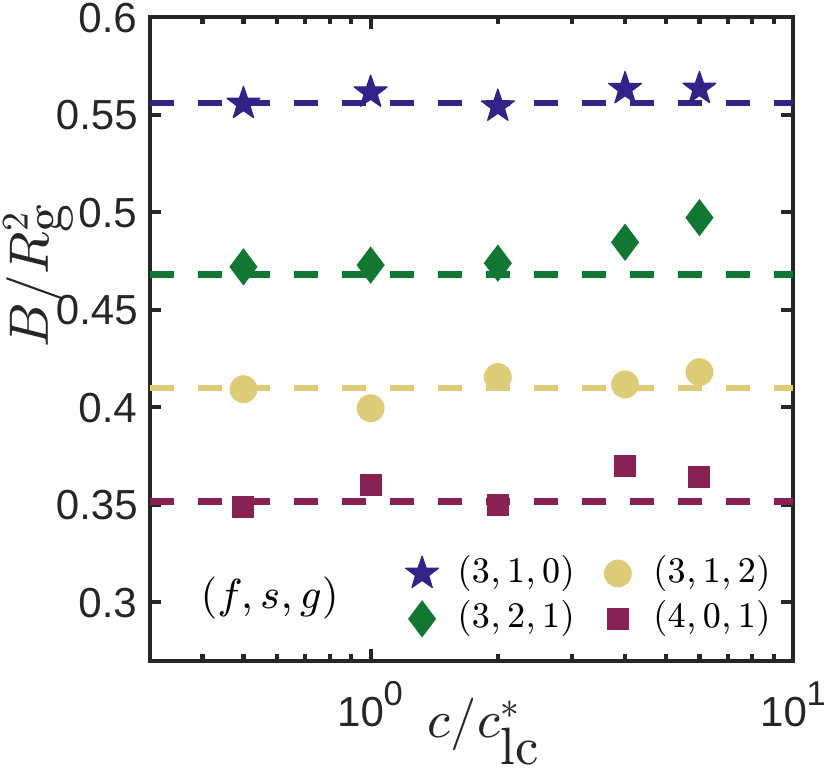}} \\
			(b) \\
			\resizebox{7cm}{!} {\includegraphics[width=7cm]{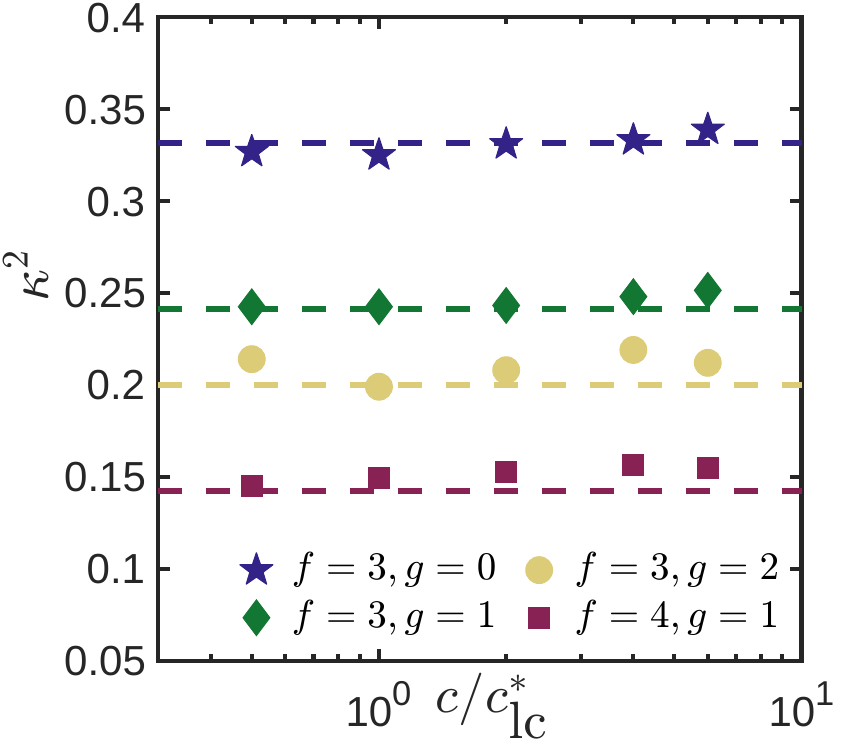}} \\
			 (c) \\
		\end{tabular}
	\end{center}	
	\vspace{-15pt}
	\caption{ (Color online) Effect of architecture and solution concentration on the shape of dendrimers. (a) Asphericity of dendrimers of different architectures in dilute solution as a function of the number of beads in a dendrimer molecule. The symbols represent dendrimer topology given by the combination $f, s, g$. (b) Effect of concentration of the solution on asphericity of dendrimers. All $f=3$ architectures considered have one spacer bead ($s=1$). The dashed lines are the asphericity in the dilute solution given in (a). (c) Effect of concentration of the solution on relative shape anisotropy of dendrimers. The dashed lines are the $\kappa^2$ in the dilute solution given in Fig.6(a) of the main text. }
    \label{fig:Asphericity}
\end{figure*}

Dendrimers are soft colloids and can have shape and size fluctuations based on the forces experienced by each of its constituent beads. This is evident from the distribution of its radius of gyration, $\psi(R_g^d)$. It is calculated by obtaining $R_g^d$ from many trajectories at different instances in time and binning them. The variance of the distribution of $R_g^d$ gives a measure of the size fluctuations. It is clear from Fig.~\ref{fig:rg_dist}(a) that increasing functionality reduces the softness of dendrimers resulting in lesser fluctuations. The $f=4$ dendrimer has a significant decrease in variance compared to the $f=3$ dendrimers, especially the $(f,s,g)=(3,1,1)$ that has a similar number of beads per molecule to that of the $f=4$ dendrimer. The fluctuations decrease at higher concentrations, but the distribution is similar at all concentrations irrespective of the architecture (shown in Fig.~\ref{fig:rg_dist}(b)).  

\section{Asphericity}

The asphericity $B$ of a molecule is defined as~\cite{theodorou1985shape,bishop1986polymer,kumari2020computing}:
\begin{align} \label{eq:Asphericity}
    B = \langle \lambda_3^2 \rangle - \dfrac{1}{2} \left[ \langle \lambda_1^2 \rangle + \langle \lambda_2^2 \rangle \right]
\end{align}
\noindent Molecules with tetrahedral or greater symmetry have $B=0$, otherwise $B>0$. In Fig.~\ref{fig:Asphericity}(a), the asphericity of dendrimers of different architectures in dilute solution, calculated using eqn~\eqref{eq:Asphericity}, is plotted as a function of the number of beads per molecule. In a dilute solution, an increase in generation number decreases $B$, implying a more spherically symmetric arrangement of beads within the molecule. The functionality four dendrimer has the lowest $B$ value. A similar trend is observed in the case of the relative shape anisotropy. The concentration of the solution seems to not affect the asphericity and shape anisotropy of dendrimer molecules as seen in Fig.~\ref{fig:Asphericity}(b) and (c).

\section{Bead density distribution}

\begin{figure*}[th]
	\begin{center}
		\begin{tabular}{ccc}
                \resizebox{5.5cm}{!} {\includegraphics[width=3cm, height=!]{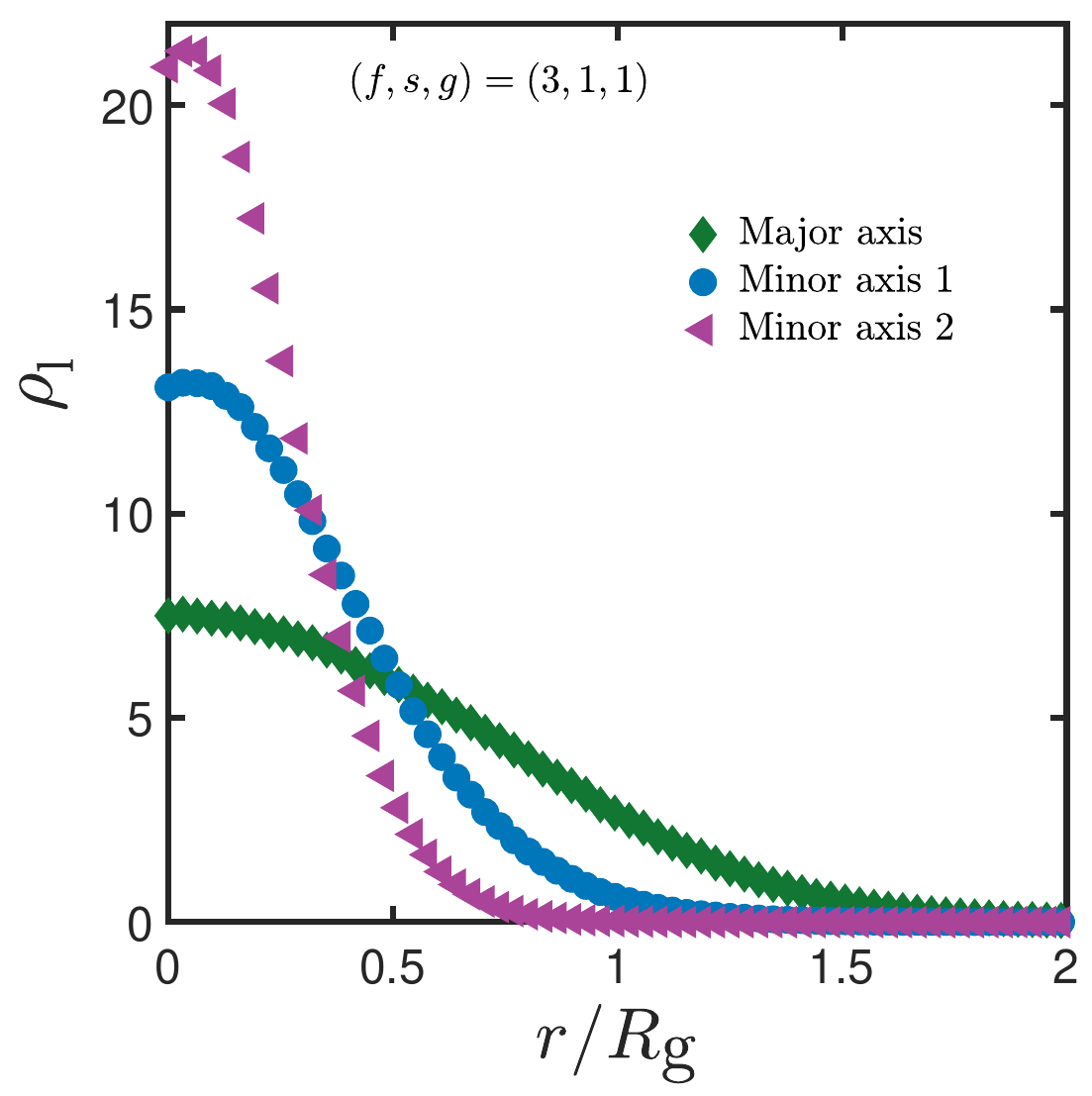}} &
              	\resizebox{5.5cm}{!} {\includegraphics[width=3cm, height=!]{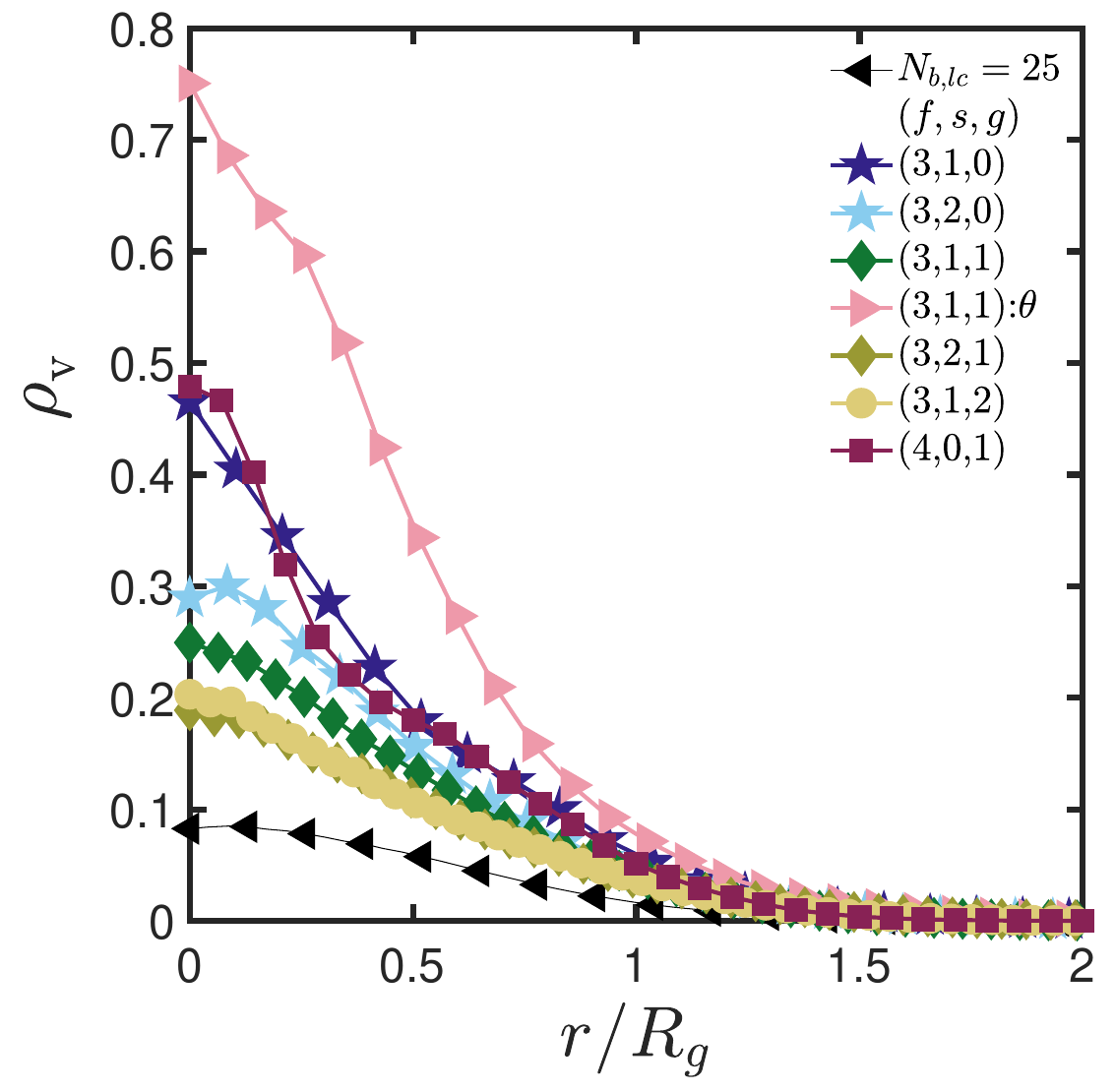}} &			
		\resizebox{5.5cm}{!} {\includegraphics[width=3cm, height=!]{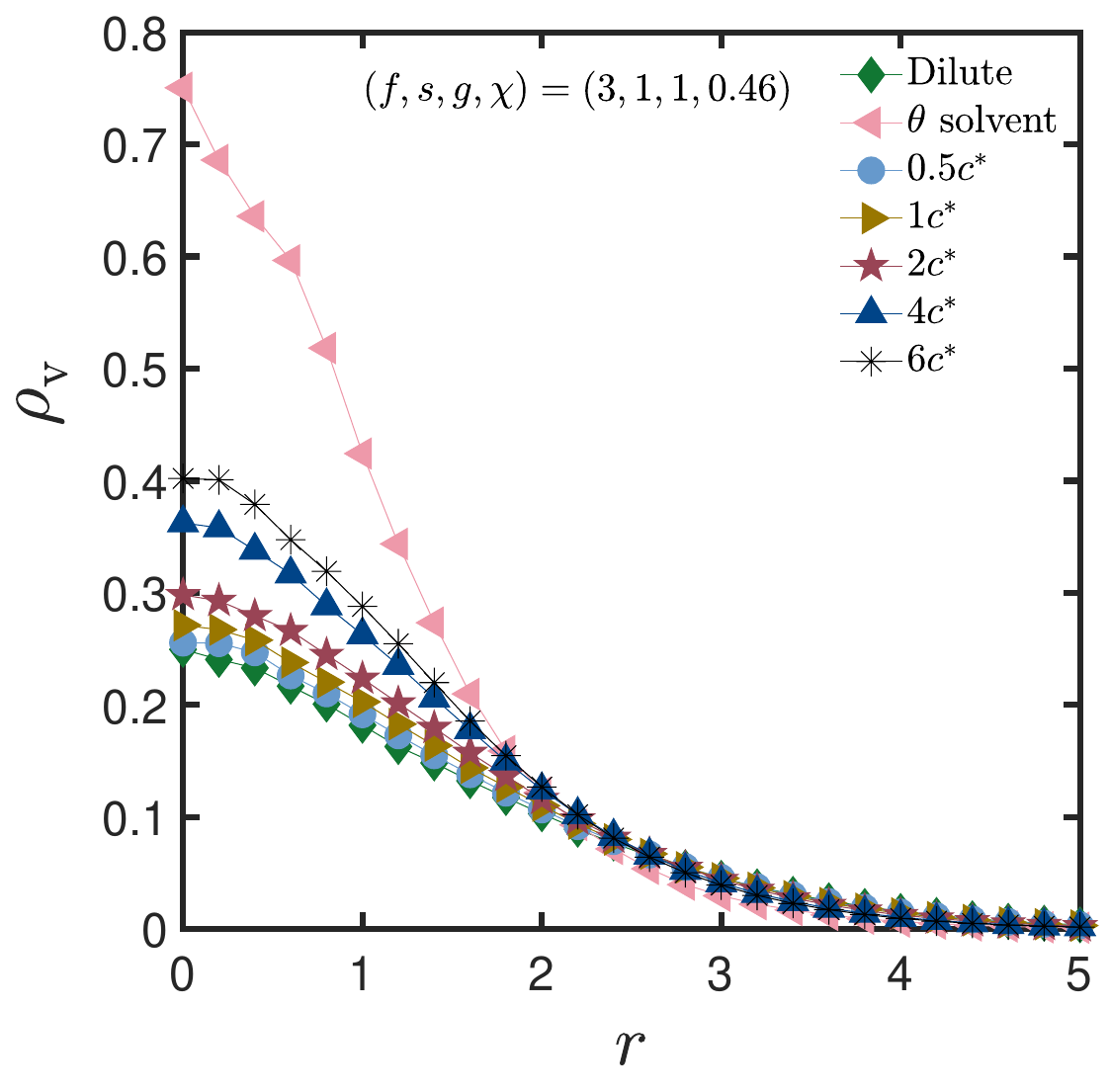}}\\
			(a) & (b)  & (c)
		\end{tabular}
	\end{center}	
	\vspace{-15pt}
        \caption{ (Color online) Bead density distribution (a) Linear bead density distribution along the 3 axes for a $f=3, s=1, g=1$ dendrimer in athermal solvent condition. (b)  Effect of architecture on the volumetric bead densities of dendrimers in dilute solution. Black triangles represent linear chain with 25 beads and dendrimers are represented using the combination $f, s, g$. (c) Effect of concentration on the volumetric bead density distribution of $f=3,s=1,g=1$ dendrimer in a semidilute solution of linear chains with $\chi=0.46$. Pink triangles in both figures represent $f=3,s=1,g=1$ dendrimer in theta solvent conditions. }
    \label{fig:density_dist}
\end{figure*}

The linear bead density distribution along the 3 axes of the gyration tensor of $(f,s,g)=(3,1,1)$ dendrimer in dilute solution is shown in Fig.~\ref{fig:density_dist}(a). The characteristic features of the distribution are observed along the major axis while it is Gaussian-like along the minor axes.

An alternative estimate of the internal density can also be obtained by counting the number of beads within concentric shells of constant thickness $\Delta r$ starting from the centre of mass of the molecule. The volumetric bead density is given by:
\begin{align}
    \rho_{\textrm{V}}(r)=\dfrac{n_{\textrm{b}}(r+\Delta r)-n_{\textrm{b}}(r)}{\Delta \textrm{V}}
\end{align}
where $n_{\textrm{b}}(r)$ is the number of beads of a molecule in a sphere of radius $r$, $\Delta \textrm{V}$ is the volume of a spherical shell of thickness $\Delta r$ given by $\dfrac{4}{3}\pi(r+ \Delta r)^3-\dfrac{4}{3}\pi r^3$. In Fig.~\ref{fig:density_dist}(b), the volumetric bead densities in concentric shells starting from the centre of the molecule have been plotted for dendrimers in dilute solution. The density is maximum at the centre for all dendrimer architectures, which drops monotonically towards the periphery of the molecule. It is important to note that the dendrimers with a higher number of beads have lower core volumetric density. This might be because of the higher effective repulsion the individual beads near the centre feel, causing them to occupy regions towards the periphery. The $f=3,s=1,g=1$ molecule in theta solvent has a significantly higher density at the core due to its collapsed state. In semidilute solution, the dendrimer volumetric bead density transitions from that in dilute solution to theta solvent conditions as solution concentration increases (shown in Fig.~\ref{fig:density_dist}(c)). This is consistent with the observations of linear bead density along the major axis and can be attributed to the onset of screening of EV at higher concentrations.

\begin{figure*}[th]
	\begin{center}
		\begin{tabular}{ccc}
                \resizebox{5.5cm}{!} {\includegraphics[width=3cm, height=!]{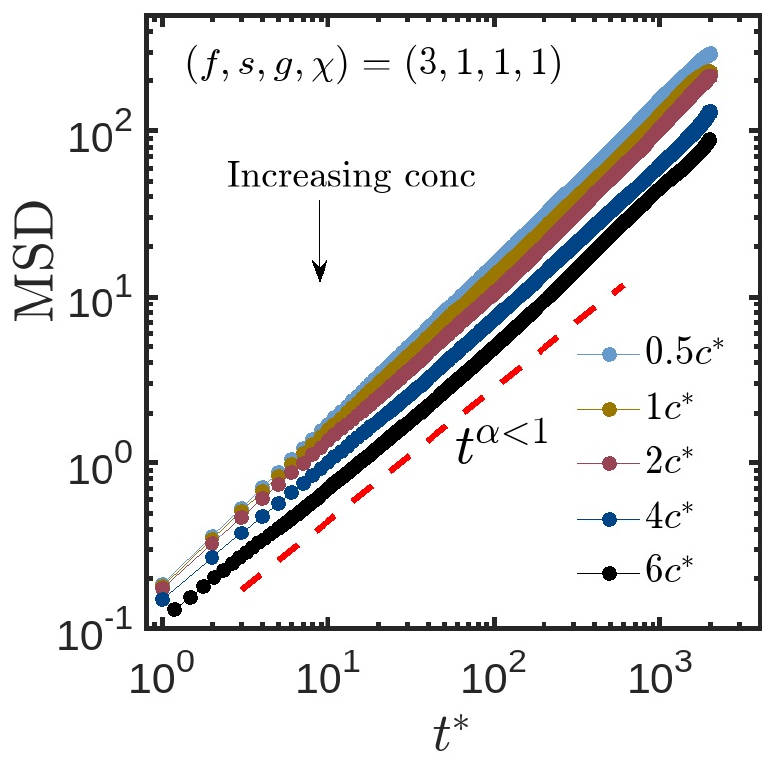}} &
              	\resizebox{5.45cm}{!} {\includegraphics[width=3cm, height=!]{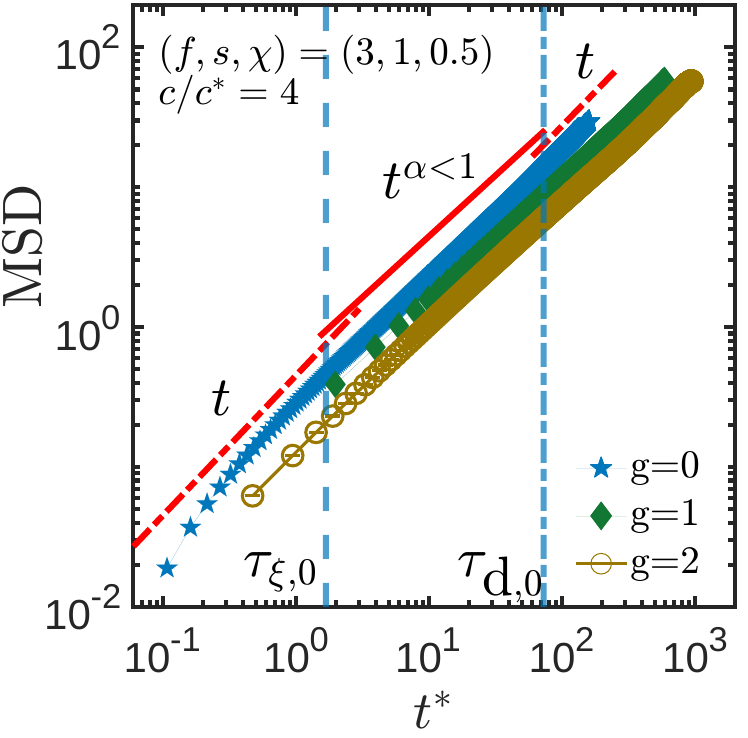}} &			
		\resizebox{5.5cm}{!} {\includegraphics[width=3cm, height=!]{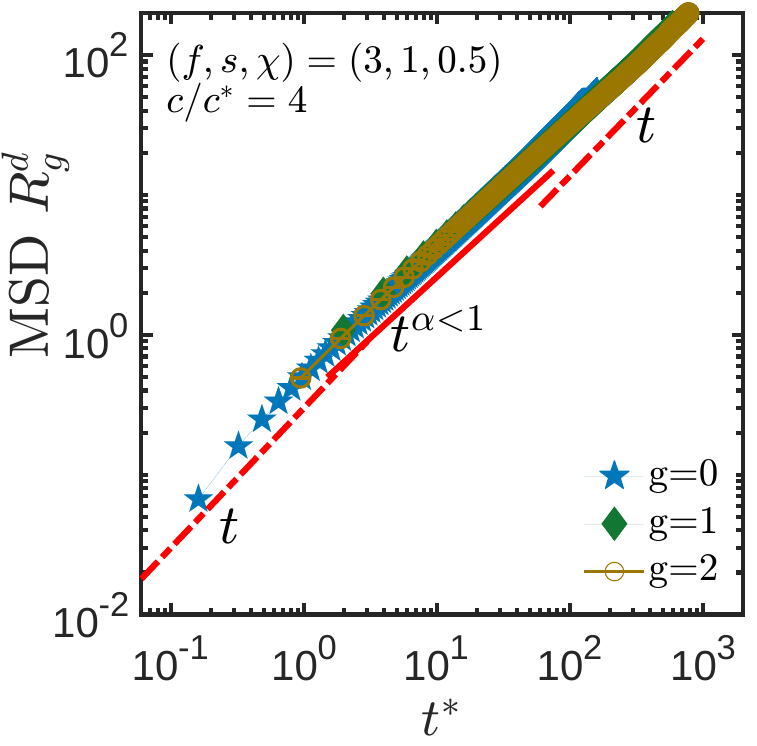}}\\
			(a) & (b)  & (c)
		\end{tabular}
	\end{center}	
	\vspace{-15pt}
	\caption{(Color online) Mean squared displacement of dendrimers as a function of time (a) Effect of solution concentration on the mean squared displacement of generation one ($g=1$) dendrimers with functionality $f=3$ and one spacer bead ($s=1$) in a solution of linear chains with 43 beads ($N_{\textrm{b}}^{\textrm{lc}}=43$). The size ratio between the dendrimers and linear chains is $0.46$ ($\chi=0.46$). (b) Effect of generation number of dendrimers on its mean squared displacement. The concentration is $4c^{\ast}$ and dendrimer parameters are $f=3,s=1,\chi=0.5$. The vertical dashed and dashed-dotted lines are $\tau_{\xi}$ and $\tau_{d}$ for the generation zero dendrimer respectively. (c) Collapse of mean squared displacement of dendrimers of different architectures considered in (b).}
\label{fig:MSD}
\end{figure*}

\section{Mean squared displacement}
The effect of solution concentration on the mean squared displacement of dendrimers is shown in Fig.~\ref{fig:MSD}(a). Subdiffusion is completely absent at low concentrations because the dendrimer size is much smaller than the solution correlation length. As concentration increases, due to the interaction between dendrimer molecules and the increased number of linear chains, a subdiffusive period exists ($\tau_{\xi} < t < \tau_d$) and the mean squared displacement of dendrimers decreases. Thus, with increased concentration, the dendrimers become less diffusive due to hindered motion.  The values of $\tau_{\xi}$ and $\tau_{\xi}$ changes with concentration as reported in Tables~\ref{tab:tau_xi},~\ref{tab:tau_d} and ~\ref{tab:f4}. Table~\ref{tab:tau_xi} contains the values of $\tau_{\xi}$ and Table~\ref{tab:tau_d} contains the values of $\tau_{d}$ for all $f=3$ dendrimers at all concentrations. Table~\ref{tab:f4} contains $\tau_{\xi}$ and $\tau_{d}$ for the $f=4,g=1$ dendrimer. When the size of the dendrimer is smaller than the correlation length ($\xi$) of the solution, it exhibits normal diffusion at all time scales and the concept of the time scales $\tau_{\xi}$ and $\tau_{d}$ does not exist.

\begin{table}[tbph]
\small
\caption{The time scale $\tau_{\xi}$ (given by eqn 17 in the main text) for the various architectures at different concentrations.} 
\vskip5pt
\label{tab:tau_xi}
\setlength{\tabcolsep}{12pt}
\renewcommand{\arraystretch}{1.25}
\begin{tabular*}{0.6\textwidth}{@{\extracolsep{\fill}}lllllll}
\hline
 \multirow{2}{*}{$(\chi,f,s,g)$}  &\multicolumn{5}{c}{$c/c^{\ast}$} \\\cline{2-6}
                                  & $0.5$ & $1$ & $2$  & $4$ & $6$      
\\                
\hline
\hline
            $(0.5,3,1,0)$
            & $-$
            & $-$
            & $8.4$  
            & $1.68$
            & $1.1$
\\
            $(0.5,3,2,0)$
            & $-$
            & $-$
            & $14.6$ 
            & $2.9$
            & $1.9$
\\
            $(0.46,3,1,1)$
            & $-$
            & $-$
            & $33.5$ 
            & $7.02$
            & $4.7$          
\\
            $(0.46,3,2,1)$
            & $-$
            & $-$
            & $65.1$ 
            & $13.62$
            & $9.1$          
\\
            $(0.5,3,1,2)$
            & $-$
            & $-$
            & $82.5$ 
            & $17.34$
            & $6.96$          
\\
            $(1.0,3,1,1)$
            & $-$
            & $18.3$
            & $3.96$ 
            & $0.96$
            & $0.6$          
\\
\hline
\end{tabular*}
\end{table}

\begin{table}[tbph]
\small
\caption{The time scale $\tau_{d}$ (defined by eqn 18 in the main text) for the various architectures at different concentrations.}
\label{tab:tau_d}
\vskip5pt
\setlength{\tabcolsep}{12pt}
\renewcommand{\arraystretch}{1.25}
\begin{tabular*}{0.6\textwidth}{@{\extracolsep{\fill}}lllllll}
\hline
 \multirow{2}{*}{$(\chi,f,s,g)$}  &\multicolumn{5}{c}{$c/c^{\ast}$} \\\cline{2-6}
                                  & $0.5$ & $1$ & $2$  & $4$ & $6$      
\\                
\hline
\hline
            $(0.5,3,1,0)$
            & $-$
            & $-$
            & $48.9$  
            & $72.9$
            & $93.1$
\\
            $(0.5,3,2,0)$
            & $-$
            & $-$
            & $92.1$ 
            & $136.3$
            & $175.6$
\\
            $(0.46,3,1,1)$
            & $-$
            & $-$
            & $168$ 
            & $248.6$
            & $302.4$          
\\
            $(0.46,3,2,1)$
            & $-$
            & $-$
            & $304.3$ 
            & $480.3$
            & $543.3$          
\\
            $(0.5,3,1,2)$
            & $-$
            & $-$
            & $556.8$ 
            & $864$
            & $1017.6$          
\\
            $(1.0,3,1,1)$
            & $-$
            & $245.7$
            & $369.6$ 
            & $517.4$
            & $676.8$          
\\
\hline
\end{tabular*}
\end{table}

\begin{table}[tbph]
\small
\caption{The time scales $\tau_{\xi}$ and $\tau_{d}$ (defined by eqns 17 and 18 in the main text) for the $\left( \chi,f,s,g \right) = \left( 1,4,0,1\right)$ dendrimer.}
\label{tab:f4}
\vskip5pt
\setlength{\tabcolsep}{12pt}
\renewcommand{\arraystretch}{1.25}
\begin{tabular*}{0.48\textwidth}{@{\extracolsep{\fill}}lll} 
\hline
        $c/c^{\ast}$
        & $\tau_{\xi}$
        & $\tau_{d}$
\\                
\hline
\hline
            $0.5$
            & $-$
            & $-$ 

\\
            $1.0$
            & $9.2$
            & $124.8$
\\
            $2.0$
            & $1.9$
            & $181.4$          
\\
            $4.0$
            & $0.6$
            & $247.6$        
\\
            $5.0$
            & $0.4$
            & $273.6$          
\\
            $6.0$
            & $0.3$
            & $305.2$ 
\\
            $6.5$
            & $0.2$
            & $313.9$ 
\\
\hline
\end{tabular*}
\end{table}

The size of dendrimers varies with their generation number. Therefore, at a constant concentration, $f$ and $s$, a generation 2 dendrimer will experience more hindrance to its motion compared to a simple star polymer ($g=0$), causing the former to diffuse slower than the latter as shown in Fig.~\ref{fig:MSD}(b). The mean squared displacement of dendrimers is given by $\textrm{MSD}=6 \,D \,t$, where $D$ is the diffusivity. The Stokes-Einstein relation for a particle of radius $R$ is $D=k_BT/6\pi \eta R$. Therefore, $\textrm{MSD} \approx k_BTt/\eta R$, and $\textrm{MSD} \, R \approx k_BTt/\eta$. Using this simple scaling argument, we have collapsed $\textrm{MSD}$ data for dendrimers of different generations as shown in Fig.~\ref{fig:MSD}(c).
	
\begin{figure}[t]
	\begin{center}
			\resizebox{9.0cm}{!}{\includegraphics[width=0.5\linewidth,height=!]{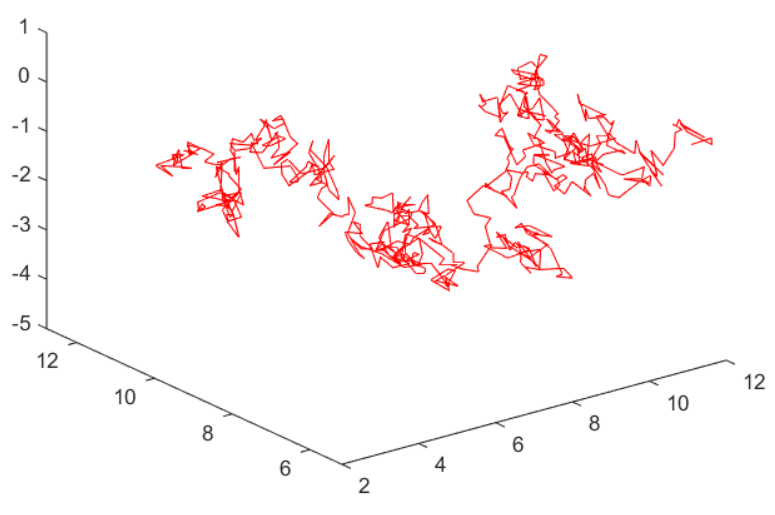}} 		
	\end{center}
	\vspace{-20pt}
	\caption{(Color online) 3D plot of the movement of the centre of mass of the dendrimer. It is a $(f,s,g,\chi)=(3,1,1,1)$ dendrimer molecule at $c/c^{\ast}=6$ in the presence of hydrodynamic interactions.}
\label{fig:cm_3D}
\end{figure}

\begin{figure*}[b]
	\begin{center}
		\begin{tabular}{cc}
			\resizebox{7cm}{!} {\includegraphics[width=0.3\textwidth]{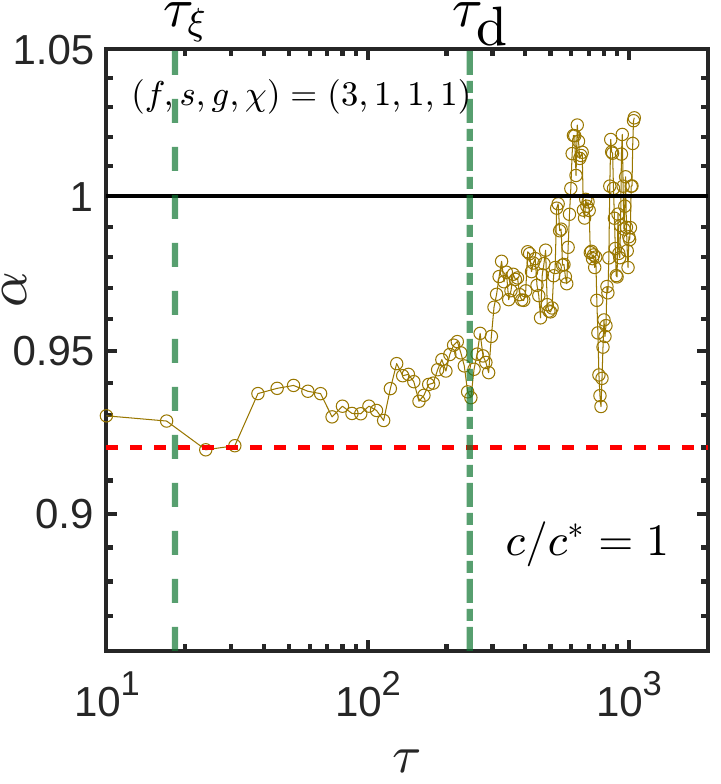}} &
			\resizebox{7cm}{!} {\includegraphics[width=0.3\textwidth]{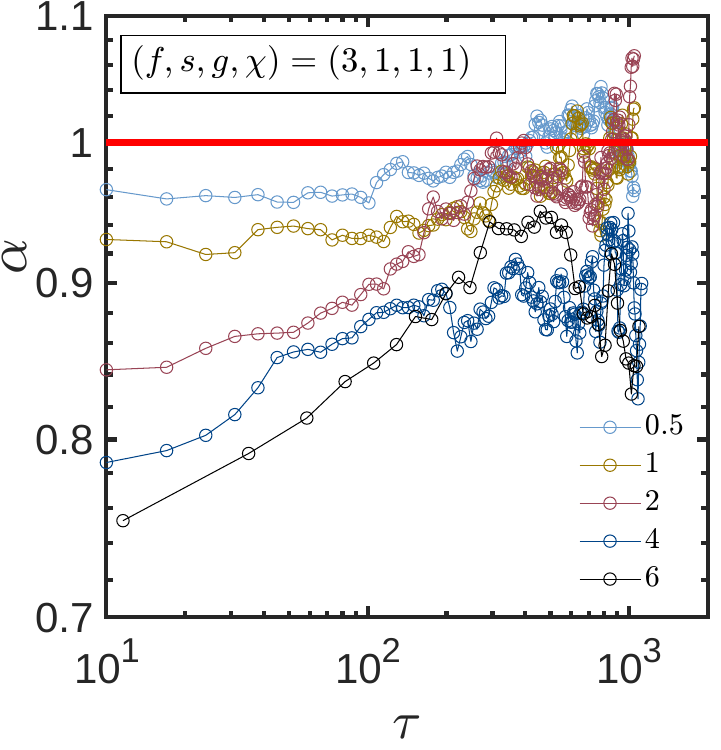}} \\               
			(a) & (b)
		\end{tabular}
	\end{center}
	\vspace{-15pt}
        \caption{ (Color online) Diffusion exponent of $(f,s,g,\chi)=(3,1,1,1)$ dendrimer as a function of time (a) The concentration considered is $c/c^{\ast}=1$. Vertical lines represent $\tau_{\xi}$ (dashed) and $\tau_d$ (dashed-dotted) for $c/c^{\ast}=1$. The horizontal dashed line is the value obtained from mean squared displacement versus time plots and the solid line is unity. (b) Diffusion exponent for different concentrations.}
    \label{fig:alpha_time}
\end{figure*}

An examination of the position of the centre of mass of a dendrimer molecule in our simulations as a function of time shows that it does not have long waiting times or hopping motion (Fig.~\ref{fig:cm_3D}). Rather the molecule moves in space via smooth random movements even at high concentrations.

\section{Diffusion exponent as a function of time}

The subdiffusion exponent can also be obtained by taking instantaneous derivatives in the mean squared displacement as given below:
\begin{align}\label{eq:alpha_time}
    \alpha= \frac{d \log (\textrm{MSD}(\tau))}{d \log \tau}
\end{align}

Fig.~\ref{fig:alpha_time}(a) shows the diffusion exponents of $(f,s,g,\chi)=(3,1,1,1)$ dendrimer as a function of time at $c/c^{\ast}=1$. It is clear that the dendrimer becomes subdiffusive when $\tau_{\xi} < \tau < \tau_d$ with the value of $\alpha$ in the subdiffusive regime similar to that obtained from the time exponent in the mean squared displacement plots. It transitions to a diffusive regime beyond $\tau_d$. Similar behaviour is observed for all concentrations as shown in Fig.~\ref{fig:alpha_time}(b).

\begin{figure*}[tbph]
	\begin{center}
		\begin{tabular}{ccc}
		    \hspace{-0.5cm}
			\resizebox{8.75cm}{!} {\includegraphics[width=0.28\linewidth]{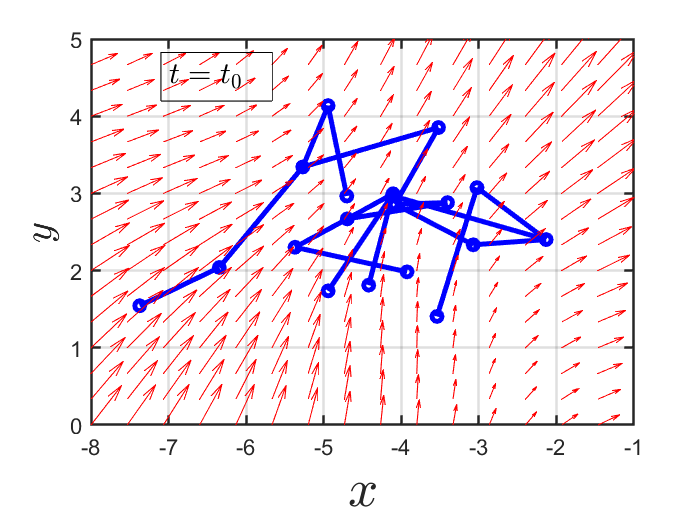}} &
			\hspace{-0.5cm}
			\resizebox{8.5cm}{!} {\includegraphics[width=0.28\linewidth]{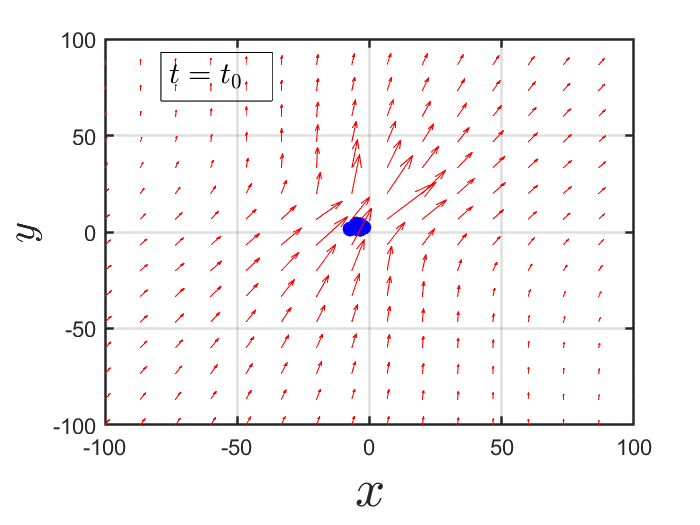}}\\
			(a) & (b)  \\
			\hspace{-0.5cm}
			\resizebox{8.5cm}{!} {\includegraphics[width=0.28\linewidth]{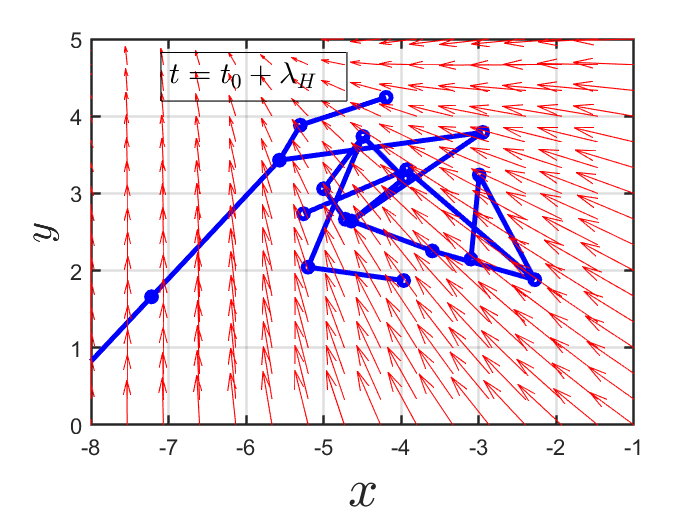}} &
            \hspace{-0.5cm}
			\resizebox{8.5cm}{!} {\includegraphics[width=0.28\linewidth]{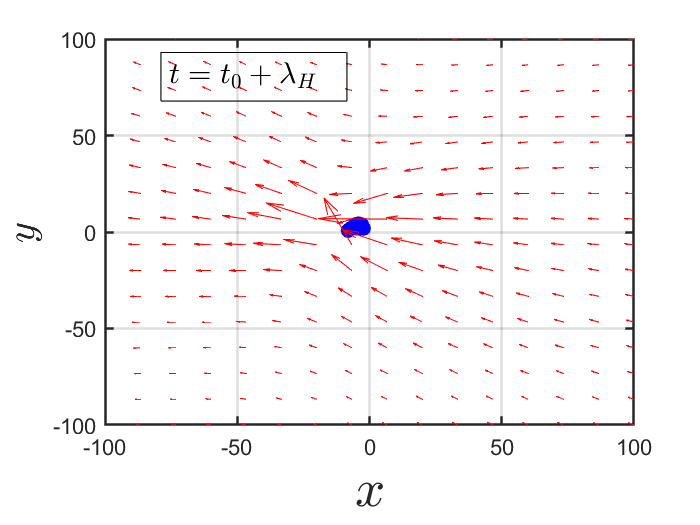}}\\
			(c) & (d)  \\
                \hspace{-0.5cm}
			\resizebox{8.5cm}{!} {\includegraphics[width=0.28\linewidth]{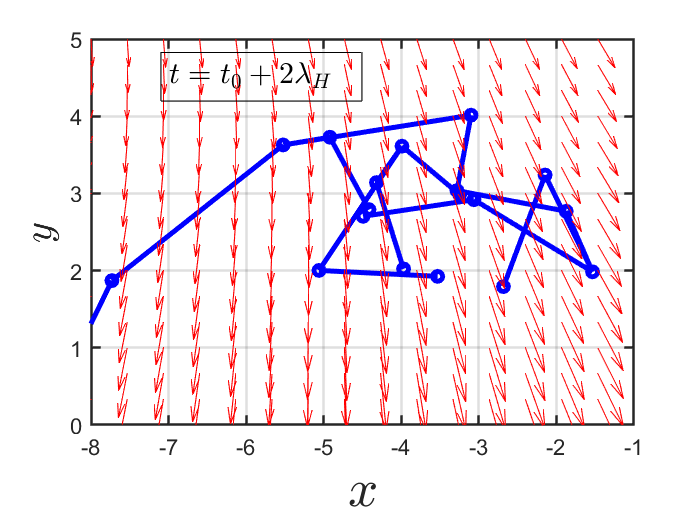}} &
            \hspace{-0.5cm}
			\resizebox{8.5cm}{!} {\includegraphics[width=0.28\linewidth]{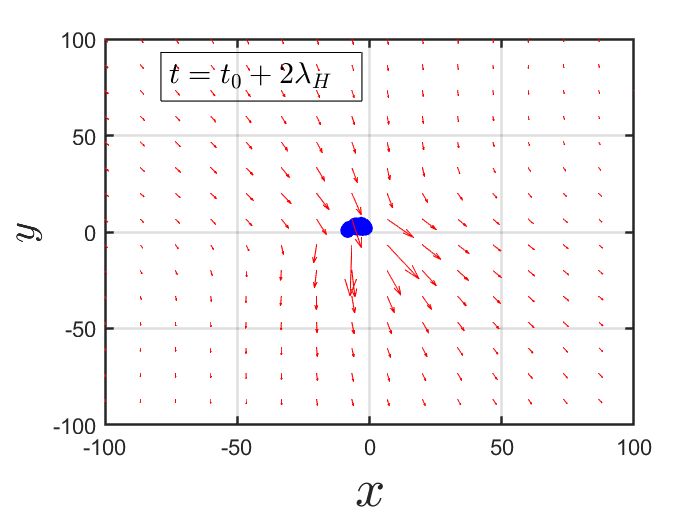}}\\
			(e) & (f)  \\
		\end{tabular}
	\end{center}
	\caption{ (Color online) The flow field about a $(f,s,g,\chi)=(3,1,1,1)$ dendrimer molecule at $c/c^{\ast}=6$ in the presence of hydrodynamic interactions. The rows in the panel represent different instances in time. The first column (Fig (a), (c) and (e)) shows the velocity field about a dendrimer at a length scale of the order of the size of the dendrimer. The grid size is equal to the correlation length of the solution. The second column (Fig (b), (d) and (f)) shows the velocity field at a length scale much larger than the dendrimer size. The grid size is $\approx 50 \xi$.}
    \label{fig:flow_field}
\end{figure*}

\begin{figure*}[t]
	\begin{center}
		\begin{tabular}{ccc}
			\resizebox{5.5cm}{!} {\includegraphics[width=0.27\textwidth]{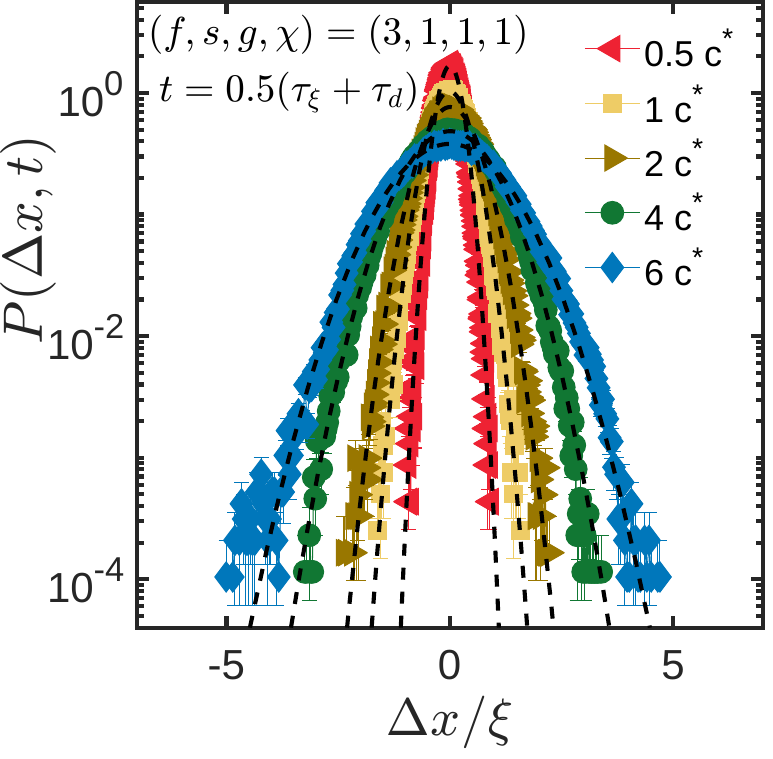}} &
			\resizebox{5.5cm}{!} {\includegraphics[width=0.27\textwidth]{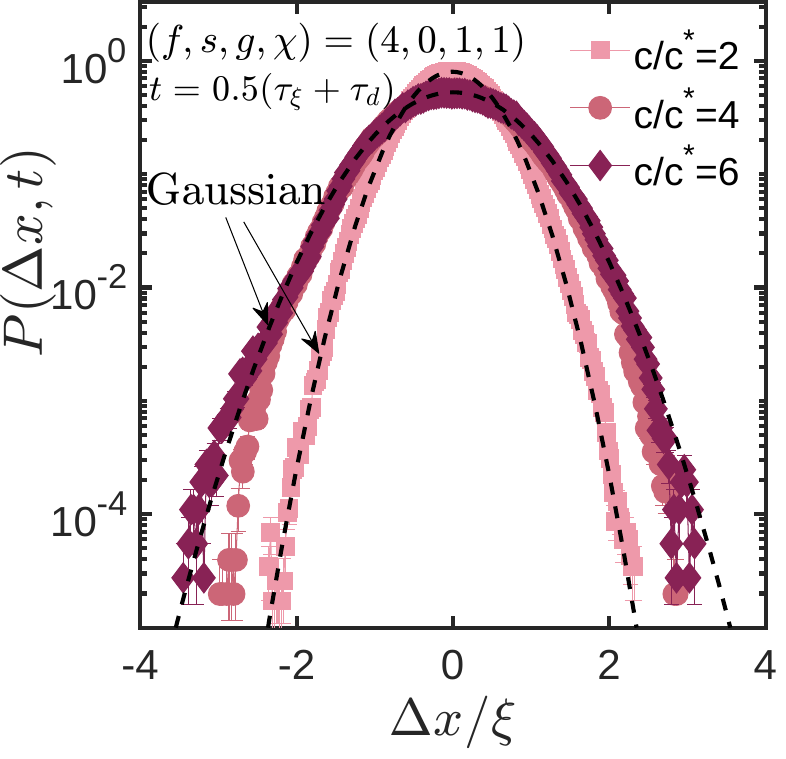}} &
			\resizebox{5.5cm}{!} {\includegraphics[width=0.27\textwidth]{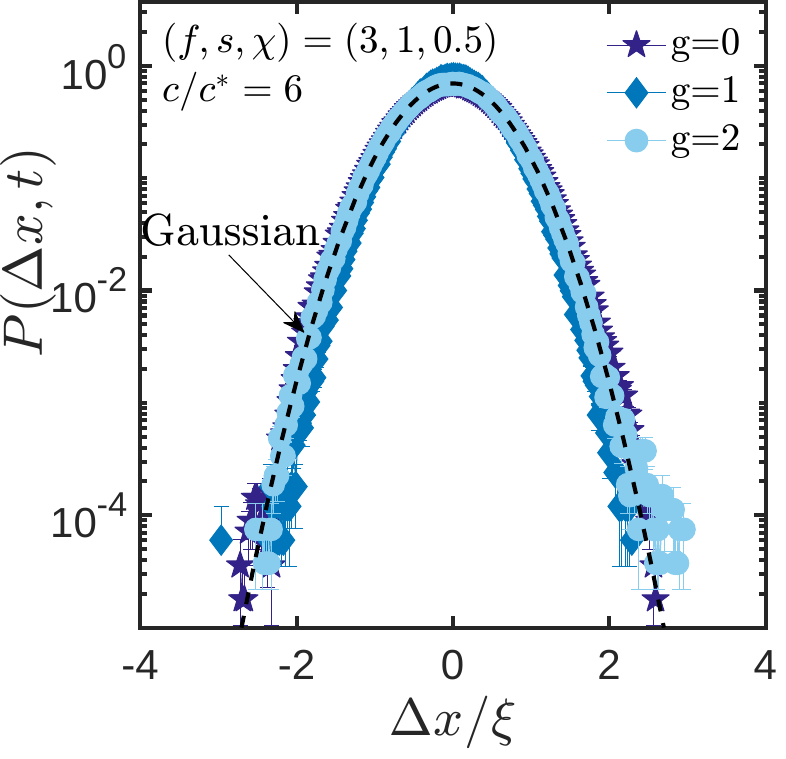}} \\                
			(a) & (b) & (c)
		\end{tabular}
	\end{center}
	\vspace{-15pt}
        \caption{ (Color online) Probability distribution function of displacement for dendrimers. (a) Generation one dendrimer ($f=3, s=1, \chi=1$) in a semidilute solution of different concentrations in the subdiffusive regime $\tau_{\xi} > t > \tau_d$ ( at $t = 0.5 (\tau_{\xi}+\tau_d)$). (b) PDD for $f=4$ dendrimer ($s=0,g=1,\chi=1$) in the subdiffusive regime at different concentrations ( at $t = 0.5 (\tau_{\xi}+\tau_d)$). (c) Functionality three dendrimers of different generations in a semidilute solution of concentration $6c^{\ast}$ in the subdiffusive regime. The dashed lines in all figures are Gaussian fits to the data.}
    \label{fig:PDD}
\end{figure*}

\section{Velocity field due to hydrodynamic interactions}
As pointed out in the manuscript, the subdiffusive motion of dendrimers is observed at intermediate times ($\tau_{\xi} < t < \tau_d$) which corresponds to the mean squared displacement of the dendrimer in the range $\xi < \textrm{MSD} < 2R_g^d$. This suggests that on such time scales, the distance covered by a dendrimer is of the order of the cage size formed by polymer strands in the solution. To examine this closely, we have now computed the velocity fields within the solution due to hydrodynamic interactions as shown in Fig.~\ref{fig:flow_field}. This velocity perturbation is obtained by calculating the product of the RPY tensor and the force exerted by each monomer at a point. Fig.~\ref{fig:flow_field}~(a), (c) and (e) show the velocity field at lengthscales comparable to dendrimer size at different instances in time while Fig.~\ref{fig:flow_field}~(b), (d) and (f) show the fields at large length scales corresponding to them. This enables us to see the effect of HI which we have observed in $\textrm{MSD}$ plots. As is clear in the zoomed-in figures, each monomer on the dendrimer feels forces of different magnitudes and directions due to its location relative to the backflow. The net effect would be to randomize the direction of motion on small length and time scales, leading to slower diffusion. At longer times, ($t > \tau_d$), the mean squared displacement of the dendrimer is much larger than its size. When observed from such large lengthscales, the entire molecule is like a single particle placed at its centre of mass with the orientation of individual beads being unimportant. The flow field set up due to HI seems to be in a particular direction and we would expect it to enhance diffusion. This explains the effect of HI at intermediate and long timescales.

\section{Probability distribution functions}

The distribution of a $f=3,s=1,g=1$ at different concentrations (Fig.~\ref{fig:PDD}(a)) and dendrimers of different architectures (Fig.~\ref{fig:PDD}(c)) in the subdiffusive regime were also found to be Gaussian. Even the special case of a dendrimer with functionality $f=4$, which is a denser molecule, does not exhibit a non-Gaussian behaviour at any of the concentrations in the subdiffusive regime (shown in Fig.~\ref{fig:PDD}(b)).

 \begin{figure*}[tbph]
	\begin{center}
		\begin{tabular}{ccc}
		    \hspace{-0.5cm}
			\resizebox{7.45cm}{!} {\includegraphics[width=0.28\textwidth]{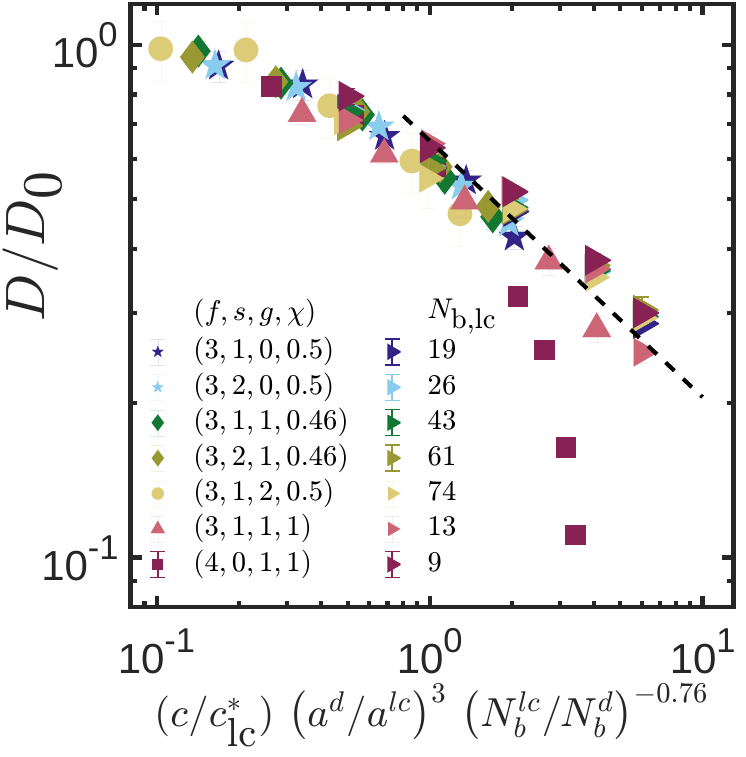}} &
			\resizebox{7.35cm}{!} {\includegraphics[width=0.28\textwidth]{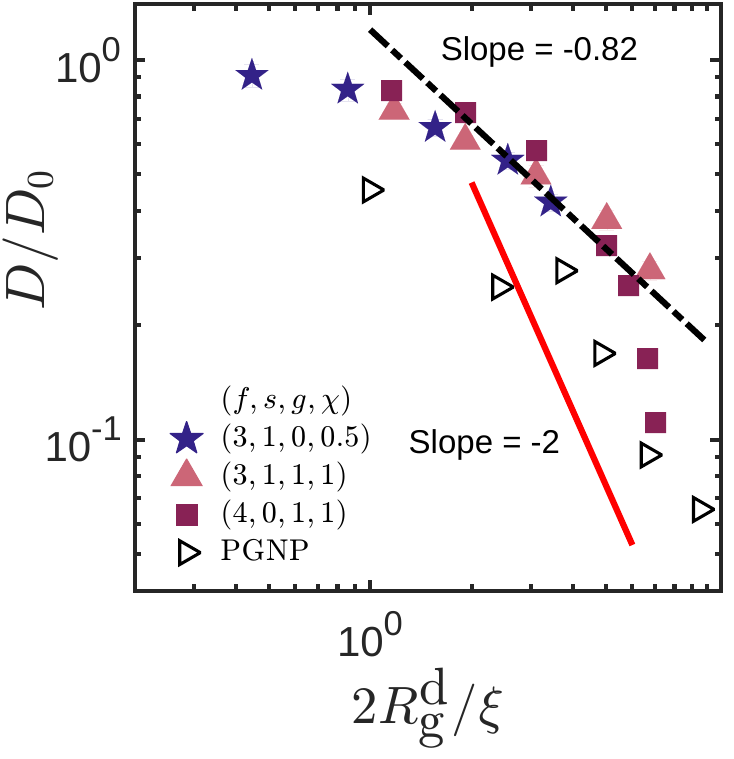}} \\   
			(a) & (b) \\
		\end{tabular}
	\end{center}
	\caption{ (Color online) Effect of solution concentration and size of tracer on its diffusivity. (a)Normalised diffusivity of dendrimers as a function scaled $c/c^{\ast}_{\textrm{lc}}$. For comparison, data for linear chains is also included with $c/c^{\ast}_{\textrm{lc}}$ in the x-axis. (b) Normalised diffusivity for dendrimers and PGNPs as a function of its size relative to the solution correlation length. The dashed line is the proposed scaling law for dendrimers and the solid line is the predictions of scaling theory.}
    \label{fig:hairy_colloids}
\end{figure*}

\section{Comparison with diffusivity of other soft colloids}

 We compared the dynamics of polymer grafted nanoparticles (PGNP), also called hairy colloids, in a solution of free polymer from work done by \citet{poling2019soft} with our simulations. According to them, the normalised diffusivity for various PGNP-free polymer systems can be collapsed if the normalised concentration ($c/c^{\ast}$) of free polymers is scaled with the ratio $\left( M_{w,f}/M_{w,g}\right)^{-1/8}$. In our system, the radius of gyration of dendrimers and linear chains in dilute solution are related by the following equation:

\begin{align}
    \chi=\dfrac{R_{\textrm{g0}}^{\textrm{d}}}{R_{\textrm{g0}}^{\textrm{lc}}}
    \label{eq:sr}
\end{align}

\noindent If $\beta = N_b^{\textrm{d}}/N_b^{\textrm{lc}}$, then using eqns \eqref{eq:Rg0_d} and \eqref{eq:Rg0_lc}, 
\begin{align}
    \chi=\dfrac{a^{\textrm{d}}}{a^{\textrm{lc}}} \beta^{\nu}
    \label{eq:sr_beta}
\end{align}

\noindent The overlap concentration of dendrimers is  
\begin{align}
    c^{\ast}_{\textrm{d}}  &= \dfrac{N_{\textrm{b}}^{\textrm{d}}}{\left( 4/3 \right)\pi (R_{\textrm{g0}}^{\textrm{d}})^3} \label{eq:cstar_d}
\end{align}

\noindent Substituting eqns~\eqref{eq:sr} and \eqref{eq:sr_beta} in \ref{eq:cstar_d} gives

\begin{align}
    c^{\ast}_{\textrm{d}} &= c^{\ast}_{\textrm{lc}} \left( \frac{N_b^{\textrm{d}}}{N_b^{\textrm{lc}}}\right)^{1-3\nu} \left( \frac{a^{lc}}{a^d} \right)^3
    \label{eq:new_cstard}
\end{align}
 
 Using eqn~\eqref{eq:new_cstard}, we have shown that dendrimer diffusivity for all architectures can be collapsed if we use $c/c^{\ast}_{\textrm{lc}}$ scaled by $\left ( N_{\textrm{b}}^{\textrm{lc}}/N_{\textrm{b}}^{\textrm{d}} \right )^{-0.76}$ instead of $c/c^{\ast}_{\textrm{d}}$  (Fig.\ref{fig:hairy_colloids}(a)). Fig .~\ref{fig:hairy_colloids}(b) shows the normalised diffusivity as a function of the size of the dendrimer or PGNP ($ M_{w,f}=15000$kDa, $ M_{w,g}=355$ kDa)\citep{poling2019soft} relative to the system correlation length. Clearly, PGNP does not follow the same scaling behaviour as our simulated dendrimers. Rather they follow the nanoparticle scaling at intermediate values of $2R/\xi$. PGNP can reduce its size with increasing concentration similar to dendrimers. However, due to the presence of the nanoparticle which occupies almost $25\%$ of the internal space, there is a limit beyond which it cannot shrink. Therefore, PGNP becomes nanoparticle-like at higher concentrations. Dendrimers reduce size with concentration and even start showing screening of excluded volume interactions as seen in the internal bead distribution plots.

\section{Scaling law for dendrimer diffusivity}

Also, the correlation length ($\xi$) of the solution, radius of gyration of dendrimers ($R_{\textrm{g}}^{\textrm{d}}$) and linear chains ($R_{\textrm{g}}^{\textrm{lc}}$) in semidilute solutions depends on concentration as follows:
\begin{align}\label{eq:xi_scaling}
    \xi= R_{\textrm{g0}}^{\textrm{lc}} \left( \dfrac{c}{c^{\ast}_{\textrm{lc}}}\right)^{\mu}
\end{align}
where $\mu=\left( -\nu \right)/\left( 3\nu -1 \right)$.  
\begin{align}\label{eq:rg_scaling}
    R_{\textrm{g}}^{\textrm{d}} = R_{\textrm{g0}}^{\textrm{d}}\left(\frac{c}{c^{\ast}_{\textrm{d}}}\right)^{\delta}
\end{align}
where $\delta=\left({1-2\nu}\right)/\left( 2\left(3\nu -1 \right) \right)$.
\begin{align}\label{eq:D_scaling}
    D^{\textrm{d}} = D_{\textrm{0}}^{\textrm{d}} \left( \dfrac{c}{c^{\ast}_{\textrm{d}}}\right) ^{\omega}
\end{align}
where $\omega=\left( \nu-1 \right)/ \left( 3\nu -1 \right)$.
\noindent Dividing eqn ~\eqref{eq:rg_scaling} by eqn ~\eqref{eq:xi_scaling},
\begin{align}\label{eq:rgbyxi}
    \dfrac{2R_{\textrm{g}}^{\textrm{d}}}{\xi} = \dfrac{2R_{\textrm{g0}}^{\textrm{d}}}{R_{\textrm{g0}}^{\textrm{lc}}}\left(\dfrac{c}{c^{\ast}_{\textrm{d}}}\right)^{\delta}  \left( \dfrac{c}{c^{\ast}_{\textrm{lc}}}\right)^{-\mu}
\end{align}
\noindent We also have the overlap concentration of linear chains given by:
\begin{align}
    c^{\ast}_{\textrm{lc}}  &= \dfrac{N_{\textrm{b}}^{\textrm{lc}}}{\left( 4/3 \right)\pi (R_{\textrm{g0}}^{\textrm{lc}})^3} \label{eq:cstar_lc}
\end{align}
Taking the ratio of equations~\eqref{eq:cstar_d} and ~\eqref{eq:cstar_lc} and substituting the values of $\chi$ and $\beta$ give:
\begin{align}
    c^{\ast}_{\textrm{lc}} &= \left( \chi^3 / \beta \right) c^{\ast}_{\textrm{d}}   \label{eq:cstar_rel}
\end{align}
Therefore,
\begin{align}
    \dfrac{2R_{\textrm{g}}^{\textrm{d}}}{\xi} = \dfrac{2\chi^{\left(1+3\mu\right)}}{\beta^{\mu} } \left(\frac{c}{c^{\ast}_{\textrm{d}}}\right)^{\delta-\mu} \label{eq:rg_by_xi} \\ \nonumber \\
    \frac{c}{c^{\ast}_{\textrm{d}}} = \left[ \left(\dfrac{2R_{\textrm{g}}^{\textrm{d}}}{\xi} \right) \left(\dfrac{\beta^{\mu} }{2\chi^{\left(1+3\mu\right)}}\right) \right] ^ { 1 / \left(\delta-\mu \right)} \label{eq:cbycstd_rg}
\end{align}

From equation~\ref{eq:D_scaling},
\begin{align}\label{eq:cbycst_D}
    \dfrac{c}{c^{\ast}_{\textrm{d}}} &= \left[ \dfrac{D^{\textrm{d}}} {D_{\textrm{0}}^{\textrm{d}}} \right] ^{\left( 1/\omega \right)}
\end{align}

Equating equation~\ref{eq:cbycstd_rg} and~\ref{eq:cbycst_D}, we get
\begin{align}\label{eq:den_scaling}
    \dfrac{D^{\textrm{d}}} {D_{\textrm{0}}^{\textrm{d}}} &= \gamma \left( \dfrac{2R_{\textrm{g}}^{\textrm{d}}}{\xi} \right) ^ { \omega/\left( \delta-\mu \right)}
\end{align}
where $\gamma= \left(\dfrac{\beta^{\mu} }{2\chi^{\left(1+3\mu\right)}}\right) ^ {\omega/\left( \delta-\mu \right)}$ and $\dfrac{\omega}{\delta-\mu}=2(\nu-1)$.

\section{The Mittag-Leffler function}

The in-built MATLAB routine for evaluating the Mittag-Leffler function with two parameters was used for the fitting the modified Mittag-leffler function through the simulation data. The code can be found at: \url{https://www.mathworks.com/matlabcentral/fileexchange/8738-mittag-leffler-function}.

\section{Diffusivity as a function of radius of gyration}

\begin{figure}[tbph]
	\begin{center}
			\resizebox{9.0cm}{!} {\includegraphics[width=8cm,height=!]{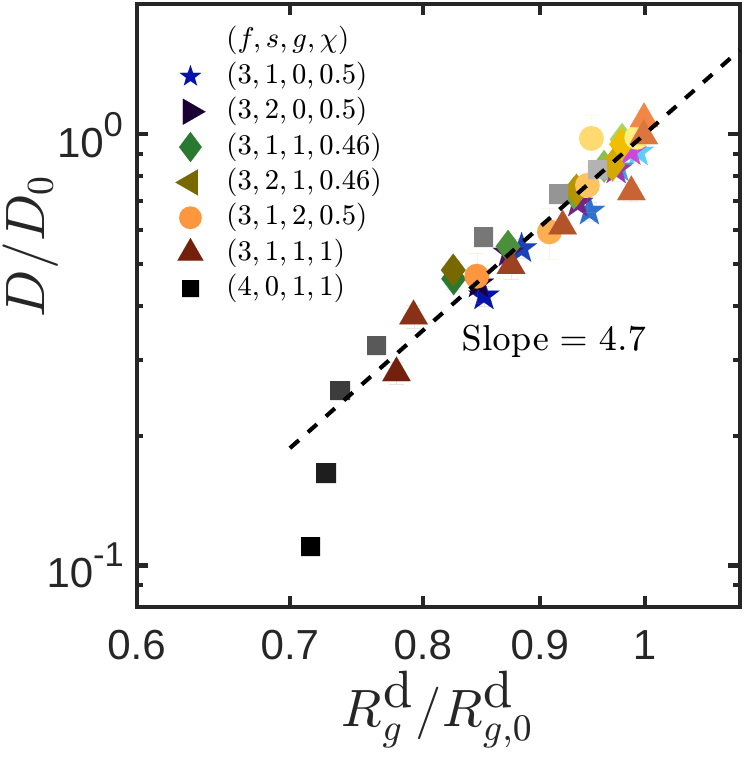}} 		
	\end{center}
	\vspace{-10pt}
	\caption{(Color online) Normalised diffusivity as a function of the normalised radius of gyration for different dendrimer architectures. The dashed line has a slope equal to 4.7.}
\label{fig:diffusivity_vs_Rg}
	\vspace{-15pt}
\end{figure}

According to the predictions of the scaling theory \cite{cai2011mobility}, the long-time diffusivity of a nanoparticle in semidilute polymer solution is given by 
\begin{align}\label{eq:deff_scaling}
    D_{\textrm{t}} \approx \frac{k_BT}{\eta_{\textrm{eff}}(\tau_d)d}
\end{align} 
where $\eta_{\textrm{eff}}(\tau_d)=\eta_{s}(d/\xi)^2$ is the effective viscosity experienced by a nanoparticle of size $d$ in a solution with solvent viscosity $\eta_s$ and correlation length $\xi$. On substituting this in eqn~\eqref{eq:deff_scaling}, the result $D_{\textrm{t}} \propto 1/d^3$ is obtained as pointed out by the reviewer. The size of a nanoparticle remains constant at all polymer concentrations. However, this is not true in the case of dendrimers. The radius of dendrimers decreases with concentration as follows:
\begin{align}\label{eq:rg_scaling2}
    R_{\textrm{g}}^{\textrm{d}} = R_{\textrm{g0}}^{\textrm{d}}\left(\frac{c}{c^{\ast}_{\textrm{d}}}\right)^{\delta}
\end{align}
where $\delta=(1-2\nu)/(2\left(3\nu -1 \right))$. Therefore,
\begin{align}\label{eq:cbycstar}
    \frac{c}{c^{\ast}_{\textrm{d}}}=\left( \frac{R_{\textrm{g}}^{\textrm{d}}}{R_{\textrm{g0}}^{\textrm{d}}} \right) ^ {(1/\delta)}
\end{align}
The diffusivity of most of the dendrimer architectures follows the scaling law for linear chains as given below:
\begin{align}\label{eq:D_scaling2}
    D^{\textrm{d}} = D_{\textrm{0}}^{\textrm{d}} \left( \frac{c}{c^{\ast}_{\textrm{d}}}\right) ^{\omega}
\end{align}
where $\omega=(\nu-1)/(3\nu -1)$. Substituting eqn~\eqref{eq:cbycstar} in eqn~\eqref{eq:D_scaling2}, we have 
\begin{align}\label{eq:D_vs_R}
    \frac{D^{\textrm{d}}}{D_{\textrm{0}}^{\textrm{d}}} =  \left( \frac{R_{\textrm{g}}^{\textrm{d}}}{R_{\textrm{g0}}^{\textrm{d}}} \right) ^ {(\omega/\delta)}
\end{align}
Substituting for $\nu=0.588$ in athermal conditions, $\omega/\delta = 4.7$. Fig~\ref{fig:diffusivity_vs_Rg} shows the variation of normalised diffusivity as a function of normalised dendrimer size. All the dendrimer architectures, except the $f=4$ dendrimer collapse to a universal curve. This shows that the diffusivity scales as $(R_g^d)^{4.7}$ instead of the $-3$ scaling of nanoparticles.


\bibliography{supplement}
\bibliographystyle{aapmrev4-2}